# Path Integral Molecular Dynamics for Exact Quantum Statistics of Multi-Electronic-State Systems

Xinzijian Liu[1] and Jian Liu[1, a)]

1. Beijing National Laboratory for Molecular Sciences, Institute of Theoretical and Computational Chemistry, College of Chemistry and Molecular Engineering, Peking University, Beijing 100871, China

(Invited contribution to the JCP Special Topic Issue on "Nuclear Quantum Effects")

a) Electronic mail: jianliupku@pku.edu.cn

## Abstract

An exact approach to compute physical properties for general multi-electronic-state (MES) systems in thermal equilibrium is presented. The approach is extended from our recent progress on path integral molecular dynamics (PIMD) [J. Chem. Phys. 145, 024103 (2016); 147, 034109 (2017)] for quantum statistical mechanics when a single potential energy surface is involved. We first define an effective potential function that is numerically favorable for MES-PIMD, and then derive corresponding estimators in MES-PIMD for evaluating various physical properties. Its application to several representative one-dimensional and multi-dimensional models demonstrates that MES-PIMD in principle offers a practical tool in either of the diabatic and adiabatic representations for studying exact quantum statistics of complex/large MES systems when the Born-Oppenheimer approximation, Condon approximation, and harmonic bath approximation are broken.

# I. Introduction

Since Feynman's pioneering work[1] in 1953, imaginary time path integral has provided an intriguing physical picture on quantum statistical mechanics[2-4]. When imaginary time path integral is integrated with state-of-art Monte Carlo (MC) or molecular dynamics (MD) techniques [namely path integral MC (PIMC) or path integral MD (PIMD)], it also offers powerful computational tools for studying "real" molecular systems where (nuclear) quantum effects play important roles[5-17]. Since it is not convenient to adjust moves of PIMC for general complex molecular systems, PIMD often offers a more practical approach for "real" systems where quantum exchange effects are not important.

Most imaginary time path integral studies have focused on systems where the Born-Oppenheimer separation of electronic and nuclear degrees of freedom is valid. When nonadiabtic effects become significant, the theoretical framework of imaginary time path integral should be reformulated for multi-electronic-state systems. Most previous investigations used the diabatic representation of the Hamiltonian, because it is convenient to have a diagonal form of the kinetic energy operator. The trace operation of the quantum Boltzmann operator may be expressed as a sum over electronically diabatic states and an integration over the configurational space for the nuclear degrees of freedom, as done by Wolynes[18] and by Cao, Minichino, and Voth[19]. Such a strategy has later been extensively used in evaluating the partition function for nonadiabatic systems with path integral[20-24]. Ananth and Miller employed the Meyer-Miller-Stock-Thoss mapping approach[25-27] to propose an imaginary time path integral method with continuous variables for both the electronic and nuclear degrees of freedom[28]. Instead of the diabatic representation, the adiabatic one was employed by Schmidt and Tully in 2007 to express the Boltzmann operator, where each path integral bead was associated with a surface index that represents which adiabatic potential energy surface the bead lies on[29]. Lu and Zhou further extended the idea to combine PIMD with surface hopping for sampling thermal equilibrium nonadiabatic systems[30].

We have recently proposed a unified thermostat scheme (the "middle" scheme) that offers a simple, robust, efficient, and accurate approach for PIMD, irrespective of whether the thermostat is stochastic or deterministic, when a single potential energy surface is considered (i.e., only an electronic state is involved)[7, 17]. The purpose of the paper is to extend our recent progress on PIMD to develop an indeed practical approach that in principle leads to exact quantum statistics for general multi-electronic-state systems, regardless of whether the diabatic or adiabatic representation is used. The paper is organized as follows: Section II first describes three splitting schemes for the Boltzmann operator in the diabatic representation, proposes an effective potential that is well-defined for general systems, and then derives multi-electronic-state PIMD (MES-PIMD) and corresponding estimators for various physical properties. Section III demonstrates numerical examples with several one-dimensional or multi-dimensional benchmark models. Conclusion remarks are given in Section IV. The adiabatic version of the MES-PIMD approach is also derived in Appendix A.

## II. Theory

### 1. Three splitting schemes in the diabatic representation

Consider an $N$-electronic-state system with Hamiltonian operator in the diabatic representation

$$\hat{\mathbf{H}} = \hat{\mathbf{T}} + \hat{\mathbf{V}} = \frac{1}{2}\hat{\mathbf{P}}^T\mathbf{M}^{-1}\hat{\mathbf{P}} + \mathbf{V}\left(\hat{\mathbf{R}}\right) = \frac{1}{2}\hat{\mathbf{P}}^T\mathbf{M}^{-1}\hat{\mathbf{P}} + \begin{pmatrix} V_{11}\left(\hat{\mathbf{R}}\right) & \cdots & V_{1N}\left(\hat{\mathbf{R}}\right) \\ \vdots & \ddots & \vdots \\ V_{1N}\left(\hat{\mathbf{R}}\right) & \cdots & V_{NN}\left(\hat{\mathbf{R}}\right) \end{pmatrix} \tag{1}$$

where $\mathbf{V}\left(\hat{\mathbf{R}}\right)$ is the symmetric potential energy matrix, $\mathbf{M}$ is the diagonal mass matrix, $\hat{\mathbf{R}}$ and $\hat{\mathbf{P}}$ are the nuclear coordinate and momentum operators, respectively. If the system has $N_{atom}$ atoms, then $3N_{atom}$ is the total number of nuclear degrees of freedom. The canonical partition function is

$$Z = \mathrm{Tr}_{ne}\left[ e^{-\beta \hat{\mathbf{H}}} \right] \tag{2}$$

and a specific thermodynamic property of interest is

$$\left\langle \hat{B} \right\rangle = \frac{1}{Z} \mathrm{Tr}_{ne}\left[ \hat{B} e^{-\beta \hat{\mathbf{H}}} \right] \quad . \tag{3}$$

The trace is over both the nuclear and electronic degrees of freedom in Eqs. (2)-(3). I.e.,

$$\mathrm{Tr}_{ne}\left[ \hat{B} \right] = \mathrm{Tr}_{n}\left[ \mathrm{Tr}_{e}\left[ \hat{B} \right] \right] \tag{4}$$

with

$$\mathrm{Tr}_{e}\left[ \hat{B} \right] = \sum_{n=1}^{N} \langle n | \hat{B} | n \rangle \tag{5}$$

and

$$\mathrm{Tr}_{n}\left[ \hat{B} \right] = \int d\mathbf{R} \langle \mathbf{R} | \hat{B} | \mathbf{R} \rangle \quad . \tag{6}$$

Inserting the resolution of the identity

$$\hat{\mathbf{1}} = \int d\mathbf{R} \sum_{n=1}^{N} |\mathbf{R}, n\rangle \langle \mathbf{R}, n| \tag{7}$$

in Eq. (2) leads to

$$Z = \lim_{P\to\infty} \int d\mathbf{R}_1 \cdots d\mathbf{R}_P \sum_{n_1,\cdots n_P=1}^{N} \prod_{i=1}^{P} \left\langle n_i, \mathbf{R}_i \right| e^{-\beta \hat{\mathbf{H}}/P} \left| n_{i+1}, \mathbf{R}_{i+1} \right\rangle \tag{8}$$

where $n_{P+1} \equiv n_1$ and $\mathbf{R}_{P+1} \equiv \mathbf{R}_1$.

Various splitting schemes may be introduced to evaluate the matrix element $\left\langle n_i, \mathbf{R}_i \right| e^{-\beta \hat{\mathbf{H}}/P} \left| n_{i+1}, \mathbf{R}_{i+1} \right\rangle$ of Eq. (8). In the paper we focus on three types of splitting schemes (for $e^{-\beta \hat{\mathbf{H}}/P}$) that are feasible for general molecular systems. When the splitting

$$e^{-\beta \hat{\mathbf{H}}/P} \approx e^{-\beta \hat{\mathbf{V}}/2P} e^{-\beta \hat{\mathbf{T}}/P} e^{-\beta \hat{\mathbf{V}}/2P} + O\left( \frac{\beta^3}{P^3} \right) \tag{9}$$

is used, the evaluation of the term $e^{-\beta\hat{\mathbf{V}}/2P}$ can be implemented through the diagonalization of the potential energy matrix $\mathbf{V}$. We note it the "diagonalization" method. Alternatively, one may decompose the term $e^{-\beta\hat{\mathbf{V}}/2P}$ into a product of a diagonal term and an off-diagonal one, i.e.,

$$e^{-\beta\hat{\mathbf{H}}/P} \approx e^{-\beta\hat{\mathbf{V}}_{\mathrm{off}}/2P} e^{-\beta\hat{\mathbf{V}}_{\mathrm{diag}}/2P} e^{-\beta\hat{T}/P} e^{-\beta\hat{\mathbf{V}}_{\mathrm{diag}}/2P} e^{-\beta\hat{\mathbf{V}}_{\mathrm{off}}/2P} + O\left(\frac{\beta^3}{P^3}\right) \quad . \tag{10}$$

When the off-diagonal term $e^{-\beta\hat{\mathbf{V}}_{\mathrm{off}}/2P}$ is approximated by its first order Taylor expansion[24, 28, 31]

$$e^{-\beta\hat{\mathbf{H}}/P} \approx \left(\hat{\mathbf{1}} - \frac{\beta}{2P}\hat{\mathbf{V}}_{\mathrm{off}}\right) e^{-\beta\hat{\mathbf{V}}_{\mathrm{diag}}/2P} e^{-\beta\hat{T}/P} e^{-\beta\hat{\mathbf{V}}_{\mathrm{diag}}/2P} \left(\hat{\mathbf{1}} - \frac{\beta}{2P}\hat{\mathbf{V}}_{\mathrm{off}}\right) + O\left(\frac{\beta^2}{P^2}\right) \quad , \tag{11}$$

it is denoted the "first-order expansion" method. Note that the off-diagonal matrix $\hat{\mathbf{V}}_{\mathrm{off}}$ in Eq. (10) is a sum of matrices

$$\hat{\mathbf{V}}_{\mathrm{off}} = \sum_{i=1}^{N} \sum_{j=1}^{i-1} \hat{\mathbf{D}}^{(ij)} \tag{12}$$

Here the element in the *i*-th row and *j*-th column of matrix $\mathbf{D}^{(ij)}$ is equal to $V_{ij}\left(\hat{\mathbf{R}}\right)$, so is that in the *j*-th row and *i*-th column, while all other elements are zero. It is trivial to verify the elements of the matrix $\mathbf{W}^{(ij)} = e^{-\beta\hat{\mathbf{D}}^{(ij)}/2P}$ are

$$\mathbf{W}_{mn}^{(ij)} = \begin{cases} 1 & (m = n \neq i \ \text{ and } \ m = n \neq j) \\ \cosh\left[-\dfrac{\beta}{2P} V_{ij}\left(\hat{\mathbf{R}}\right)\right] & (m = n = i \ \text{ or } \ m = n = j) \\ \sinh\left[-\dfrac{\beta}{2P} V_{ij}\left(\hat{\mathbf{R}}\right)\right] & (m = i, n = j \ \text{ or } \ m = j, n = i) \\ 0 & \text{otherwise} \end{cases} \quad . \tag{13}$$

The off-diagonal term $e^{-\beta\hat{\mathbf{V}}_{\mathrm{off}}/2P}$ may then be decomposed into a product of matrices $\left\{\mathbf{W}^{(ij)}\right\}$. Eq. (10) now becomes

$$e^{-\beta\hat{\mathbf{H}}/P} \approx \prod_{i=1}^{N} \prod_{j=1}^{i-1} \mathbf{W}^{(ij)}\left(\hat{\mathbf{R}}\right) e^{-\beta\hat{\mathbf{V}}_{\mathrm{diag}}/2P} e^{-\beta\hat{T}/P} e^{-\beta\hat{\mathbf{V}}_{\mathrm{diag}}/2P} \prod_{i=N}^{1} \prod_{j=i-1}^{1} \mathbf{W}^{(ij)}\left(\hat{\mathbf{R}}\right) + O\left(\frac{\beta^3}{P^3}\right) \quad , \tag{14}$$

which is denoted the "hyperbolic function" method. The decomposition in Eq. (14) was first used in Ref. [30].

One may show the ascendant order for the error of the splitting is

$$\text{“diagonalization”} < \text{“hyperbolic function”} < \text{“first-order expansion”} \quad . \tag{15}$$

In the nonadiabatic limit where all off-diagonal elements $V_{ij}(\mathbf{R}) \to 0$, the three splitting methods are expected to demonstrate similar numerical performance as the number of beads $P$ changes. This is because the error in the splitting scheme [Eq. (9), (10), or (11)] is also related to the values of the off-diagonal elements.

It is straightforward to see that the ascendant order for the numerical cost of the splitting is

$$\text{“first-order expansion”} < \text{“hyperbolic function”} < \text{“diagonalization”} \quad . \tag{16}$$

Because the calculation of the force often takes the dominant effort for simulating “real” molecular systems, the difference among the three splitting methods on the computation cost for these systems is expected to be marginal.

Substituting Eq. (9), Eq. (11) or Eq. (14) into Eq. (8), one obtains a unified form for the partition function

$$\begin{aligned} Z = \lim_{P\to\infty} \left| \frac{P\mathbf{M}}{2\pi\beta\hbar^2} \right|^{P/2} \int d\mathbf{R}_1 \cdots d\mathbf{R}_P \exp\left[ -\frac{\beta}{2}\omega_P^2 \sum_{i=1}^{P} (\mathbf{R}_i - \mathbf{R}_{i+1})^T \mathbf{M} (\mathbf{R}_i - \mathbf{R}_{i+1}) \right] \\ \times \mathrm{Tr}_e \left[ \prod_{i=1}^{P} \mathbf{O}(\mathbf{R}_i)^T \mathbf{O}(\mathbf{R}_i) \right] \end{aligned} \tag{17}$$

with $\omega_P = \sqrt{P} / \beta\hbar$. When the diagonalization method is employed, we have

$$\mathbf{O}(\mathbf{R}) = e^{-\beta \mathbf{V}(\mathbf{R})/2P} = \mathbf{T}(\mathbf{R}) e^{-\beta \mathbf{\Lambda}(\mathbf{R})/2P} \mathbf{T}(\mathbf{R})^T \quad , \tag{18}$$

where the orthogonal matrix $\mathbf{T}(\mathbf{R})$ and the diagonal matrix $\mathbf{\Lambda}(\mathbf{R})$ are given by the eigen-decomposition of $\mathbf{V}(\mathbf{R})$,

$$\mathbf{V}(\mathbf{R}) = \mathbf{T}(\mathbf{R}) \mathbf{\Lambda}(\mathbf{R}) \mathbf{T}(\mathbf{R})^T \quad . \tag{19}$$

I.e., the diagonal elements $\{\lambda_k(\mathbf{R}), k = 1, \cdots, N\}$ of $\mathbf{\Lambda}(\mathbf{R})$ are the eigenvalues of $\mathbf{V}(\mathbf{R})$, which is the adiabatic state potential energy.

In Eq. (17) we obtain

$$\mathbf{O}(\mathbf{R}) = \left[\mathbf{1} - \frac{\beta}{2P}\mathbf{V}_{\mathrm{off}}(\mathbf{R})\right] e^{-\beta \mathbf{V}_{\mathrm{diag}}(\mathbf{R})/2P} \tag{20}$$

for the first-order expansion method, and

$$\mathbf{O}(\mathbf{R}) = \prod_{i=1}^{N}\prod_{j=1}^{i-1} \mathbf{W}^{(ij)}(\mathbf{R})\, e^{-\beta \mathbf{V}_{\mathrm{diag}}(\mathbf{R})/2P} \tag{21}$$

for the hyperbolic function method.

For convenience we define

$$\mathbf{\Theta}(\mathbf{R}) = \mathbf{O}(\mathbf{R})^{T}\, \mathbf{O}(\mathbf{R}) \quad . \tag{22}$$

We then have

$$\mathbf{\Theta}(\mathbf{R}) = \mathbf{T}(\mathbf{R}) e^{-\beta \mathbf{\Lambda}(\mathbf{R})/P} \mathbf{T}(\mathbf{R})^{T} \tag{23}$$

for the diagonalization method,

$$\mathbf{\Theta}(\mathbf{R}) = e^{-\beta \mathbf{V}_{\mathrm{diag}}(\mathbf{R})/2P} \left[\mathbf{1} - \frac{\beta}{2P}\mathbf{V}_{\mathrm{off}}(\mathbf{R})\right]^{2} e^{-\beta \mathbf{V}_{\mathrm{diag}}(\mathbf{R})/2P} \tag{24}$$

for the first-order expansion method, and

$$\mathbf{\Theta}(\mathbf{R}) = e^{-\beta \mathbf{V}_{\mathrm{diag}}(\mathbf{R})/2P} \left[\prod_{i=1}^{N}\prod_{j=1}^{i-1} \mathbf{W}^{(ij)}(\mathbf{R})\right]^{T} \left[\prod_{i=1}^{N}\prod_{j=1}^{i-1} \mathbf{W}^{(ij)}(\mathbf{R})\right] e^{-\beta \mathbf{V}_{\mathrm{diag}}(\mathbf{R})/2P} \tag{25}$$

for the hyperbolic function method.

**2. Effective potential function in the diabatic representation**

$\mathrm{Tr}_e\left[\prod_{i=1}^{P} \mathbf{\Theta}(\mathbf{R}_i)\right]$ is often *not* positive-definite for general multi-electronic-state systems, regardless of which one of the three splitting methods is employed. If an effective potential function $\phi(\mathbf{R}_1, \cdots, \mathbf{R}_P)$ is given by

$$e^{-\beta\phi(\mathbf{R}_1,\cdots,\mathbf{R}_P)} \equiv \left| \mathrm{Tr}_e \left[ \prod_{i=1}^{P} \mathbf{\Theta}(\mathbf{R}_i) \right] \right| \quad , \tag{26}$$

the coordinates that are close to the region where $\mathrm{Tr}_e \left[ \prod_{i=1}^{P} \mathbf{\Theta}(\mathbf{R}_i) \right]$ changes the sign lead to $\phi(\mathbf{R}_1,\cdots,\mathbf{R}_P) \to \infty$. This presents severe numerical problems for the use of MD for performing imaginary time path integral. (For the same reason it is also challenging to implement MC when Eq. (26) is used for general systems.)

Because $\mathrm{Tr}_e \left[ \prod_{i=1}^{P} e^{-\beta \mathbf{V}_{\mathrm{diag}}(\mathbf{R}_i)/P} \right]$ is always positive-definite, an effective (real-valued) potential function $\phi^{(\mathrm{dia})}(\mathbf{R}_1,\cdots,\mathbf{R}_P)$ may be defined by

$$e^{-\beta\phi^{(\mathrm{dia})}(\mathbf{R}_1,\cdots,\mathbf{R}_P)} \equiv \mathrm{Tr}_e \left[ \prod_{i=1}^{P} e^{-\beta \mathbf{V}_{\mathrm{diag}}(\mathbf{R}_i)/P} \right] \quad . \tag{27}$$

[$\phi^{(\mathrm{dia})}(\mathbf{R}_1,\cdots,\mathbf{R}_P)$ has no singularity.] The partition function [Eq. (17)] may then be expressed as

$$Z = \lim_{P\to\infty} \left| \frac{P\mathbf{M}}{2\pi\beta\hbar^2} \right|^{P/2} \int d\mathbf{R}_1 \cdots d\mathbf{R}_P \exp\left[ -\beta U_{eff}^{(\mathrm{dia})}(\mathbf{R}_1,\cdots,\mathbf{R}_P) \right] \tilde{Z}^{(\mathrm{dia})}(\mathbf{R}_1,\cdots,\mathbf{R}_P) \tag{28}$$

with the estimator for the partition function

$$\tilde{Z}^{(\mathrm{dia})}(\mathbf{R}_1,\cdots,\mathbf{R}_P) = \frac{\mathrm{Tr}_e \left[ \prod_{i=1}^{P} \mathbf{\Theta}(\mathbf{R}_i) \right]}{\mathrm{Tr}_e \left[ \prod_{i=1}^{P} e^{-\beta \mathbf{V}_{\mathrm{diag}}(\mathbf{R}_i)/P} \right]} \tag{29}$$

and

$$U_{eff}^{(\mathrm{dia})}(\mathbf{R}_1,\cdots,\mathbf{R}_P) = \frac{1}{2}\omega_P^2 \sum_{i=1}^{P} (\mathbf{R}_i - \mathbf{R}_{i+1})^T \mathbf{M} (\mathbf{R}_i - \mathbf{R}_{i+1}) + \phi^{(\mathrm{dia})}(\mathbf{R}_1,\cdots,\mathbf{R}_P) \quad . \tag{30}$$

When the couplings between different diabatic states vanish, the estimator $\tilde{Z}^{(\mathrm{dia})}(\mathbf{R}_1,\cdots,\mathbf{R}_P) = 1$ and the partition function for the multi-state system Eq. (88) is reduced to

$$Z_{(\mathrm{dia})}^{(\mathrm{no-coup})} = \lim_{P\to\infty} \left| \frac{P\mathbf{M}}{2\pi\beta\hbar^2} \right|^{P/2} \int d\mathbf{R}_1 \cdots d\mathbf{R}_P \exp\left[ -\beta U_{eff}^{(\mathrm{dia})}(\mathbf{R}_1,\cdots,\mathbf{R}_P) \right] \quad , \tag{31}$$

which is a *well-defined* physical quantity

$$Z_{(\text{dia})}^{(\text{no-coup})} = \prod_{j=1}^{N} Z_j^{(\text{dia})} \quad , \tag{32}$$

where $Z_j^{(\text{dia})}$ is the single-electronic-state partition function for the *j*-th electronically diabatic state. That is, the effective potential in Eq. (30) leads to a well-defined (physically meaningful) canonical ensemble at the inverse temperature $\beta$.

## 3. Staging path integral molecular dynamics for multi-electronic-state systems

Applying the staging transformation[13, 14, 32, 33]

$$\begin{aligned} &\mathbf{R}_1 = \boldsymbol{\xi}_1 \\ &\mathbf{R}_j = \boldsymbol{\xi}_j + \frac{j-1}{j}\mathbf{R}_{j+1} + \frac{1}{j}\boldsymbol{\xi}_1 \quad (j = 2, \cdots, P) \end{aligned} \tag{33}$$

to Eq. (28) yields

$$\begin{aligned} Z = \lim_{P\to\infty} \left| \frac{P\mathbf{M}}{2\pi\beta\hbar^2} \right|^{P/2} \int d\boldsymbol{\xi}_1 \cdots d\boldsymbol{\xi}_P \exp\left[ -\frac{\beta}{2}\omega_P^2 \sum_{j=1}^{P} \boldsymbol{\xi}_j^T \bar{\mathbf{M}}_j \boldsymbol{\xi}_j - \beta\phi^{(\text{dia})}(\boldsymbol{\xi}_1, \cdots, \boldsymbol{\xi}_P) \right] \\ \times \tilde{Z}^{(\text{dia})}(\boldsymbol{\xi}_1, \cdots, \boldsymbol{\xi}_P) \end{aligned} \tag{34}$$

where

$$\begin{aligned} &\bar{\mathbf{M}}_1 = 0 \\ &\bar{\mathbf{M}}_j = \frac{j}{j-1}\mathbf{M} \quad (j = 2, \cdots, P) \end{aligned} \quad . \tag{35}$$

Any thermodynamic property in Eq. (3) is expressed as

$$\langle \hat{B} \rangle = \lim_{P\to\infty} \frac{\int d\boldsymbol{\xi}_1 \cdots d\boldsymbol{\xi}_P \tilde{B}^{(\text{dia})}(\boldsymbol{\xi}_1, \cdots, \boldsymbol{\xi}_P) \exp\left[ -\frac{\beta}{2}\omega_P^2 \sum_{j=1}^{P} \boldsymbol{\xi}_j^T \bar{\mathbf{M}}_j \boldsymbol{\xi}_j - \beta\phi^{(\text{dia})}(\boldsymbol{\xi}_1, \cdots, \boldsymbol{\xi}_P) \right]}{\int d\boldsymbol{\xi}_1 \cdots d\boldsymbol{\xi}_P \tilde{Z}^{(\text{dia})}(\boldsymbol{\xi}_1, \cdots, \boldsymbol{\xi}_P) \exp\left[ -\frac{\beta}{2}\omega_P^2 \sum_{j=1}^{P} \boldsymbol{\xi}_j^T \bar{\mathbf{M}}_j \boldsymbol{\xi}_j - \beta\phi^{(\text{dia})}(\boldsymbol{\xi}_1, \cdots, \boldsymbol{\xi}_P) \right]} \quad . \tag{36}$$

with $\tilde{B}^{(\text{dia})}(\boldsymbol{\xi}_1, \cdots, \boldsymbol{\xi}_P)$ as the estimator for operator $\hat{B}$ in the diabatic representation.

One may employ the MD or MC scheme to perform the integrals in Eq. (36). For instance, inserting fictitious momenta $(\mathbf{p}_1, \cdots, \mathbf{p}_P)$ into Eq. (34) leads to

$$Z=\lim_{P\to\infty}\left|\frac{P\mathbf{M}}{2\pi\beta\hbar^2}\right|^{P/2}\prod_{j=1}^{P}\left|\frac{2\pi}{\beta}\tilde{\mathbf{M}}_j\right|^{-1/2}\int d\xi_1\cdots d\xi_P d\mathbf{p}_1\cdots d\mathbf{p}_P\ \tilde{Z}^{(\mathrm{dia})}\left(\xi_1,\cdots,\xi_P\right)$$
$$\times\exp\left\{-\beta\left[\frac{1}{2}\sum_{j=1}^{P}\mathbf{p}_j^T\tilde{\mathbf{M}}_j^{-1}\mathbf{p}_j+\frac{1}{2}\omega_P^2\sum_{j=1}^{P}\xi_j^T\bar{\mathbf{M}}_j\xi_j+\phi^{(\mathrm{dia})}\left(\xi_1,\cdots,\xi_P\right)\right]\right\} \quad (37)$$

where the fictitious masses are chosen as

$$\begin{aligned}\tilde{\mathbf{M}}_1&=\mathbf{M}\\ \tilde{\mathbf{M}}_j&=\bar{\mathbf{M}}_j\quad\left(j=2,\cdots,P\right)\end{aligned}\ . \quad (38)$$

The effective Hamiltonian for Eq. (37) is

$$H_{eff}^{(\mathrm{dia})}\left(\xi_1,\cdots,\xi_P;\mathbf{p}_1,\cdots,\mathbf{p}_P\right)=\frac{1}{2}\sum_{j=1}^{P}\mathbf{p}_j^T\tilde{\mathbf{M}}_j^{-1}\mathbf{p}_j+\frac{1}{2}\omega_P^2\sum_{j=1}^{P}\xi_j^T\bar{\mathbf{M}}_j\xi_j+\phi^{(\mathrm{dia})}\left(\xi_1,\cdots,\xi_P\right) \quad (39)$$

Thus the equations of motion are given by

$$\begin{aligned}\dot{\xi}_j&=\tilde{\mathbf{M}}_j^{-1}\mathbf{p}_j\\ \dot{\mathbf{p}}_j&=-\omega_P^2\bar{\mathbf{M}}_j\xi_j-\frac{\partial\phi^{(\mathrm{dia})}}{\partial\xi_j}\end{aligned}\quad\left(j=1,\cdots,P\right)\ . \quad (40)$$

The term $\partial\phi^{(\mathrm{dia})}/\partial\xi_j$ in Eq. (40) is obtained by the chain rule

$$\begin{aligned}\frac{\partial\phi^{(\mathrm{dia})}}{\partial\xi_1}&=\sum_{i=1}^{P}\frac{\partial\phi^{(\mathrm{dia})}}{\partial\mathbf{R}_i}\\ \frac{\partial\phi^{(\mathrm{dia})}}{\partial\xi_j}&=\frac{\partial\phi^{(\mathrm{dia})}}{\partial\mathbf{R}_j}+\frac{j-2}{j-1}\frac{\partial\phi^{(\mathrm{dia})}}{\partial\xi_{j-1}}\quad\left(j=2,\cdots,P\right)\end{aligned} \quad (41)$$

and

$$\frac{\partial\phi^{(\mathrm{dia})}}{\partial\mathbf{R}_j}=\frac{1}{P}\frac{\mathrm{Tr}_e\left[\frac{\partial\mathbf{V}_{\mathrm{diag}}\left(\mathbf{R}_j\right)}{\partial\mathbf{R}_j}\prod_{i=1}^{P}e^{-\beta\mathbf{V}_{\mathrm{diag}}\left(\mathbf{R}_i\right)/P}\right]}{\mathrm{Tr}_e\left[\prod_{i=1}^{P}e^{-\beta\mathbf{V}_{\mathrm{diag}}\left(\mathbf{R}_i\right)/P}\right]}\ . \quad (42)$$

Eq. (36) then becomes

$$\left\langle\hat{B}\right\rangle=\lim_{P\to\infty}\frac{\int d\xi_1\cdots d\xi_P d\mathbf{p}_1\cdots d\mathbf{p}_P\tilde{B}^{(\mathrm{dia})}\left(\xi_1,\cdots,\xi_P\right)\exp\left[-\beta H_{eff}^{(\mathrm{dia})}\left(\xi_1,\cdots,\xi_P;\mathbf{p}_1,\cdots,\mathbf{p}_P\right)\right]}{\int d\xi_1\cdots d\xi_P d\mathbf{p}_1\cdots d\mathbf{p}_P\tilde{Z}^{(\mathrm{dia})}\left(\xi_1,\cdots,\xi_P\right)\exp\left[-\beta H_{eff}^{(\mathrm{dia})}\left(\xi_1,\cdots,\xi_P;\mathbf{p}_1,\cdots,\mathbf{p}_P\right)\right]}\ . \quad (43)$$

Use the bracket $\langle\ \rangle_{H_{eff}^{(\mathrm{dia})}}$ to represent the phase space average with the probability distribution

$\exp\left[-\beta H_{eff}^{(\mathrm{dia})}\left(\boldsymbol{\xi}_1,\cdots,\boldsymbol{\xi}_P;\mathbf{p}_1,\cdots,\mathbf{p}_P\right)\right]$. E. g., the denominator and numerator of Eq. (43) are denoted $\left\langle \tilde{Z}^{(\mathrm{dia})}\left(\boldsymbol{\xi}_1,\cdots,\boldsymbol{\xi}_P\right)\right\rangle_{H_{eff}^{(\mathrm{dia})}}$ and $\left\langle \tilde{B}^{(\mathrm{dia})}\left(\boldsymbol{\xi}_1,\cdots,\boldsymbol{\xi}_P\right)\right\rangle_{H_{eff}^{(\mathrm{dia})}}$, respectively. Eq. (43) is then recast as

$$\left\langle \hat{B}\right\rangle = \lim_{P\to\infty}\frac{\left\langle \tilde{B}^{(\mathrm{dia})}\left(\boldsymbol{\xi}_1,\cdots,\boldsymbol{\xi}_P\right)\right\rangle_{H_{eff}^{(\mathrm{dia})}}}{\left\langle \tilde{Z}^{(\mathrm{dia})}\left(\boldsymbol{\xi}_1,\cdots,\boldsymbol{\xi}_P\right)\right\rangle_{H_{eff}^{(\mathrm{dia})}}} \quad . \tag{44}$$

Below we show the expression of the estimator $\tilde{B}^{(\mathrm{dia})}$ in Eq. (36) or Eq. (43) for various physical properties. When $\hat{B}$ is an operator dependent of the nuclear coordinate and the electronic state, the estimator is

$$\tilde{B}^{(\mathrm{dia})}\left(\mathbf{R}_1,\cdots,\mathbf{R}_P\right)=\frac{\frac{1}{P}\mathrm{Tr}_e\left(\sum_{k=1}^{P}\left\{\left[\prod_{i=1}^{k-1}\boldsymbol{\Theta}\left(\mathbf{R}_i\right)\right]\mathbf{O}\left(\mathbf{R}_k\right)^T\mathbf{B}\left(\mathbf{R}_k\right)\mathbf{O}\left(\mathbf{R}_k\right)\left[\prod_{i=k+1}^{P}\boldsymbol{\Theta}\left(\mathbf{R}_i\right)\right]\right\}\right)}{\mathrm{Tr}_e\left[\prod_{j=1}^{P}e^{-\beta\mathbf{V}_{\mathrm{diag}}\left(\mathbf{R}_j\right)/P}\right]} \tag{45}$$

where $\mathbf{B}\left(\mathbf{R}\right)$ is an $N\times N$ matrix-valued function of the nuclear coordinate $\mathbf{R}$. For instance, the estimator for the potential energy is

$$\tilde{V}^{(\mathrm{dia})}\left(\mathbf{R}_1,\cdots,\mathbf{R}_P\right)=\frac{\frac{1}{P}\mathrm{Tr}_e\left(\sum_{k=1}^{P}\left\{\left[\prod_{i=1}^{k-1}\boldsymbol{\Theta}\left(\mathbf{R}_i\right)\right]\mathbf{O}\left(\mathbf{R}_k\right)^T\mathbf{V}\left(\mathbf{R}_k\right)\mathbf{O}\left(\mathbf{R}_k\right)\left[\prod_{i=k+1}^{P}\boldsymbol{\Theta}\left(\mathbf{R}_i\right)\right]\right\}\right)}{\mathrm{Tr}_e\left[\prod_{j=1}^{P}e^{-\beta\mathbf{V}_{\mathrm{diag}}\left(\mathbf{R}_j\right)/P}\right]} \quad . \tag{46}$$

When $\hat{B}=\frac{1}{2}\hat{\mathbf{P}}^T\mathbf{M}^{-1}\hat{\mathbf{P}}$ is the nuclear kinetic energy operator, the primitive estimator is

$$\tilde{K}_{\mathrm{prim}}^{(\mathrm{dia})}\left(\mathbf{R}_1,\cdots,\mathbf{R}_P\right)=\tilde{Z}^{(\mathrm{dia})}\left[\frac{3N_{atom}P}{2\beta}-\frac{1}{2}\omega_P^2\sum_{i=1}^{P}\left(\mathbf{R}_i-\mathbf{R}_{i+1}\right)^T\mathbf{M}\left(\mathbf{R}_i-\mathbf{R}_{i+1}\right)\right] \tag{47}$$

and the virial version is

$$\tilde{K}_{\mathrm{vir}}^{(\mathrm{dia})}\left(\mathbf{R}_1,\cdots,\mathbf{R}_P\right)=\frac{3N_{atom}}{2\beta}\tilde{Z}^{(\mathrm{dia})}-\frac{1}{2\beta}\sum_{j=1}^{P}\left(\mathbf{R}_j-\mathbf{R}^*\right)^T\frac{\frac{\partial}{\partial\mathbf{R}_j}\left\{\mathrm{Tr}_e\left[\prod_{i=1}^{P}\left[\mathbf{O}\left(\mathbf{R}_i\right)^T\mathbf{O}\left(\mathbf{R}_i\right)\right]\right]\right\}}{\mathrm{Tr}_e\left[\prod_{i=1}^{P}e^{-\beta\mathbf{V}_{\mathrm{diag}}\left(\mathbf{R}_i\right)/P}\right]}, \tag{48}$$

where

$$\mathbf{R}^*=\mathbf{R}_c\equiv\frac{1}{P}\sum_{i=1}^{P}\mathbf{R}_i \tag{49}$$

or $\mathbf{R}^*$ can be any one of the $P$ beads, i.e.,

$$\mathbf{R}^* = \mathbf{R}_k \tag{50}$$

where $k \in \{1, 2, \cdots, P\}$ and $k$ is a fixed number in Eq. (48).

In Eq. (48) it is easy to obtain the analytic expression of $\partial \mathbf{O} / \partial \mathbf{R}$ for the first-order expansion method from Eq. (20) or for the hyperbolic function method from Eq. (21). When the diagonalization method is used, we may numerically evaluate $\partial \mathbf{O} / \partial \mathbf{R}$ with the Taylor series

$$\begin{aligned} \frac{\partial \mathbf{O}}{\partial \mathbf{R}} &= \sum_{m=1}^{\infty} \frac{1}{m!} \left( -\frac{\beta}{2P} \right)^m \sum_{j=0}^{m-1} \mathbf{V}(\mathbf{R})^j \frac{\partial \mathbf{V}}{\partial \mathbf{R}} \mathbf{V}(\mathbf{R})^{m-j-1} \\ &= \mathbf{T}(\mathbf{R}) \left\{ \sum_{m=1}^{\infty} \frac{1}{m!} \left( -\frac{\beta}{2P} \right)^m \sum_{j=0}^{m-1} \mathbf{\Lambda}(\mathbf{R})^j \left[ \mathbf{T}(\mathbf{R})^T \frac{\partial \mathbf{V}}{\partial \mathbf{R}} \mathbf{T}(\mathbf{R}) \right] \mathbf{\Lambda}(\mathbf{R})^{m-j-1} \right\} \mathbf{T}(\mathbf{R})^T \end{aligned} . \tag{51}$$

It is, however, difficult to numerically converge the calculation when $\{ |\beta \lambda_k(\mathbf{R}) / 2P|, \ k = 1, \cdots, N \}$ are large. A trick to solve this problem is to first evaluate $\frac{\partial}{\partial \mathbf{R}} e^{-\beta \mathbf{V}(\mathbf{R})/2P/2^L}$ instead, where $L$ is an integer that is large enough such that $\{ |\beta \lambda_k(\mathbf{R}) / 2P / 2^L|, k = 1, \cdots, N \}$ are small enough to guarantee the numerical convergence of Eq. (51) for $\frac{\partial}{\partial \mathbf{R}} e^{-\beta \mathbf{V}(\mathbf{R})/2P/2^L}$. We then employ

$$\frac{\partial}{\partial \mathbf{R}} e^{-\varepsilon \mathbf{V}(\mathbf{R})} = \left[ \frac{\partial}{\partial \mathbf{R}} e^{-\varepsilon \mathbf{V}(\mathbf{R})/2} \right] e^{-\varepsilon \mathbf{V}(\mathbf{R})/2} + e^{-\varepsilon \mathbf{V}(\mathbf{R})/2} \left[ \frac{\partial}{\partial \mathbf{R}} e^{-\varepsilon \mathbf{V}(\mathbf{R})/2} \right] \tag{52}$$

recursively (i.e., $L$ times) to obtain $\frac{\partial \mathbf{O}}{\partial \mathbf{R}} = \frac{\partial}{\partial \mathbf{R}} e^{-\beta \mathbf{V}(\mathbf{R})/2P}$.

The primitive and virial estimators for the heat capacity $C_V = \frac{\partial}{\partial T} \langle \hat{\mathbf{H}} \rangle$ may be derived by extending the work in Ref. [34] for multi-electronic-state systems. It is straightforward to show the primitive estimator is

$$
\begin{aligned}
C_V^{\text{prim}} &= -\frac{3}{2} N_{atom} P k_B \\
&+ \frac{k_B \beta^2}{\left\langle \tilde{Z}^{(\text{dia})} \right\rangle_{H_{eff}^{(\text{dia})}}} \left( \frac{2}{\beta} \left\langle \tilde{K}_{\text{prim}}^{(\text{dia})} \right\rangle_{H_{eff}^{(\text{dia})}} + \left\langle \tilde{K}_{\text{prim}} \tilde{K}_{\text{prim}}^{(\text{dia})} \right\rangle_{H_{eff}^{(\text{dia})}} + 2 \left\langle \tilde{K}_{\text{prim}} \tilde{V}^{(\text{dia})} \right\rangle_{H_{eff}^{(\text{dia})}} - \left\langle \tilde{E}_{\text{prim}}^{(\text{dia})} \right\rangle_{H_{eff}^{(\text{dia})}}^2 \right) \\
&+ \frac{k_B \beta^2}{\left\langle \tilde{Z}^{(\text{dia})} \right\rangle_{H_{eff}^{(\text{dia})}}} \left\langle \frac{\theta_1(\mathbf{R}_1, \cdots, \mathbf{R}_P) + \theta_2(\mathbf{R}_1, \cdots, \mathbf{R}_P)}{\text{Tr}_e \left[ \prod_{j=1}^{P} e^{-\beta \mathbf{V}_{\text{diag}}(\mathbf{R}_j)/P} \right]} \right\rangle_{H_{eff}^{(\text{dia})}}
\end{aligned} \tag{53}
$$

where $\tilde{K}_{\text{prim}}$ represents

$$
\tilde{K}_{\text{prim}}(\mathbf{R}_1, \cdots, \mathbf{R}_P) = \frac{3 N_{atom} P}{2\beta} - \frac{1}{2} \omega_P^2 \sum_{i=1}^{P} (\mathbf{R}_i - \mathbf{R}_{i+1})^T \mathbf{M} (\mathbf{R}_i - \mathbf{R}_{i+1}) \ , \tag{54}
$$

$\tilde{E}_{\text{prim}}^{(\text{dia})}$ is the primitive estimator for the total energy

$$
\tilde{E}_{\text{prim}}^{(\text{dia})} = \tilde{K}_{\text{prim}}^{(\text{dia})} + \tilde{V}^{(\text{dia})} \ , \tag{55}
$$

and $\theta_1$ and $\theta_2$ are given by

$$
\begin{aligned}
\theta_1(\mathbf{R}_1, \cdots, \mathbf{R}_P) = \frac{2}{P^2} \text{Tr}_e \Bigg( \sum_{j=1}^{P} \sum_{k=j+1}^{P} \Bigg\{ \left[ \prod_{i=1}^{j-1} \mathbf{\Theta}(\mathbf{R}_i) \right] \mathbf{O}(\mathbf{R}_j)^T \mathbf{V}(\mathbf{R}_j) \mathbf{O}(\mathbf{R}_j) \left[ \prod_{i=j+1}^{k-1} \mathbf{\Theta}(\mathbf{R}_i) \right] \\
\times \mathbf{O}(\mathbf{R}_k)^T \mathbf{V}(\mathbf{R}_k) \mathbf{O}(\mathbf{R}_k) \left[ \prod_{i=k+1}^{P} \mathbf{\Theta}(\mathbf{R}_i) \right] \Bigg\} \Bigg)
\end{aligned} \tag{56}
$$

and

$$
\theta_2(\mathbf{R}_1, \cdots, \mathbf{R}_P) = \frac{1}{P^2} \text{Tr}_e \left( \sum_{k=1}^{P} \left\{ \left[ \prod_{i=1}^{k-1} \mathbf{\Theta}(\mathbf{R}_i) \right] \mathbf{O}(\mathbf{R}_k)^T \mathbf{V}(\mathbf{R}_k)^2 \mathbf{O}(\mathbf{R}_k) \left[ \prod_{i=j+1}^{k-1} \mathbf{\Theta}(\mathbf{R}_i) \right] \right\} \right) \ . \tag{57}
$$

The virial estimator for the heat capacity is

$$
\begin{aligned}
C_V^{\text{vir}} &= \frac{k_B \beta^2}{\left\langle \tilde{Z}^{(\text{dia})} \right\rangle_{H_{eff}^{(\text{dia})}}} \Bigg( \left\langle \tilde{K}_{\text{prim}} \tilde{K}_{\text{vir}}^{(\text{dia})} \right\rangle_{H_{eff}^{(\text{dia})}} + \left\langle \tilde{K}_{\text{prim}} \tilde{V}^{(\text{dia})} \right\rangle_{H_{eff}^{(\text{dia})}} - \left\langle \tilde{E}_{\text{vir}}^{(\text{dia})} \right\rangle_{H_{eff}^{(\text{dia})}} \left\langle \tilde{E}_{\text{prim}}^{(\text{dia})} \right\rangle_{H_{eff}^{(\text{dia})}} \\
&\qquad + \frac{3 N_{atom}}{2\beta} \left\langle \tilde{V}^{(\text{dia})} \right\rangle_{H_{eff}^{(\text{dia})}} + \frac{1}{\beta} \left\langle \tilde{K}_{\text{vir}}^{(\text{dia})} \right\rangle_{H_{eff}^{(\text{dia})}} \Bigg) \\
&+ \frac{k_B \beta^2}{\left\langle \tilde{Z}^{(\text{dia})} \right\rangle_{H_{eff}^{(\text{dia})}}} \left\langle \frac{\theta_1 + \theta_2 + \theta_3(\mathbf{R}_1, \cdots, \mathbf{R}_P) + \theta_4(\mathbf{R}_1, \cdots, \mathbf{R}_P) + \theta_5(\mathbf{R}_1, \cdots, \mathbf{R}_P)}{\text{Tr}_e \left[ \prod_{j=1}^{P} e^{-\beta \mathbf{V}_{\text{diag}}(\mathbf{R}_j)/P} \right]} \right\rangle_{H_{eff}^{(\text{dia})}}
\end{aligned} \tag{58}
$$

where $\tilde{E}_{\text{vir}}^{(\text{dia})}$ is the virial estimator for the total energy

$$\tilde{E}_{\text{vir}}^{(\text{dia})} = \tilde{K}_{\text{vir}}^{(\text{dia})} + \tilde{V}^{(\text{dia})} \quad , \tag{59}$$

$\theta_1$ and $\theta_2$ are given by Eqs. (56)-(57), $\theta_3$, $\theta_4$ and $\theta_5$ are

$$\begin{aligned}\theta_3\left(\mathbf{R}_1,\cdots,\mathbf{R}_P\right) = -\frac{1}{2\beta P}\text{Tr}_e\Bigg\{\sum_{j=1}^{P}\sum_{k=1}^{j-1}\Bigg(\left[\prod_{i=1}^{k-1}\mathbf{\Theta}\left(\mathbf{R}_i\right)\right]\mathbf{O}\left(\mathbf{R}_k\right)^T\mathbf{V}\left(\mathbf{R}_k\right)\mathbf{O}\left(\mathbf{R}_k\right)\left[\prod_{i=k+1}^{j-1}\mathbf{\Theta}\left(\mathbf{R}_i\right)\right]\Bigg)\\ \times\left[\left(\mathbf{R}_j-\mathbf{R}^*\right)^T\frac{\partial\mathbf{\Theta}\left(\mathbf{R}_j\right)}{\partial\mathbf{R}_j}\right]\left[\prod_{i=j+1}^{P}\mathbf{\Theta}\left(\mathbf{R}_i\right)\right]\Bigg\}\end{aligned} \quad , \tag{60}$$

$$\begin{aligned}\theta_4\left(\mathbf{R}_1,\cdots,\mathbf{R}_P\right) = -\frac{1}{2\beta P}\text{Tr}_e\Bigg(\sum_{j=1}^{P}\left[\prod_{i=1}^{j-1}\mathbf{\Theta}\left(\mathbf{R}_i\right)\right]\Bigg\{\left(\mathbf{R}_j-\mathbf{R}^*\right)^T\frac{\partial}{\partial\mathbf{R}_j}\left[\mathbf{O}\left(\mathbf{R}_j\right)^T\mathbf{V}\left(\mathbf{R}_j\right)\mathbf{O}\left(\mathbf{R}_j\right)\right]\Bigg\}\\ \times\left[\prod_{i=j+1}^{P}\mathbf{\Theta}\left(\mathbf{R}_i\right)\right]\Bigg)\end{aligned} \quad , \tag{61}$$

and

$$\begin{aligned}\theta_5\left(\mathbf{R}_1,\cdots,\mathbf{R}_P\right) = -\frac{1}{2\beta P}\text{Tr}_e\Bigg(\sum_{j=1}^{P}\sum_{k=j+1}^{P}\Bigg\{\left[\prod_{i=1}^{j-1}\mathbf{\Theta}\left(\mathbf{R}_i\right)\right]\left[\left(\mathbf{R}_j-\mathbf{R}^*\right)^T\frac{\partial\mathbf{\Theta}\left(\mathbf{R}_j\right)}{\partial\mathbf{R}_j}\right]\\ \times\left[\prod_{i=j+1}^{k-1}\mathbf{\Theta}\left(\mathbf{R}_i\right)\right]\mathbf{O}\left(\mathbf{R}_k\right)^T\mathbf{V}\left(\mathbf{R}_k\right)\mathbf{O}\left(\mathbf{R}_k\right)\left[\prod_{i=k+1}^{P}\mathbf{\Theta}\left(\mathbf{R}_i\right)\right]\Bigg\}\Bigg)\end{aligned} \quad . \tag{62}$$

We then consider the electronic state density matrix, which is a kind of reduced density matrix because the nuclear degrees of freedom are integrated out in Eq. (36). The estimator for the electronic state density matrix element in the diabatic representation $\rho_{ij}^{(\text{ds})}$ is

$$\tilde{\rho}_{ij}^{(\text{dia-ds})}\left(\mathbf{R}_1,\cdots,\mathbf{R}_P\right) = \frac{\frac{1}{P}\text{Tr}_e\left(\sum_{k=1}^{P}\left\{\left[\prod_{l=1}^{k-1}\mathbf{\Theta}\left(\mathbf{R}_l\right)\right]\mathbf{O}\left(\mathbf{R}_k\right)^T\boldsymbol{\rho}_{ij}^{(\text{ds})}\mathbf{O}\left(\mathbf{R}_k\right)\left[\prod_{l=k+1}^{P}\mathbf{\Theta}\left(\mathbf{R}_l\right)\right]\right\}\right)}{\text{Tr}_e\left[\prod_{l=1}^{P}e^{-\beta\mathbf{V}_{\text{diag}}\left(\mathbf{R}_l\right)/P}\right]} \tag{63}$$

where

$$\boldsymbol{\rho}_{ij}^{(\text{ds})} = \frac{1}{2}\left(|i\rangle\langle j| + |j\rangle\langle i|\right) \tag{64}$$

with $|i\rangle$ and $|j\rangle$ as the *i*-th and *j*-th electronically diabatic states in the representation for the Hamiltonian Eq. (1).

When the diagonalization method is employed, it is possible to evaluate the electronic state density matrix element in the adiabatic representation. The transformation between the density matrix in the diabatic representation and that in the adiabatic representation is

$$\boldsymbol{\rho}^{(\mathrm{ads})}(\mathbf{R}) = \mathbf{T}(\mathbf{R})^T \boldsymbol{\rho}^{(\mathrm{ds})}\mathbf{T}(\mathbf{R}) \tag{65}$$

The estimator for the electronic state density matrix element in the adiabatic representation $\rho_{mn}^{(\mathrm{ads})}$ is then given by

$$\begin{aligned} \tilde{\rho}_{mn}^{(\mathrm{dia-ads})}(\mathbf{R}_1,\cdots,\mathbf{R}_P) &= \frac{1}{P}\left(\mathrm{Tr}_e\left[\prod_{j=1}^{P} e^{-\beta \mathbf{V}_{\mathrm{diag}}(\mathbf{R}_j)/P}\right]\right)^{-1} \\ &\times \mathrm{Tr}_e\left(\sum_{k=1}^{P}\left\{\left[\prod_{i=1}^{k-1}\boldsymbol{\Theta}(\mathbf{R}_i)\right]\mathbf{T}(\mathbf{R}_k)e^{-\beta\left[\lambda_m(\mathbf{R}_k)+\lambda_n(\mathbf{R}_k)\right]/2P}\boldsymbol{\rho}_{mn}^{(\mathrm{ads})}(\mathbf{R}_k)\mathbf{T}(\mathbf{R}_k)^T\left[\prod_{i=k+1}^{P}\boldsymbol{\Theta}(\mathbf{R}_i)\right]\right\}\right) \end{aligned} \tag{66}$$

where

$$\boldsymbol{\rho}_{mn}^{(\mathrm{ads})}(\mathbf{R}) = \frac{1}{2}\left(\left|m(\mathbf{R})\right\rangle\left\langle n(\mathbf{R})\right| + \left|n(\mathbf{R})\right\rangle\left\langle m(\mathbf{R})\right|\right) \tag{67}$$

with $\left|m(\mathbf{R})\right\rangle$ and $\left|n(\mathbf{R})\right\rangle$ as the $m$-th and $n$-th electronically adiabatic states when the nuclear coordinate is $\mathbf{R}$.

The coherence length of the electronic state density[35, 36] may be defined by

$$L_{coh} = \frac{\left(\sum_{i,j=1}^{N}\left|\rho_{ij}\right|\right)^2}{N\sum_{i,j=1}^{N}\left|\rho_{ij}\right|^2} \quad . \tag{68}$$

It measures the extent of the off-diagonal elements of the electronic state density matrix[35-37]. It is easy to verify that $L_{coh} \to 1$ in the high-temperature limit, $L_{coh} \to N$ in the complete coherence limit, and $L_{coh} \to 1/N$ for a pure basis state of the employed representation[37]. The value of $L_{coh}$ depends on whether the diabatic or adiabatic representation is used. (See Appendix C.)

In Appendix A we derive the adiabatic version of MES-PIMD and the corresponding estimators for the same physical properties. (Although we employ staging PIMD for demonstration in the paper, it is trivial to follow the same procedure to develop normal-mode PIMD for multi-electronic-

state systems.)

## 4. Thermostat schemes for PIMD

Note that Eq. (40) must be coupled to a thermostatting method to ensure a proper canonical distribution. Eq. (40) may be decomposed into

$$\begin{pmatrix} \dot{\xi}_j \\ \dot{\mathbf{p}}_j \end{pmatrix} = \underbrace{\begin{pmatrix} \tilde{\mathbf{M}}_j^{-1}\mathbf{p}_j \\ \mathbf{0} \end{pmatrix}} + \underbrace{\begin{pmatrix} \mathbf{0} \\ -\omega_P^2 \bar{\mathbf{M}}_j \xi_j - \dfrac{\partial \phi^{(\mathrm{dia})}}{\partial \xi_j} \end{pmatrix}} \quad (j = 1, \cdots, P) \tag{69}$$

The phase space propagator $e^{\mathcal{L}\Delta t}$ (for a finite time interval $\Delta t$) for PIMD with the effective Hamiltonian Eq. (39) is approximated as

$$e^{\mathcal{L}\Delta t} \approx e^{\mathcal{L}^{\mathrm{Middle}}\Delta t} = e^{\mathcal{L}_{\mathbf{p}}\Delta t/2} e^{\mathcal{L}_{\xi}\Delta t/2} e^{\mathcal{L}_T \Delta t} e^{\mathcal{L}_{\xi}\Delta t/2} e^{\mathcal{L}_{\mathbf{p}}\Delta t/2} \quad , \tag{70}$$

in the "middle" scheme[7, 17]. Here the phase space propagator $e^{\mathcal{L}_T \Delta t}$ for the thermostat part is placed in the middle of the conventional velocity-Verlet algorithm. The relevant Kolmogorov operators of $e^{\mathcal{L}_{\xi}\Delta t}$ and $e^{\mathcal{L}_{\mathbf{p}}\Delta t}$ are

$$\mathcal{L}_{\xi}\mathcal{P} = -\mathbf{p}^T \tilde{\mathbf{M}}^{-1} \frac{\partial \mathcal{P}}{\partial \xi} \quad , \tag{71}$$

$$\mathcal{L}_{\mathbf{p}}\mathcal{P} = \left(\frac{\partial U}{\partial \xi}\right)^T \frac{\partial \mathcal{P}}{\partial \mathbf{p}} \quad , \tag{72}$$

Where the phase space density distribution $\mathcal{P} \equiv \mathcal{P}(\xi, \mathbf{p})$, the mass matrix $\tilde{\mathbf{M}} \equiv \left\{\tilde{\mathbf{M}}_j, \, j = \overline{1, P}\right\}$, and $U(\xi) = \frac{1}{2}\omega_P^2 \sum_{j=1}^{P} \xi_j^T \bar{\mathbf{M}}_j \xi_j + \phi^{(\mathrm{dia})}(\xi_1, \cdots, \xi_P)$ with $\xi \equiv \left\{\xi_j, \, j = \overline{1, P}\right\}$. When Langevin dynamics is employed as the thermostat in Eq. (70), its relevant Kolmogorov operator $\mathcal{L}_T$ is

$$\mathcal{L}_T \mathcal{P} = \frac{\partial}{\partial \mathbf{p}} \cdot (\gamma \mathbf{p} \mathcal{P}) + \frac{1}{\beta} \frac{\partial}{\partial \mathbf{p}} \cdot \left(\gamma \mathbf{M} \frac{\partial \mathcal{P}}{\partial \mathbf{p}}\right) \quad , \tag{73}$$

where $\gamma$ is the friction coefficient[7, 17, 38]. It was derived in Ref. [7] that the relevant Kolmogorov operator for the Andersen thermostat[39] is

$$\mathcal{L}_T \mathcal{P} = \nu \left[ \rho_{\mathrm{MB}}(\mathbf{p}) \int_{-\infty}^{\infty} \mathcal{P}(\boldsymbol{\xi}, \mathbf{p}) d\mathbf{p} - \mathcal{P}(\boldsymbol{\xi}, \mathbf{p}) \right] . \tag{74}$$

Here $\rho_{\mathrm{MB}}(\mathbf{p})$ is the Maxwell (or Maxwell-Boltzmann) momentum distribution and $\nu$ is the collision frequency that specifies the coupling strength between the system and the heat bath. All phase space propagators in Eq. (70) may then be exactly obtained[7].

In addition to the decomposition Eq. (69), another decomposition of Eq. (40) often used for PIMD[15, 17, 33, 40] is

$$\begin{pmatrix} \dot{\boldsymbol{\xi}}_j \\ \dot{\mathbf{p}}_j \end{pmatrix} = \underbrace{\begin{pmatrix} \tilde{\mathbf{M}}_j^{-1} \mathbf{p}_j \\ -\omega_P^2 \bar{\mathbf{M}}_j \boldsymbol{\xi}_j \end{pmatrix}} + \underbrace{\begin{pmatrix} \mathbf{0} \\ -\dfrac{\partial \phi^{(\mathrm{dia})}}{\partial \boldsymbol{\xi}_j} \end{pmatrix}} \quad (j = 1, \cdots, P) . \tag{75}$$

$\mathcal{L}_{\mathrm{harm}}$ and $\mathcal{L}_{\phi}$ are the relevant Kolmogorov operators for the first and second terms in the right-hand side (RHS) of Eq. (75), respectively. The phase space propagator $e^{\mathcal{L}\Delta t}$ (for a finite time interval $\Delta t$) for PIMD with the effective Hamiltonian Eq. (39) is then approximated as

$$e^{\mathcal{L}\Delta t} \approx e^{\mathcal{L}_{(\mathrm{harm})}^{\mathrm{Middle}} \Delta t} = e^{\mathcal{L}_{\phi} \Delta t/2} e^{\mathcal{L}_{\mathrm{harm}} \Delta t/2} e^{\mathcal{L}_T \Delta t} e^{\mathcal{L}_{\mathrm{harm}} \Delta t/2} e^{\mathcal{L}_{\phi} \Delta t/2} \tag{76}$$

in the "middle" thermostat scheme[17], and

$$e^{\mathcal{L}\Delta t} \approx e^{\mathcal{L}^{\mathrm{Side}} \Delta t} = e^{\mathcal{L}_T \Delta t/2} e^{\mathcal{L}_{\phi} \Delta t/2} e^{\mathcal{L}_{\mathrm{harm}} \Delta t} e^{\mathcal{L}_{\phi} \Delta t/2} e^{\mathcal{L}_T \Delta t/2} . \tag{77}$$

in the conventional "side" thermostat scheme[15, 17, 33, 40]. Here the two phase space propagators $e^{\mathcal{L}_{\phi}\Delta t}$ and $e^{\mathcal{L}_{\mathrm{harm}}\Delta t}$ may also be analytically solved. It has been demonstrated that the conventional "side" scheme Eq. (77) is less accurate and less efficient than the "middle" scheme Eq. (70) [or Eq. (76)] when only one electronic state is involved[7, 17]. [Eq. (70) is exact in the harmonic limit, while Eq. (76) or Eq. (77) only does so in the free particle limit[17].] In addition to stochastic thermostats, the conclusion is similar for deterministic thermostats such as Nosé-Hoover chains[41] or Nosé-Hoover thermostat[42, 43] for performing PIMD[7].

It is trivial to extend the thermostat algorithms of the "middle" or "side" scheme in Refs. [7, 17] for the effective Hamiltonian Eq. (39) for multi-electronic-state systems, for which we will compare the performance of the two thermostat schemes [i.e., Eq. (70) and Eq. (77)] for MES-PIMD.

## III. Numerical examples

In this section we use several benchmark models to investigate the performance of the three splitting methods of diabatic MES-PIMD (proposed in the previous section) and that of the adiabatic version of MES-PIMD (derived in Appendix A).

### 1. Multi-electronic-state system coupled to a single nuclear degree of freedom

The simplest non-trivial benchmark model is perhaps a multi-electronic-state system coupled to a one-dimensional Morse oscillator. ($3N_{atom}=1$ in this case.) The diabatic potential matrix elements of the Hamiltonian Eq. (1) are

$$\begin{aligned} V_{nn}(R) &= D_{(n)}\left(1-\exp\left[-\sqrt{\frac{M\omega_{(n)}^2}{2D_{(n)}}}\left(R-R_{eq}^{(n)}\right)\right]\right)^2+V_0^{(n)} \\ V_{mn}(R) &= c^{(mn)}\exp\left[-\alpha_1^{(mn)}\left(R-R^{(mn)}\right)^2\right]\cos\left[\alpha_2^{(mn)}\left(R-R^{(mn)}\right)\right] \quad (m\neq n) \end{aligned} \tag{78}$$

where $M$ is the mass, $\left\{\omega_{(n)}, D_{(n)}, R_{eq}^{(n)}, V_0^{(n)}\right\}$ are the frequency, relative dissociation energy, displacement, and minimum potential value of the Morse oscillator on the *n*-th electronically diabatic state, respectively, $R^{(mn)}$ is the intersection of the *m*-th and *n*-th diabatic potential energy surfaces, $\left\{c^{(mn)}, \alpha_1^{(mn)}, \alpha_2^{(mn)}\right\}$ are three parameters for the potential coupling term $V_{mn}$. Because there is *no guarantee* in general systems that the coupling term $V_{mn}$ should maintain the same sign for any values of the nuclear coordinate, a cosine term is used in Eq. (78). When $V_{mn}(\mathbf{R})$ changes the sign as $\mathbf{R}$ varies, $\mathrm{Tr}_e\left[\prod_{i=1}^{P}\mathbf{\Theta}\left(\mathbf{R}_i\right)\right]$ is not always positive-definite (even in a two-state system), leading to the failure of the definition of an effective potential term by using Eq. (26).

By diagonalizing the diabatic potential matrix Eq. (78), the adiabatic potential energy surfaces

$\{\lambda_m(\mathbf{R})\}$ and the overlap matrix $\mathbf{C}(\mathbf{R}_l, \mathbf{R}_{l+1})$ [defined in Appendix A] may be analytically obtained in one-dimensional systems before performing adiabatic MES-PIMD. Note that the element in the $l$-th row and $m$-th column of $\mathbf{C}(\mathbf{R}_i, \mathbf{R}_{i+1})$ is the overlap between the $l$-th adiabatic state with the nuclear coordinate $\mathbf{R}_i$ and the $m$-th adiabatic state with the nuclear coordinate $\mathbf{R}_{i+1}$.

The first suite of examples involve a two-electronic-state system with Eq. (78). Table 1 presents the three typical sets of parameters, for which the corresponding potential matrix elements in Eq. (78) are plotted as functions of the nuclear coordinate in Fig. 1. We first compare the "middle" and "side" thermostat schemes for the first-order expansion method of diabatic MES-PIMD. As demonstrated in Fig. 2, the "middle" scheme increases the accuracy by an order of magnitude over the conventional "side" one, or the "middle" scheme increases the time interval $\Delta t$ by a factor of 5~10 for the same accuracy. Similar behaviors exist in the other methods of diabatic MES-PIMD and in adiabatic MES-PIMD. (Results not shown.)

Fig. 3 demonstrates the results (for the average potential, kinetic energy, heat capacity, and coherence length) of adiabatic MES-PIMD and the three splitting methods of diabatic MES-PIMD as functions of the number of the path integral beads $P$. (Use Model b of Table 1 as an example.) When $P$ is sufficiently large, all four approaches lead to converged results that reproduce the exact data obtained by discrete variable representation (DVR). In the three splitting methods of diabatic MES-PIMD for evaluating most physical properties, the diagonalization method converges the fastest when $P$ is reasonably large, while the first-order expansion method the slowest. This is consistent with the ascendant order of the error in Eq. (15). Adiabatic MES-PIMD yields similar results to those produced by the diagonalization method of diabatic MES-PIMD because they share the same splitting scheme of the Boltzmann operator [Eq. (9)], irrespective of whether the diabatic or adiabatic effective potential term [Eq. (27) or Eq. (87)] is employed for performing PIMD. Interestingly adiabatic MES-PIMD and the three splitting methods of diabatic MES-PIMD demonstrate very similar performance as the number of beads $P$ varies when evaluating the kinetic energy.

Fig. 4 or Fig. 5 demonstrates the results as the temperature or the difference between the minimum of that ground state and that of the excited state $\Delta R = R_{eq}^{(2)} - R_{eq}^{(1)}$ varies. It is shown that MES-PIMD is capable to reproduce the exact results.

The second suite of examples involve a seven-state system with Eq. (78). While the sets of parameters are described in Table 2, the diabatic potential matrix elements are plotted as functions of the nuclear coordinate in Fig. 6. Use the first-order expansion method (of diabatic PIMD) for demonstration for the comparison between the "middle" and "side" thermostat schemes in Fig. 7. It is shown that the "middle" thermostat scheme greatly improves over the "side" one, consistent with what Fig. 2 suggests. Since it is rather tedious to obtain the electronically adiabatic states and potential energy surfaces for the seven-state system, we focus on the numerical performance of diabatic PIMD. Fig. 8 demonstrates that all three splitting methods behave similarly for this system in evaluating the potential energy, the kinetic energy, and the heat capacity. [This suggests that in the seven-state system the off-diagonal elements of the potential matrix (in diabatic representation) are relatively small, i.e., in the nonadiabatic limit.] The coherence length of the electronic state density is an exception—the diagonalization method always converges faster than the other two methods. We then investigate the seven-electronic-state system at different temperatures. Fig. 9 shows that the MES-PIMD results agree well with the exact ones produced by DVR. E. g., the turn-over behavior in the heat capacity as a function of the temperature is well-reproduced by MES-PIMD in Fig. 9c.

Figs. 2 and 7 demonstrate that the “middle” thermostat scheme also performs much better than the “side” one when the multi-electronic-state system is investigated. This is mainly because the effective potential term behaves like a single potential energy surface—the conclusion for PIMD in Refs. [17] and [7] should hold for the MES-PIMD approach as well. We then focus on employing the “middle” scheme for MES-PIMD to study the multi-electronic-state system coupled to many nuclear degrees of freedom.

## 2. Multi-electronic-state system coupled to many nuclear degrees of freedom

A simple case for the multi-electronic-state system coupled to many nuclear degrees of freedom is the popular spin-boson (SB) model of a two-electronic-state nonadiabatic process in a condensed phase environment[44]. In the SB model each nuclear degree of freedom is described by an independent harmonic oscillator. The diabatic potential matrix elements are

$$\begin{aligned} V_{nn}(\mathbf{R}) &= V_0^{(n)} + \sum_j \frac{1}{2} M_{(j)} \omega_{(j,n)}^2 \left( R_{(j)} - R_{eq,(j)}^{(n)} \right)^2 \\ V_{mn}(\mathbf{R}) &= \Delta^{(mn)} \qquad (m \neq n) \end{aligned} \tag{79}$$

where $R_{(j)}$ is the $j$-th element of the nuclear coordinate vector $\mathbf{R}$, $M_{(j)}$ is the mass for the $j$-th nuclear degree of freedom, $\left\{ \omega_{(j,n)}, R_{eq,(j)}^{(n)} \right\}$ are the frequency and displacement of the $j$-th harmonic oscillator on the $n$-th electronically diabatic state, respectively, $\Delta^{(mn)}$ is the constant coupling between the $m$-th and $n$-th diabatic states, and $V_0^{(n)}$ is the minimum of the $n$-th diabatic potential energy surface. We focus on a suite of six SB models that were used to show the subtle difference between the diabatic and adiabatic versions of the symmetrical quasiclassical Meyer-Miller approach[45] when the second-derivative coupling terms are ignored. (The parameters are listed in Table 3.) For the SB models the adiabatic potential energy surfaces $\left\{ \lambda_m(\mathbf{R}) \right\}$ and the overlap matrix $\mathbf{C}(\mathbf{R}_l, \mathbf{R}_{l+1})$ may be analytically obtained.

Figs. 10-15 then employ the six SB models to compare the adiabatic version with the three methods in the diabatic representation on the performance of convergence as a function of the number of beads $P$. Adiabatic MES-PIMD and the three methods of diabatic MES-PIMD behave similarly on evaluating the kinetic energy. When thermodynamic properties such as the potential energy and the heat capacity are estimated, the first-order expansion and hyperbolic function methods (in the diabatic version) perform analogously to the diagonalization method and the adiabatic version. This is because that the nonadiabatic coupling constant $\Delta^{(mn)} = 0.1 \quad (m \neq n)$ in the six SB models is relatively small (i.e., in the nonadiabatic limit). In the nonadiabatic limit the differences between the three splitting methods are greatly reduced. Panels (d) of Figs. 10-15 suggest that adiabatic MES-PIMD and the diagonalization method (of the diabatic version) always perform significantly better than the first-order expansion and hyperbolic function methods in evaluating the coherence length. The converged results generated by the adiabatic version (with the effective Hamiltonian Eq. (94)) agree well with those obtained by the diabatic version (with the effective Hamiltonian Eq. (39)) of MES-PIMD in Figs. 10-15. This indicates that adiabatic MES-PIMD offers a practical tool without necessitating the second-derivative coupling terms.

We then study a multidimensional seven-state model, in which each nuclear degree of freedom is described by an independent Morse oscillator, i.e., the diabatic potential matrix elements are

$$
\begin{aligned}
V_{nn}(R) &= V_0^{(n)} + \sum_j \left\{ D_{(j,n)} \left( 1 - \exp\left[ -\sqrt{\frac{M_{(j)}\omega_{(j,n)}^2}{2D_{(j,n)}}} \left( R_{(j)} - R_{eq,(j)}^{(n)} \right) \right] \right)^2 \right\} \\
V_{mn}(\mathbf{R}) &= \sum_j \left\{ c_{(j)}^{(mn)} \exp\left[ -\alpha_{1,(j)}^{(mn)} \left( R_{(j)} - R_{(j)}^{(mn)} \right)^2 \right] \cos\left[ \alpha_{2,(j)}^{(mn)} \left( R_{(j)} - R_{(j)}^{(mn)} \right) \right] \right\} \quad (m \neq n)
\end{aligned}
\quad . \qquad (80)
$$

Here $V_0^{(n)}$ is the minimum of the $n$-th diabatic potential energy surface, $M_{(j)}$ is the mass for the $j$-th nuclear degree of freedom, $\left\{ \omega_{(j,n)}, D_{(j,n)}, R_{eq,(j)}^{(n)}, V_{0,(j)}^{(n)} \right\}$ are the frequency, relative dissociation energy, displacement, and minimum potential value of the Morse oscillator on the $n$-th electronically diabatic state, respectively, $R_{(j)}^{(mn)}$ is the intersection of the $m$-th and $n$-th diabatic potential energy surfaces in the $j$-th nuclear degree of freedom, $\left\{c_{(j)}^{(mn)}, \alpha_{1,(j)}^{(mn)}, \alpha_{2,(j)}^{(mn)}\right\}$ are three parameters for the potential coupling term $V_{mn}$. The parameters are listed in Table 4. As it involves tedious work to yield the electronically adiabatic states and potential energy surfaces for the seven-state system from its diabatic representation Eq. (80), we concentrate on the three splitting methods of diabatic PIMD in the investigation.

Fig. 16 compares the three methods of diabatic MES-PIMD on the performance of convergence as a function of the number of beads $P$. It is demonstrated that the three methods of diabatic MES-PIMD lead to the same converged results when $P$ is sufficiently large. As the number of beads $P$ increases, the three methods yield similar results for the kinetic energy. The results for other physical properties of the seven-electronic-state system suggest that the diagonalization method demonstrates the best numerical performance in converging the calculations.

In the nonadiabatic limit where all off-diagonal elements $\left\{V_{ij}\left(\mathbf{R}\right)\right\}$ (in the diabatic representation) are relatively small, the difference among the three splitting methods (first-order expansion, hyperbolic function, and diagonalization) of diabatic MES-PIMD is expected to be small as well. Some typical examples are demonstrated in Figs. 8 and 10-15. In MES systems where off-diagonal elements $\left\{V_{ij}\left(\mathbf{R}\right)\right\}$ are reasonably large (e.g., in Figs. 3 and 16), the three splitting methods may show considerably different numerical performances. We may present another example to demonstrate this. Consider the seven-state model system of Fig. 8. We increase the off-diagonal elements in matrix $H_{\mathrm{FMO}}$ of Eq. (123) (i.e., $c^{(mn)} = H_{\mathrm{FMO}}^{(mn)}$ $(m \neq n)$) by a factor of 10 while keeping the diagonal elements and other parameters unchanged in the model. We then plot the results for the modified seven-state system as functions of the number of the path integral beads $P$ in Fig. 17. In contrast to Fig. 8 (in the nonadiabatic limit), Fig. 17 shows that the behaviors of the three splitting methods are more different when the couplings between different electronic states are larger.

When the number of path integral beads $P$ is fixed, the three splitting methods often yield different estimators for the same physical property as well as different configurational distributions of the beads. Use the two-electronic-state model system of Fig. 3 as an example. In Fig. 18 we first use the diagonalization method to sample the path integral beads, and then employ the different estimators derived in the three splitting methods to evaluate the same physical property. Fig. 18 shows that the different estimators lead to significantly different results even when the same configurational distribution of the beads is used. The kinetic energy is an exception. This is because its estimators derived in the three splitting methods are the same. The same numerical behaviors are observed in Fig. 19, where the first-order expansion method is employed to sample the beads while the different estimators yielded by the three splitting methods are used to compute the same physical property. Comparison of Fig. 19 to Fig. 18 suggests that the numerical performance is much more sensitive to the estimator rather than to the configurational distribution of the beads (at least for the model system in Figs. 18-19). Regardless of whether the most inaccurate configurational distribution of the beads (produced by the first-order expansion method) or the most accurate one (obtained by the diagonalization method) is used, the same estimator yields similar results for a physical property (e.g., the potential energy, heat capacity, or coherence length of the electronic state density). As shown in most test cases [Figs. 3b, 8b, and 10b-15b], the three splitting methods produce similar results for the kinetic energy (when the same parameters are used). This is because the estimators (for the kinetic energy) in the three methods share the same expression. It should be stressed that the configurational distribution of the beads may also become important. For instance, the three splitting methods lead to noticeably different results for the kinetic energy in Fig. 16b.

Because the coherence length involves the elements of the electronic state density, it is very sensitive to the accuracy of the splitting scheme (including both the accuracy of the estimator and that of the configurational distribution of the beads). In all cases (even those in the nonadiabatic limit) the diagonalization method is significantly superior to the first-order expansion or hyperbolic function method.

## IV. Conclusion remarks

In the paper we extend the unified theoretical framework for MD/PIMD thermostats[7, 17] to derive a novel practical PIMD approach for studying exact quantum statistics of general multi-electronic-state systems in thermal equilibrium. Both diabatic and adiabatic versions of MES-PIMD are presented. (See Appendix A.) We propose the effective potential term given in Eq. (27) in the diabatic representation [or Eq. (87) in the adiabatic representation] which avoids singularity and is then always well-defined. (See more discussion in Appendix B.) This yields numerically stable PIMD trajectories for sampling the canonical ensample when two or more electronic states are involved. It is shown that in the MES-PIMD approach the "middle" thermostat scheme is much more efficient than the conventional schemes such as the "side" one.

Three splitting methods are proposed for diabatic MES-PIMD. While the estimators for the same physical property in the three splitting methods are often different, their estimators for the kinetic energy are the same. This is the main reason why in most of the benchmark examples the three splitting methods perform similarly in evaluating the kinetic energy but may show considerably different behaviors in converging the results for other physical properties. While the first-order expansion method often converges slowly and the hyperbolic function method improves the convergence performance, the diagonalization method employs the least number of path integral beads to obtain converged results. Since the major computational efforts come from the evaluation of the forces $\partial \mathbf{V}(\mathbf{R})/\partial \mathbf{R}$ in "real" systems, the diagonalization method is the most economic as long as the number of electronic states is not large.

Adiabatic MES-PIMD performs similarly to the diagonalization method, because they involve the same splitting scheme of the Boltzmann operator. While propagation in adiabatic MES-PIMD does not need even the first-derivative coupling terms, evaluation of physical properties through adiabatic MES-PIMD does not require the second-derivative coupling terms even for the virial estimators. It is then expected that adiabatic MES-PIMD offers a useful approach for studying thermal equilibrium MES molecular systems on the fly with modern quantum chemistry calculations.

While a few theoretical approaches have been successfully developed for studying quantum statistics for specific types of multi-electronic-state models (e.g., especially when the nuclear degrees of freedom are described by harmonic bath models)[37, 46-48], the diabatic or adiabatic version of MES-PIMD offers a feasible exact approach for more general complex/large systems in thermal equilibrium when the Born-Oppenheimer approximation, Condon approximation, and harmonic bath approximation are broken. The MES-PIMD approach will also be useful for preparing the necessary initial condition (e.g., initial nuclear configurations) for various multi-state real time dynamics methods[25-27, 45, 48-52, 53, 54-64], as well as for a number of nonadibatic reaction rate theories[18-21, 23, 65], for studying nonadiabatic processes in condensed phase.

**Acknowledgements**

This work was supported by the Ministry of Science and Technology of China (MOST) Grant No. 2016YFC0202803, by the National Natural Science Foundation of China (NSFC) Grants No. 21373018 and No. 21573007, by the Recruitment Program of Global Experts, by Specialized Research Fund for the Doctoral Program of Higher Education No. 20130001110009, and by Special Program for Applied Research on Super Computation of the NSFC-Guangdong Joint Fund (the second phase) under Grant No. U1501501. We acknowledge the Beijing and Tianjin supercomputer centers and the High-performance Computing Platform of Peking University for providing computational resources. This research also used resources of the National Energy Research Scientific Computing Center, a DOE Office of Science User Facility supported by the Office of Science of the U.S. Department of Energy under Contract No. DE-AC02-05CH11231.

## Appendix A: Multi-electronic-state PIMD in the adiabatic representation

When the diagonalization method (in the diabatic representation) is employed, we have

$$\mathrm{Tr}_e\left[\prod_{i=1}^{P}\mathbf{\Theta}\left(\mathbf{R}_i\right)\right]=\mathrm{Tr}_e\left[\prod_{i=1}^{P}e^{-\beta\mathbf{\Lambda}(\mathbf{R}_i)/P}\mathbf{C}\left(\mathbf{R}_i,\mathbf{R}_{i+1}\right)\right] \tag{81}$$

with $\mathbf{C}\left(\mathbf{R}_i,\mathbf{R}_{i+1}\right)\equiv\mathbf{T}\left(\mathbf{R}_i\right)^T\mathbf{T}\left(\mathbf{R}_{i+1}\right)$. It is easy to show that the element in the *l*-th row and *m*-th column of matrix $\mathbf{C}\left(\mathbf{R}_i,\mathbf{R}_{i+1}\right)$ is simply the overlap between the two electronically adiabatic states $\left\langle l\left(\mathbf{R}_i\right)\middle|m\left(\mathbf{R}_{i+1}\right)\right\rangle$. Here $\left|l\left(\mathbf{R}_i\right)\right\rangle$ and $\left|m\left(\mathbf{R}_{i+1}\right)\right\rangle$ are the *l*-th (electronically) adiabatic state with the nuclear coordinate $\mathbf{R}_i$ and *m*-th adiabatic state with the nuclear coordinate $\mathbf{R}_{i+1}$, respectively. When the adiabatic representation is employed, it is possible to evaluate matrix $\mathbf{C}\left(\mathbf{R}_i,\mathbf{R}_{i+1}\right)$ without the knowledge of the orthogonal transformation matrix $\mathbf{T}\left(\mathbf{R}\right)$. In the adiabatic representation the trace over the electronic degrees of freedom is

$$\mathrm{Tr}_e\left[\hat{B}\right]=\sum_{m=1}^{N}\left\langle m(\mathbf{R})\middle|\hat{B}\middle|m(\mathbf{R})\right\rangle \tag{82}$$

and the resolution of the identity is

$$\hat{\mathbf{1}}=\int d\mathbf{R}\sum_{m=1}^{N}\left|\mathbf{R},m(\mathbf{R})\right\rangle\left\langle\mathbf{R},m(\mathbf{R})\right| \quad . \tag{83}$$

Eq. (81) now becomes

$$\mathrm{Tr}_e\left[\prod_{i=1}^{P}\mathbf{\Theta}\left(\mathbf{R}_i\right)\right]=\sum_{m_1,\cdots m_P=1}^{N}\left[\prod_{i=1}^{P}e^{-\beta\lambda_{m_i}(\mathbf{R}_i)/P}\left\langle m_i\left(\mathbf{R}_i\right)\middle|m_{i+1}\left(\mathbf{R}_{i+1}\right)\right\rangle\right] \quad . \tag{84}$$

Note that the summations in the RHS of Eq. (84) are over the electronically adiabatic states instead. The partition function [Eq. (17)] may then be expressed in the adiabatic representation as

$$\begin{aligned}Z=\lim_{P\to\infty}\left|\frac{P\mathbf{M}}{2\pi\beta\hbar^2}\right|^{P/2}\int d\mathbf{R}_1\cdots d\mathbf{R}_P\exp\left[-\frac{\beta}{2}\omega_P^2\sum_{i=1}^{P}\left(\mathbf{R}_i-\mathbf{R}_{i+1}\right)^T\mathbf{M}\left(\mathbf{R}_i-\mathbf{R}_{i+1}\right)\right]\\ \times\sum_{m_1,\cdots m_P=1}^{N}\left[\prod_{i=1}^{P}e^{-\beta\lambda_{m_i}(\mathbf{R}_i)/P}\left\langle m_i\left(\mathbf{R}_i\right)\middle|m_{i+1}\left(\mathbf{R}_{i+1}\right)\right\rangle\right]\end{aligned} \quad . \tag{85}$$

This is equivalent to the expression obtained by Schmidt and Tully[29].

A more compact form of Eq. (85) is

$$Z = \lim_{P\to\infty} \left| \frac{P\mathbf{M}}{2\pi\beta\hbar^2} \right|^{P/2} \int d\mathbf{R}_1 \cdots d\mathbf{R}_P \exp\left[ -\frac{\beta}{2}\omega_P^2 \sum_{i=1}^{P} \left(\mathbf{R}_i - \mathbf{R}_{i+1}\right)^T \mathbf{M}\left(\mathbf{R}_i - \mathbf{R}_{i+1}\right) \right] \times \mathrm{Tr}_e\left[ \prod_{i=1}^{P} e^{-\beta\mathbf{\Lambda}(\mathbf{R}_i)/P} \mathbf{C}\left(\mathbf{R}_i, \mathbf{R}_{i+1}\right) \right] . \quad (86)$$

For the same reason discussed in Section II-2, it is doomed to fail for general systems when the absolute value of $\mathrm{Tr}_e\left[ \prod_{i=1}^{P} e^{-\beta\mathbf{\Lambda}(\mathbf{R}_i)/P} \mathbf{C}\left(\mathbf{R}_i, \mathbf{R}_{i+1}\right) \right]$ is used to define an effective potential term for PIMD (or PIMC). Instead, $\mathrm{Tr}_e\left[ \prod_{i=1}^{P} e^{-\beta\mathbf{\Lambda}(\mathbf{R}_i)/P} \right]$ is always positive-definite, which leads to an effective (real-valued) potential function $\phi^{(\text{adia})}\left(\mathbf{R}_1, \cdots, \mathbf{R}_P\right)$ defined by

$$e^{-\beta\phi^{(\text{adia})}(\mathbf{R}_1,\cdots,\mathbf{R}_P)} \equiv \mathrm{Tr}_e\left[ \prod_{i=1}^{P} e^{-\beta\mathbf{\Lambda}(\mathbf{R}_i)/P} \right] . \quad (87)$$

Here $\phi^{(\text{adia})}\left(\mathbf{R}_1, \cdots, \mathbf{R}_P\right)$ has no singularity. Eq. (86) then becomes

$$Z = \lim_{P\to\infty} \left| \frac{P\mathbf{M}}{2\pi\beta\hbar^2} \right|^{P/2} \int d\mathbf{R}_1 \cdots d\mathbf{R}_P \exp\left[ -\beta U_{eff}^{(\text{adia})}\left(\mathbf{R}_1, \cdots, \mathbf{R}_P\right) \right] \tilde{Z}^{(\text{adia})}\left(\mathbf{R}_1, \cdots, \mathbf{R}_P\right) \quad (88)$$

with the estimator for the partition function

$$\tilde{Z}^{(\text{adia})}\left(\mathbf{R}_1, \cdots, \mathbf{R}_P\right) = \frac{\mathrm{Tr}_e\left[ \prod_{i=1}^{P} e^{-\beta\mathbf{\Lambda}(\mathbf{R}_i)/P} \mathbf{C}\left(\mathbf{R}_i, \mathbf{R}_{i+1}\right) \right]}{\mathrm{Tr}_e\left[ \prod_{i=1}^{P} e^{-\beta\mathbf{\Lambda}(\mathbf{R}_i)/P} \right]} \quad (89)$$

and

$$U_{eff}^{(\text{adia})}\left(\mathbf{R}_1, \cdots, \mathbf{R}_P\right) = \frac{1}{2}\omega_P^2 \sum_{i=1}^{P} \left(\mathbf{R}_i - \mathbf{R}_{i+1}\right)^T \mathbf{M}\left(\mathbf{R}_i - \mathbf{R}_{i+1}\right) + \phi^{(\text{adia})}\left(\mathbf{R}_1, \cdots, \mathbf{R}_P\right) . \quad (90)$$

When the couplings between different adiabatic states vanish, the estimator $\tilde{Z}^{(\text{adia})}\left(\mathbf{R}_1, \cdots, \mathbf{R}_P\right) = 1$ and the partition function for the multi-state system Eq. (88) is reduced to

$$Z^{(\text{no-coup})} = \lim_{P\to\infty} \left| \frac{P\mathbf{M}}{2\pi\beta\hbar^2} \right|^{P/2} \int d\mathbf{R}_1 \cdots d\mathbf{R}_P \exp\left[ -\beta U_{eff}^{(\text{adia})}\left(\mathbf{R}_1, \cdots, \mathbf{R}_P\right) \right] , \quad (91)$$

which is a well-defined physical quantity

$$Z^{(\text{no-coup})} = \prod_{j=1}^{N} Z_j^{(\text{adia})} \quad , \tag{92}$$

where $Z_j^{(\text{adia})}$ is the single-electronic-state partition function for the *j*-th electronically adiabatic state.

Evaluation of any physical property [Eq. (3)] in the adiabatic representation is given by

$$\langle \hat{B} \rangle = \lim_{P\to\infty} \frac{\int d\mathbf{R}_1 \cdots d\mathbf{R}_P \exp\left[-\beta U_{eff}^{(\text{adia})}\left(\mathbf{R}_1,\cdots,\mathbf{R}_P\right)\right] \tilde{B}^{(\text{adia})}\left(\mathbf{R}_1,\cdots,\mathbf{R}_P\right)}{\int d\mathbf{R}_1 \cdots d\mathbf{R}_P \exp\left[-\beta U_{eff}^{(\text{adia})}\left(\mathbf{R}_1,\cdots,\mathbf{R}_P\right)\right] \tilde{Z}^{(\text{adia})}\left(\mathbf{R}_1,\cdots,\mathbf{R}_P\right)} \quad , \tag{93}$$

with $\tilde{B}^{(\text{adia})}\left(\mathbf{R}_1,\cdots,\mathbf{R}_P\right)$ as the estimator for operator $\hat{B}$ in the adiabatic representation. It is trivial to apply the same procedure in Section II-3 to derive MES-PIMD in the adiabatic representation for Eq. (93).

Define the effective Hamiltonian in the adiabatic representation [with the staging coordinates Eq. (33)]

$$H_{eff}^{(\text{adia})}\left(\boldsymbol{\xi}_1,\cdots,\boldsymbol{\xi}_P;\mathbf{p}_1,\cdots,\mathbf{p}_P\right) = \frac{1}{2}\sum_{j=1}^{P} \mathbf{p}_j^T \tilde{\mathbf{M}}_j^{-1} \mathbf{p}_j + \frac{1}{2}\omega_P^2 \sum_{j=1}^{P} \boldsymbol{\xi}_j^T \bar{\mathbf{M}}_j \boldsymbol{\xi}_j + \phi^{(\text{adia})}\left(\boldsymbol{\xi}_1,\cdots,\boldsymbol{\xi}_P\right) \quad . \tag{94}$$

Its equations of motion are

$$\begin{aligned} \dot{\boldsymbol{\xi}}_j &= \tilde{\mathbf{M}}_j^{-1} \mathbf{p}_j \\ \dot{\mathbf{p}}_j &= -\omega_P^2 \bar{\mathbf{M}}_j \boldsymbol{\xi}_j - \frac{\partial \phi^{(\text{adia})}}{\partial \boldsymbol{\xi}_j} \quad \left(j=1,\cdots,P\right) \quad . \end{aligned} \tag{95}$$

The term $\partial \phi^{(\text{adia})} / \partial \boldsymbol{\xi}_j$ in Eq. (40) is given by the chain rule

$$\begin{aligned} \frac{\partial \phi^{(\text{adia})}}{\partial \boldsymbol{\xi}_1} &= \sum_{i=1}^{P} \frac{\partial \phi^{(\text{adia})}}{\partial \mathbf{R}_i} \\ \frac{\partial \phi^{(\text{adia})}}{\partial \boldsymbol{\xi}_j} &= \frac{\partial \phi^{(\text{adia})}}{\partial \mathbf{R}_j} + \frac{j-2}{j-1} \frac{\partial \phi^{(\text{adia})}}{\partial \boldsymbol{\xi}_{j-1}} \quad \left(j=2,\cdots,P\right) \end{aligned} \tag{96}$$

and

$$\frac{\partial \phi^{(\text{adia})}}{\partial \mathbf{R}_i} = \frac{1}{P} \frac{\mathrm{Tr}_e\left[\frac{\partial \boldsymbol{\Lambda}\left(\mathbf{R}_i\right)}{\partial \mathbf{R}_i} \prod_{j=1}^{P} e^{-\beta \boldsymbol{\Lambda}\left(\mathbf{R}_j\right)/P}\right]}{\mathrm{Tr}_e\left[\prod_{j=1}^{P} e^{-\beta \boldsymbol{\Lambda}\left(\mathbf{R}_j\right)/P}\right]} \quad . \tag{97}$$

Eq. (93) then becomes

$$\left\langle \hat{B} \right\rangle = \lim_{P\to\infty} \frac{\left\langle \tilde{B}^{(\text{adia})}\left(\xi_1,\cdots,\xi_P\right)\right\rangle_{H_{eff}^{(\text{adia})}}}{\left\langle \tilde{Z}^{(\text{adia})}\left(\xi_1,\cdots,\xi_P\right)\right\rangle_{H_{eff}^{(\text{adia})}}} \quad . \tag{98}$$

Here the bracket $\langle\ \rangle_{H_{eff}^{(\text{adia})}}$ corresponds to the phase space average with the probability distribution $\exp\left[-\beta H_{eff}^{(\text{adia})}\left(\xi_1,\cdots,\xi_P;\mathbf{p}_1,\cdots,\mathbf{p}_P\right)\right]$, e. g.,

$$\begin{aligned}\left\langle \tilde{B}^{(\text{adia})}\left(\xi_1,\cdots,\xi_P\right)\right\rangle_{H_{eff}^{(\text{adia})}} &= \int d\xi_1\cdots d\xi_P d\mathbf{p}_1\cdots d\mathbf{p}_P \tilde{B}^{(\text{adia})}\left(\xi_1,\cdots,\xi_P\right) \\ &\quad\times\exp\left[-\beta H_{eff}^{(\text{adia})}\left(\xi_1,\cdots,\xi_P;\mathbf{p}_1,\cdots,\mathbf{p}_P\right)\right]\end{aligned} \quad . \tag{99}$$

We then consider the expression of the estimator $\tilde{B}^{(\text{adia})}$ in Eq. (93) or Eq. (98) for different physical properties. When $\hat{B}$ is an operator dependent of the nuclear coordinate and the electronic state, the estimator is

$$\begin{aligned}\tilde{B}^{(\text{adia})}\left(\mathbf{R}_1,\cdots,\mathbf{R}_P\right) = \frac{1}{P}\left(\text{Tr}_e\left[\prod_{i=1}^{P} e^{-\beta\mathbf{\Lambda}(\mathbf{R}_i)/P}\right]\right)^{-1} \text{Tr}_e\Bigg(\sum_{k=1}^{P}\Bigg\{\left[\prod_{i=1}^{k-1}\left(e^{-\beta\mathbf{\Lambda}(\mathbf{R}_i)/P}\mathbf{C}\left(\mathbf{R}_i,\mathbf{R}_{i+1}\right)\right)\right] \\ \times e^{-\beta\mathbf{\Lambda}(\mathbf{R}_k)/2P}\mathbf{B}\left(\mathbf{R}_k\right)e^{-\beta\mathbf{\Lambda}(\mathbf{R}_k)/2P}\mathbf{C}\left(\mathbf{R}_k,\mathbf{R}_{k+1}\right) \\ \times\left[\prod_{i=k+1}^{P} e^{-\beta\mathbf{\Lambda}(\mathbf{R}_i)/P}\mathbf{C}\left(\mathbf{R}_i,\mathbf{R}_{i+1}\right)\right]\Bigg\}\Bigg)\end{aligned} \tag{100}$$

where $\mathbf{B}(\mathbf{R})$ is an $N\times N$ matrix-valued function of the nuclear coordinate in the adiabatic representation. For instance, the estimator for the potential energy is

$$\begin{aligned}\tilde{V}^{(\text{adia})}\left(\mathbf{R}_1,\cdots,\mathbf{R}_P\right) = \frac{1}{P}\left(\text{Tr}_e\left[\prod_{i=1}^{P} e^{-\beta\mathbf{\Lambda}(\mathbf{R}_i)/P}\right]\right)^{-1} \text{Tr}_e\Bigg(\sum_{k=1}^{P}\Bigg\{\left[\prod_{i=1}^{k-1}\left(e^{-\beta\mathbf{\Lambda}(\mathbf{R}_i)/P}\mathbf{C}\left(\mathbf{R}_i,\mathbf{R}_{i+1}\right)\right)\right] \\ \times e^{-\beta\mathbf{\Lambda}(\mathbf{R}_k)/2P}\mathbf{\Lambda}\left(\mathbf{R}_k\right)e^{-\beta\mathbf{\Lambda}(\mathbf{R}_k)/2P}\mathbf{C}\left(\mathbf{R}_k,\mathbf{R}_{k+1}\right) \\ \times\left[\prod_{i=k+1}^{P} e^{-\beta\mathbf{\Lambda}(\mathbf{R}_i)/P}\mathbf{C}\left(\mathbf{R}_i,\mathbf{R}_{i+1}\right)\right]\Bigg\}\Bigg)\end{aligned} \quad . \tag{101}$$

When $\hat{B} = \frac{1}{2}\hat{\mathbf{P}}^T\mathbf{M}^{-1}\hat{\mathbf{P}}$ is the nuclear kinetic energy operator, the primitive estimator is

$$\tilde{K}_{\text{prim}}^{(\text{adia})}\left(\mathbf{R}_1,\cdots,\mathbf{R}_P\right) = \tilde{Z}^{(\text{adia})}\left[\frac{3N_{atom}P}{2\beta} - \frac{1}{2}\omega_P^2\sum_{i=1}^{P}\left(\mathbf{R}_i-\mathbf{R}_{i+1}\right)^T\mathbf{M}\left(\mathbf{R}_i-\mathbf{R}_{i+1}\right)\right] \tag{102}$$

and the virial version is

$$\tilde{K}_{\text{vir}}^{(\text{adia})}\left(\mathbf{R}_1,\cdots,\mathbf{R}_P\right)=\frac{3N_{atom}}{2\beta}\tilde{Z}^{(\text{adia})} - \frac{1}{2\beta}\sum_{j=1}^{P}\left(\mathbf{R}_j-\mathbf{R}^*\right)^T \frac{\frac{\partial}{\partial\mathbf{R}_j}\left\{\mathrm{Tr}_e\left[\prod_{i=1}^{P}e^{-\beta\mathbf{\Lambda}(\mathbf{R}_i)/P}\mathbf{C}\left(\mathbf{R}_i,\mathbf{R}_{i+1}\right)\right]\right\}}{\mathrm{Tr}_e\left[\prod_{i=1}^{P}e^{-\beta\mathbf{\Lambda}(\mathbf{R}_i)/P}\right]}, \quad (103)$$

where $\mathbf{R}^*$ is given by either Eq. (49) or Eq. (50). The derivative term in the RHS of Eq. (103) may be expressed as

$$\begin{aligned}&\frac{\partial}{\partial\mathbf{R}_j}\left\{\mathrm{Tr}_e\left[\prod_{i=1}^{P}e^{-\beta\mathbf{\Lambda}(\mathbf{R}_i)/P}\mathbf{C}\left(\mathbf{R}_i,\mathbf{R}_{i+1}\right)\right]\right\}\\ &=\mathrm{Tr}_e\left[\left(\prod_{i=1}^{j-2}e^{-\beta\mathbf{\Lambda}(\mathbf{R}_i)/P}\mathbf{C}\left(\mathbf{R}_i,\mathbf{R}_{i+1}\right)\right)e^{-\beta\mathbf{\Lambda}(\mathbf{R}_{j-1})/P}\mathbf{\Phi}^{(kin)}\left(\mathbf{R}_{j-1},\mathbf{R}_j,\mathbf{R}_{j+1}\right)\left(\prod_{i=j+1}^{P}e^{-\beta\mathbf{\Lambda}(\mathbf{R}_i)/P}\mathbf{C}\left(\mathbf{R}_i,\mathbf{R}_{i+1}\right)\right)\right]\end{aligned}. \quad (104)$$

Here the element in the *m*-th row and *n*-th column of matrix $\mathbf{\Phi}^{(kin)}\left(\mathbf{R}_{j-1},\mathbf{R}_j,\mathbf{R}_{j+1}\right)$ is

$$\begin{aligned}\Phi_{mn}^{(kin)}\left(\mathbf{R}_{j-1},\mathbf{R}_j,\mathbf{R}_{j+1}\right)=\sum_{k=1}^{N}\Bigg\{&\left\langle m\left(\mathbf{R}_{j-1}\right)\middle|\frac{\partial k\left(\mathbf{R}_j\right)}{\partial\mathbf{R}_j}\right\rangle e^{-\beta\lambda_k(\mathbf{R}_j)/P}\left\langle k\left(\mathbf{R}_j\right)\middle|n\left(\mathbf{R}_{j+1}\right)\right\rangle\\ &-\frac{\beta}{P}\left\langle m\left(\mathbf{R}_{j-1}\right)\middle|k\left(\mathbf{R}_j\right)\right\rangle\frac{\partial\lambda_k\left(\mathbf{R}_j\right)}{\partial\mathbf{R}_j}e^{-\beta\lambda_k(\mathbf{R}_j)/P}\left\langle k\left(\mathbf{R}_j\right)\middle|n\left(\mathbf{R}_{j-1}\right)\right\rangle\\ &+\left\langle m\left(\mathbf{R}_{j-1}\right)\middle|k\left(\mathbf{R}_j\right)\right\rangle e^{-\beta\lambda_k(\mathbf{R}_j)/P}\left\langle\frac{\partial k\left(\mathbf{R}_j\right)}{\partial\mathbf{R}_j}\middle|n\left(\mathbf{R}_{j-1}\right)\right\rangle\Bigg\}\end{aligned}. \quad (105)$$

The primitive estimator for the heat capacity $C_V=\frac{\partial}{\partial T}\left\langle\hat{\mathbf{H}}\right\rangle$ is

$$\begin{aligned}C_V^{\text{prim}}=&-\frac{3}{2}N_{atom}Pk_B\\ &+\frac{k_B\beta^2}{\left\langle\tilde{Z}^{(\text{adia})}\right\rangle_{H_{eff}^{(\text{adia})}}}\left(\frac{2}{\beta}\left\langle\tilde{K}_{\text{prim}}^{(\text{adia})}\right\rangle_{H_{eff}^{(\text{adia})}}+\left\langle\tilde{K}_{\text{prim}}\tilde{K}_{\text{prim}}^{(\text{adia})}\right\rangle_{H_{eff}^{(\text{adia})}}+2\left\langle\tilde{K}_{\text{prim}}\tilde{V}^{(\text{adia})}\right\rangle_{H_{eff}^{(\text{adia})}}-\left\langle\tilde{E}_{\text{prim}}^{(\text{adia})}\right\rangle_{H_{eff}^{(\text{adia})}}^2\right)\\ &+\frac{k_B\beta^2}{\left\langle\tilde{Z}^{(\text{adia})}\right\rangle_{H_{eff}^{(\text{adia})}}}\left\langle\frac{\theta_1^{(\text{adia})}\left(\mathbf{R}_1,\cdots,\mathbf{R}_P\right)+\theta_2^{(\text{adia})}\left(\mathbf{R}_1,\cdots,\mathbf{R}_P\right)}{\mathrm{Tr}_e\left[\prod_{j=1}^{P}e^{-\beta\mathbf{\Lambda}(\mathbf{R}_j)/P}\right]}\right\rangle_{H_{eff}^{(\text{adia})}}\end{aligned} \quad (106)$$

where $\tilde{K}_{\text{prim}}$ is given in Eq. (54) and $\tilde{E}_{\text{prim}}^{(\text{adia})}$ is the primitive estimator for the total energy

$$\tilde{E}_{\text{prim}}^{(\text{adia})}=\tilde{K}_{\text{prim}}^{(\text{adia})}+\tilde{V}^{(\text{adia})} , \quad (107)$$

and $\theta_1^{(\text{adia})}$ and $\theta_2^{(\text{adia})}$ are given by

$$\theta_1^{(\mathrm{adia})}\left(\mathbf{R}_1,\cdots,\mathbf{R}_P\right)=\frac{2}{P^2}\mathrm{Tr}_e\left(\sum_{j=1}^{P}\sum_{k=j+1}^{P}\left\{\left[\prod_{i=1}^{j-1}\left(e^{-\beta\mathbf{\Lambda}(\mathbf{R}_i)/P}\mathbf{C}\left(\mathbf{R}_i,\mathbf{R}_{i+1}\right)\right)\right]\mathbf{\Lambda}\left(\mathbf{R}_j\right)e^{-\beta\mathbf{\Lambda}\left(\mathbf{R}_j\right)/P}\mathbf{C}\left(\mathbf{R}_j,\mathbf{R}_{j+1}\right)\right.\right.$$
$$\times\left[\prod_{i=j+1}^{k-1}\left(e^{-\beta\mathbf{\Lambda}(\mathbf{R}_i)/P}\mathbf{C}\left(\mathbf{R}_i,\mathbf{R}_{i+1}\right)\right)\right]\mathbf{\Lambda}\left(\mathbf{R}_k\right)e^{-\beta\mathbf{\Lambda}(\mathbf{R}_k)/P}\mathbf{C}\left(\mathbf{R}_k,\mathbf{R}_{k+1}\right)\tag{108}$$
$$\left.\left.\times\left[\prod_{i=k+1}^{P}\left(e^{-\beta\mathbf{\Lambda}(\mathbf{R}_i)/P}\mathbf{C}\left(\mathbf{R}_i,\mathbf{R}_{i+1}\right)\right)\right]\right\}\right)$$

and

$$\theta_2^{(\mathrm{adia})}\left(\mathbf{R}_1,\cdots,\mathbf{R}_P\right)=\frac{1}{P^2}\mathrm{Tr}_e\left(\sum_{k=1}^{P}\left\{\left[\prod_{i=1}^{k-1}\left(e^{-\beta\mathbf{\Lambda}(\mathbf{R}_i)/P}\mathbf{C}\left(\mathbf{R}_i,\mathbf{R}_{i+1}\right)\right)\right]\mathbf{\Lambda}\left(\mathbf{R}_k\right)^2e^{-\beta\mathbf{\Lambda}(\mathbf{R}_k)/P}\mathbf{C}\left(\mathbf{R}_k,\mathbf{R}_{k+1}\right)\right.\right.$$
$$\left.\left.\times\left[\prod_{i=k+1}^{P}\left(e^{-\beta\mathbf{\Lambda}(\mathbf{R}_i)/P}\mathbf{C}\left(\mathbf{R}_i,\mathbf{R}_{i+1}\right)\right)\right]\right\}\right).\tag{109}$$

The virial estimator for the heat capacity is

$$C_V^{\mathrm{vir}}=\frac{k_B\beta^2}{\left\langle\tilde{Z}^{(\mathrm{adia})}\right\rangle_{H_{eff}^{(\mathrm{adia})}}}\left(\left\langle\tilde{K}_{\mathrm{prim}}\tilde{K}_{\mathrm{vir}}^{(\mathrm{adia})}\right\rangle_{H_{eff}^{(\mathrm{adia})}}+\left\langle\tilde{K}_{\mathrm{prim}}\tilde{V}^{(\mathrm{adia})}\right\rangle_{H_{eff}^{(\mathrm{adia})}}-\left\langle\tilde{E}_{\mathrm{vir}}^{(\mathrm{adia})}\right\rangle_{H_{eff}^{(\mathrm{adia})}}\left\langle\tilde{E}_{\mathrm{prim}}^{(\mathrm{adia})}\right\rangle_{H_{eff}^{(\mathrm{adia})}}\right.$$
$$\left.+\frac{3N_{atom}}{2\beta}\left\langle\tilde{V}^{(\mathrm{adia})}\right\rangle_{H_{eff}^{(\mathrm{adia})}}+\frac{1}{\beta}\left\langle\tilde{K}_{\mathrm{vir}}^{(\mathrm{adia})}\right\rangle_{H_{eff}^{(\mathrm{adia})}}\right)\tag{110}$$
$$+\frac{k_B\beta^2}{\left\langle\tilde{Z}^{(\mathrm{adia})}\right\rangle_{H_{eff}^{(\mathrm{adia})}}}\left\langle\frac{\theta_1^{(\mathrm{adia})}+\theta_2^{(\mathrm{adia})}+\theta_3^{(\mathrm{adia})}\left(\mathbf{R}_1,\cdots,\mathbf{R}_P\right)+\theta_4^{(\mathrm{adia})}\left(\mathbf{R}_1,\cdots,\mathbf{R}_P\right)+\theta_5^{(\mathrm{adia})}\left(\mathbf{R}_1,\cdots,\mathbf{R}_P\right)}{\mathrm{Tr}_e\left[\prod_{j=1}^{P}e^{-\beta\mathbf{\Lambda}\left(\mathbf{R}_j\right)/P}\right]}\right\rangle_{H_{eff}^{(\mathrm{adia})}}$$

where $\tilde{E}_{\mathrm{vir}}^{(\mathrm{adia})}$ is the virial estimator for the total energy

$$\tilde{E}_{\mathrm{vir}}^{(\mathrm{adia})}=\tilde{K}_{\mathrm{vir}}^{(\mathrm{adia})}+\tilde{V}^{(\mathrm{adia})}\quad,\tag{111}$$

$\theta_1^{(\mathrm{adia})}$ and $\theta_2^{(\mathrm{adia})}$ are given by Eq. (56)-(57), $\theta_3^{(\mathrm{adia})}$, $\theta_4^{(\mathrm{adia})}$ and $\theta_5^{(\mathrm{adia})}$ are

$$\theta_3^{(\mathrm{adia})}\left(\mathbf{R}_1,\cdots,\mathbf{R}_P\right)=-\frac{1}{2\beta P}\sum_{j=1}^{P}\left(\mathbf{R}_j-\mathbf{R}^*\right)^T\mathrm{Tr}_e\left(\sum_{k=1}^{j-1}\left\{\left[\prod_{i=1}^{k-1}\left(e^{-\beta\mathbf{\Lambda}(\mathbf{R}_i)/P}\mathbf{C}\left(\mathbf{R}_i,\mathbf{R}_{i+1}\right)\right)\right]\mathbf{\Lambda}\left(\mathbf{R}_k\right)e^{-\beta\mathbf{\Lambda}(\mathbf{R}_k)/P}\right.\right.$$
$$\times\left[\prod_{i=k}^{j-2}\left(\mathbf{C}\left(\mathbf{R}_i,\mathbf{R}_{i+1}\right)e^{-\beta\mathbf{\Lambda}(\mathbf{R}_{i+1})/P}\right)\right]\mathbf{\Phi}^{(kin)}\left(\mathbf{R}_{j-1},\mathbf{R}_j,\mathbf{R}_{j+1}\right)\quad,\tag{112}$$
$$\left.\left.\times\left[\prod_{i=j+1}^{P}\left(e^{-\beta\mathbf{\Lambda}(\mathbf{R}_i)/P}\mathbf{C}\left(\mathbf{R}_i,\mathbf{R}_{i+1}\right)\right)\right]\right\}\right)$$

$$\theta_4^{(\mathrm{adia})}\left(\mathbf{R}_1,\cdots,\mathbf{R}_P\right)=-\frac{1}{2\beta P}\sum_{j=1}^{P}\left(\mathbf{R}_j-\mathbf{R}^*\right)^T \mathrm{Tr}_e\left(\left[\prod_{i=1}^{j-2}\left(e^{-\beta\mathbf{\Lambda}(\mathbf{R}_i)/P}\mathbf{C}\left(\mathbf{R}_i,\mathbf{R}_{i+1}\right)\right)\right]e^{-\beta\mathbf{\Lambda}\left(\mathbf{R}_{j-1}\right)/P}\right.$$
$$\left.\times\mathbf{\Phi}^{(theta)}\left(\mathbf{R}_{j-1},\mathbf{R}_j,\mathbf{R}_{j+1}\right)\left[\prod_{i=j+1}^{P}\left(e^{-\beta\mathbf{\Lambda}(\mathbf{R}_i)/P}\mathbf{C}\left(\mathbf{R}_i,\mathbf{R}_{i+1}\right)\right)\right]\right) \quad , \quad (113)$$

and

$$\theta_5^{(\mathrm{adia})}\left(\mathbf{R}_1,\cdots,\mathbf{R}_P\right)=-\frac{1}{2\beta P}\sum_{j=1}^{P}\left(\mathbf{R}_j-\mathbf{R}^*\right)^T \mathrm{Tr}_e\left(\sum_{k=j+1}^{P}\left\{\left[\prod_{i=1}^{j-2}\left(e^{-\beta\mathbf{\Lambda}(\mathbf{R}_i)/P}\mathbf{C}\left(\mathbf{R}_i,\mathbf{R}_{i+1}\right)\right)\right]e^{-\beta\mathbf{\Lambda}\left(\mathbf{R}_{j-1}\right)/P}\right.\right.$$
$$\times\mathbf{\Phi}^{(kin)}\left(\mathbf{R}_{j-1},\mathbf{R}_j,\mathbf{R}_{j+1}\right)\left[\prod_{i=j+1}^{k-1}\left(e^{-\beta\mathbf{\Lambda}(\mathbf{R}_i)/P}\mathbf{C}\left(\mathbf{R}_i,\mathbf{R}_{i+1}\right)\right)\right] \quad , \quad (114)$$
$$\left.\left.\times\mathbf{\Lambda}\left(\mathbf{R}_k\right)e^{-\beta\mathbf{\Lambda}(\mathbf{R}_k)/P}\left[\prod_{i=k+1}^{P}\left(e^{-\beta\mathbf{\Lambda}(\mathbf{R}_i)/P}\mathbf{C}\left(\mathbf{R}_i,\mathbf{R}_{i+1}\right)\right)\right]\right\}\right)$$

with the element in the $m$-th row and $n$-th column of matrix $\mathbf{\Phi}^{(theta)}\left(\mathbf{R}_{i-1},\mathbf{R}_i,\mathbf{R}_{i+1}\right)$ [of Eq. (113)] given by

$$\Phi_{mn}^{(theta)}\left(\mathbf{R}_{j-1},\mathbf{R}_j,\mathbf{R}_{j+1}\right)=\sum_{k=1}^{N}\left\{\left\langle m\left(\mathbf{R}_{j-1}\right)\middle|\frac{\partial k\left(\mathbf{R}_j\right)}{\partial\mathbf{R}_j}\right\rangle\lambda_k\left(\mathbf{R}_j\right)e^{-\beta\lambda_k\left(\mathbf{R}_j\right)/P}\left\langle k\left(\mathbf{R}_j\right)\middle|n\left(\mathbf{R}_{j+1}\right)\right\rangle\right.$$
$$+\left\langle m\left(\mathbf{R}_{j-1}\right)\middle|k\left(\mathbf{R}_j\right)\right\rangle\frac{\partial\lambda_k\left(\mathbf{R}_j\right)}{\partial\mathbf{R}_j}\left[1-\frac{\beta}{P}\lambda_k\left(\mathbf{R}_j\right)\right]e^{-\beta\lambda_k\left(\mathbf{R}_j\right)/P}\left\langle k\left(\mathbf{R}_j\right)\middle|n\left(\mathbf{R}_{j+1}\right)\right\rangle . \quad (115)$$
$$\left.+\left\langle m\left(\mathbf{R}_{j-1}\right)\middle|k\left(\mathbf{R}_j\right)\right\rangle\lambda_k\left(\mathbf{R}_j\right)e^{-\beta\lambda_k\left(\mathbf{R}_j\right)/P}\left\langle\frac{\partial k\left(\mathbf{R}_j\right)}{\partial\mathbf{R}_j}\middle|n\left(\mathbf{R}_{j+1}\right)\right\rangle\right\}$$

The estimator for the element in the $m$-th row and $n$-th column of the electronic state density matrix in the adiabatic representation $\rho_{mn}^{(\mathrm{ads})}$ is then given by

$$\tilde{\rho}_{mn}^{(\mathrm{adia-ads})}\left(\mathbf{R}_1,\cdots,\mathbf{R}_P\right)=\frac{1}{P}\left(\mathrm{Tr}_e\left[\prod_{j=1}^{P}e^{-\beta\mathbf{\Lambda}\left(\mathbf{R}_j\right)/P}\right]\right)^{-1}$$
$$\times\mathrm{Tr}_e\left(\sum_{k=1}^{P}\left\{\left[\prod_{i=1}^{k-1}e^{-\beta\mathbf{\Lambda}(\mathbf{R}_i)/P}\mathbf{C}\left(\mathbf{R}_i,\mathbf{R}_{i+1}\right)\right]\right.\right.$$
$$\times\frac{e^{-\beta\lambda_m(\mathbf{R}_k)/P}+e^{-\beta\lambda_n(\mathbf{R}_k)/P}}{2}\boldsymbol{\rho}_{mn}^{(\mathrm{ads})}\left(\mathbf{R}_k\right)\mathbf{C}\left(\mathbf{R}_k,\mathbf{R}_{k+1}\right) \quad (116)$$
$$\left.\left.\times\left[\prod_{i=k+1}^{P}e^{-\beta\mathbf{\Lambda}(\mathbf{R}_i)/P}\mathbf{C}\left(\mathbf{R}_i,\mathbf{R}_{i+1}\right)\right]\right\}\right)$$

with $\boldsymbol{\rho}_{mn}^{(\mathrm{ads})}(\mathbf{R})$ defined by Eq. (67).

Although we started from Eq. (81) that was obtained in the diabatic representation, Eqs. (86)-(116) demonstrated that multi-electronic-state PIMD may be performed in the adiabatic representation without any knowledge of the diabatic states. Not less importantly, when the primitive estimator is involved evaluation of the physical properties such as the average total energy, kinetic energy, heat capacity, *etc*. requires only the overlap matrix $\mathbf{C}(\mathbf{R}_i, \mathbf{R}_{i+1})$, of which the element in the *l*-th row and *m*-th column is the overlap of the adiabatic states $\langle l(\mathbf{R}_i) | m(\mathbf{R}_{i+1}) \rangle$. Even when the virial estimator is employed, we only need $\mathbf{C}(\mathbf{R}_i, \mathbf{R}_{i+1})$ and the first-derivative coupling elements such as $\left\langle \frac{\partial l(\mathbf{R}_i)}{\partial \mathbf{R}_i} \middle| m(\mathbf{R}_{i+1}) \right\rangle$. That is, *no* second-derivative nonadiabatic coupling terms are needed in the adiabatic version of the MES-PIMD approach for studying exact quantum statistical mechanics. This then offers a practical tool to perform MES-PIMD simulations with *ab initio* electronic structure calculations for "real" molecular systems.

Note that the diagonalization method in the diabatic representation is not equivalent to the adiabatic version of MES-PIMD derived above. The orthogonal transformation matrix $\mathbf{T}(\mathbf{R})$ is necessary in Eqs. (51)-(52) for efficiently evaluating the force term Eq. (42) and the estimators in the diabatic representation. As contrast $\mathbf{T}(\mathbf{R})$ is not required in the adiabatic version. When the transformation matrix $\mathbf{T}(\mathbf{R})$ between the adiabatic and diabatic states is available, it is possible to evaluate in adiabatic MES-PIMD the diabatic state density matrix. E.g., Eq. (65) yields the estimator for the element in the *i*-th row and *j*-th column of the electronic state density matrix in the diabatic representation $\rho_{ij}^{(\mathrm{ds})}$ is

$$\tilde{\rho}_{mn}^{(\mathrm{adia-ds})}\left(\mathbf{R}_1,\cdots,\mathbf{R}_P\right)=\frac{1}{P}\left(\mathrm{Tr}_e\left[\prod_{j=1}^{P}e^{-\beta\mathbf{\Lambda}\left(\mathbf{R}_j\right)/P}\right]\right)^{-1}$$
$$\times\mathrm{Tr}_e\left(\sum_{k=1}^{P}\left\{\left[\prod_{l=1}^{k-1}e^{-\beta\mathbf{\Lambda}(\mathbf{R}_l)/P}\mathbf{C}\left(\mathbf{R}_l,\mathbf{R}_{l+1}\right)\right]\right.\right.$$
$$\times e^{-\beta\mathbf{\Lambda}(\mathbf{R}_k)/P}\mathbf{T}^T\left(\mathbf{R}_k\right)\boldsymbol{\rho}_{ij}^{(\mathrm{ds})}\left(\mathbf{R}_k\right)\mathbf{T}\left(\mathbf{R}_{k+1}\right)$$
$$\left.\left.\times\left[\prod_{l=k+1}^{P}e^{-\beta\mathbf{\Lambda}(\mathbf{R}_l)/P}\mathbf{C}\left(\mathbf{R}_l,\mathbf{R}_{l+1}\right)\right]\right\}\right) \tag{117}$$

with $\boldsymbol{\rho}_{ij}^{(\mathrm{ds})}$ given by Eq. (64).

It is straightforward to follow Section II-4 to generate such as the "middle" thermostat algorithm for the adiabatic version of MES-PIMD.

## Appendix B: Other choices for an effective potential term

For general multi-electronic-state systems in the diabatic representation $\mathrm{Tr}_e\left[\prod_{i=1}^{P}e^{-\beta\mathbf{V}_{\mathrm{diag}}\left(\mathbf{R}_i\right)/P}\right]$ is always positive-definite while $\mathrm{Tr}_e\left[\prod_{i=1}^{P}\mathbf{\Theta}\left(\mathbf{R}_i\right)\right]$ is often *not*. Instead of Eq. (27), another effective (real-valued) potential function $\phi_{\mathrm{mod}}^{(\mathrm{dia})}\left(\mathbf{R}_1,\cdots,\mathbf{R}_P\right)$ may be defined by

$$e^{-\beta\phi_{\mathrm{mod}}^{(\mathrm{dia})}\left(\mathbf{R}_1,\cdots,\mathbf{R}_P\right)}\equiv\left\{\frac{c^2\left(\mathrm{Tr}_e\left[\prod_{i=1}^{P}e^{-\beta\mathbf{V}_{\mathrm{diag}}\left(\mathbf{R}_i\right)/P}\right]\right)^2+\left(\mathrm{Tr}_e\left[\prod_{i=1}^{P}\mathbf{\Theta}\left(\mathbf{R}_i\right)\right]\right)^2}{c^2+1}\right\}^{1/2}, \tag{118}$$

where $c$ is a positive real constant. $\phi_{\mathrm{mod}}^{(\mathrm{dia})}\left(\mathbf{R}_1,\cdots,\mathbf{R}_P\right)$ is well-defined as long as $c$ is finite. As $c\to\infty$ the effective potential function $\phi_{\mathrm{mod}}^{(\mathrm{dia})}\left(\mathbf{R}_1,\cdots,\mathbf{R}_P\right)$ approaches the one defined by Eq. (27) for diabatic MES-PIMD in Section II. In another limit $c\to 0$ $\phi_{\mathrm{mod}}^{(\mathrm{dia})}\left(\mathbf{R}_1,\cdots,\mathbf{R}_P\right)$ approaches the potential term defined by Eq. (26) that may meet severe numerical problems. When the value of the parameter $c$ is chosen in a reasonable region, the potential function $\phi_{\mathrm{mod}}^{(\mathrm{dia})}\left(\mathbf{R}_1,\cdots,\mathbf{R}_P\right)$ given by Eq. (118) may lead to a more efficient sampling for diabatic MES-PIMD.

Similarly, in the adiabatic representation an effective (real-valued) potential function

$\phi_{\text{mod}}^{(\text{adia})}(\mathbf{R}_1,\cdots,\mathbf{R}_P)$ may be well-defined by

$$e^{-\beta\phi_{\text{mod}}^{(\text{adia})}(\mathbf{R}_1,\cdots,\mathbf{R}_P)} \equiv \left\{ \frac{c^2\left(\text{Tr}_e\left[\prod_{i=1}^{P} e^{-\beta\mathbf{\Lambda}(\mathbf{R}_i)/P}\right]\right)^2 + \left(\text{Tr}_e\left[\prod_{i=1}^{P} e^{-\beta\mathbf{\Lambda}(\mathbf{R}_i)/P}\mathbf{C}(\mathbf{R}_i,\mathbf{R}_{i+1})\right]\right)^2}{c^2+1} \right\}^{1/2} \quad . \quad (119)$$

In addition that $\phi_{\text{mod}}^{(\text{adia})}(\mathbf{R}_1,\cdots,\mathbf{R}_P)$ avoids the numerical instability that $\left|\text{Tr}_e\left[\prod_{i=1}^{P} e^{-\beta\mathbf{\Lambda}(\mathbf{R}_i)/P}\mathbf{C}(\mathbf{R}_i,\mathbf{R}_{i+1})\right]\right|$ leads to, it may offer more efficient adiabatic MES-PIMD sampling than the effective potential given by Eq. (87).

Besides Eq. (118) [or Eq. (119)], some other choices for the effective potential term in MES-PIMD may also be proposed. For instance, when the hyperbolic function method [Eq. (14) and Eq. (25)] in the diabatic representation is employed, a choice other than Eq. (27) for a well-defined effective potential function is

$$e^{-\beta\phi_{(\text{hyper})}^{(\text{dia})}(\mathbf{R}_1,\cdots,\mathbf{R}_P)} \equiv \text{Tr}_e\left[\prod_{i=1}^{P} e^{-\beta\mathbf{V}_{\text{diag}}(\mathbf{R}_i)/2P}\mathbf{G}_{\text{diag}}(\mathbf{R}_i)e^{-\beta\mathbf{V}_{\text{diag}}(\mathbf{R}_i)/2P}\right] \quad , \quad (120)$$

where the $m$-th diagonal element of the diagonal matrix $\mathbf{G}_{\text{diag}}(\mathbf{R}_i)$ may be given by

$$\text{G}_{\text{diag}}^{(mm)}(\mathbf{R}_i) = \prod_{j\neq m}\cosh^2\left[-\beta V_{jm}(\mathbf{R}_i)/2P\right] \quad (121)$$

or by

$$\text{G}_{\text{diag}}^{(mm)}(\mathbf{R}_i) = \prod_{j\neq m}\cosh\left[-\beta V_{jm}(\mathbf{R}_i)/P\right] \quad . \quad (122)$$

Eq. (118) [Eq. (120), or Eq. (119)] does not demonstrate noticeably better sampling performance than Eq. (27) [or Eq. (87)] as we investigate the benchmark models in Section III. (Results not shown.) It will be interesting in future to test them for more multi-electronic-state systems. Although various other choices are possible, it should be stressed that Eq. (27) [or Eq. (87)] offers the simplest well-defined effective potential for MES-PIMD that is practical for general multi-electronic-state systems and that leads to a well-defined physical quantity (as discussed in Section II).

**Appendix C: Difference between the coherence length in the diabatic representation and that in the adiabatic representation**

It is well-known that thermodynamic properties are independent of the representation. The coherence length of the electronic state density matrix defined in Eq. (68), however, depends on whether the diabatic or adiabatic representation is employed. Use the one-dimensional two-electronic-state model [of which the parameters are described in Table 1] for demonstration. Fig. 20 shows the coherence length as a function of the difference between the minimum of that ground state and that of the excited state $\Delta R = R_{eq}^{(2)} - R_{eq}^{(1)}$ with other parameters fixed for the system at the inverse temperature $\beta = 27000$. The behavior of the results shown in Fig. 20a for the diabatic representation is significantly different from that shown in Fig. 20b for the adiabatic representation. The coherence length of the electronic state density is not a well-defined physical quantity. Despite that a few choices other than Eq. (68) have been proposed for quantifying the coherence length[36], none of them are independent of the representation of the electronic states.

**Tables and Figures**

**Table 1. Parameters [in Eq. (78)] for a two-electronic-state system coupled to a Morse oscillator [unit: atomic unit (au)]**

| | $n$ | $\omega_{(n)}$ | $R_{eq}^{(n)}$ | $D_{(n)}$ | $V_0^{(n)}$ | $c^{(12)}$ | $\alpha_1^{(12)}$ | $\alpha_2^{(12)}$ | $R^{(12)}$ |
|---|---|---|---|---|---|---|---|---|---|
| (a) | 1 | $9.43\times10^{-5}$ | -1.75 | $4.71\times10^{-3}$ | 0 | $6.11\times10^{-5}$ | 0.05 | 0.05 | 1.714 |
| | 2 | $9\times10^{-5}$ | -0.4274 | $4.71\times10^{-3}$ | $9.8\times10^{-5}$ | | | | |
| (b) | 1 | $9.43\times10^{-5}$ | -1.75 | $4.71\times10^{-3}$ | 0 | $6.11\times10^{-5}$ | 0.05 | 0.05 | 0.923 |
| | 2 | $9\times10^{-5}$ | 0.9226 | $4.71\times10^{-3}$ | $9.8\times10^{-5}$ | | | | |
| (c) | 1 | $9.43\times10^{-5}$ | -1.75 | $4.71\times10^{-3}$ | 0 | $6.11\times10^{-5}$ | 0.05 | 0.05 | 1.260 |
| | 2 | $9\times10^{-5}$ | 2.4976 | $4.71\times10^{-3}$ | $9.8\times10^{-5}$ | | | | |

**Table 2. Parameters [in Eq. (78)] for a seven-electronic-state system coupled to a Morse oscillator**

| $n$ | $\omega_{(n)}$ / cm$^{-1}$ | $R_{eq}^{(n)}$ / au | $D_{(n)}$ / au | $\alpha_1^{(ij)}$ / au | $\alpha_2^{(ij)}$ / au |
|---|---|---|---|---|---|
| 1 | 212.2 | 0 | $7.28\times10^{-2}$ | | |
| 2 | 209.0 | 5 | $7.17\times10^{-2}$ | | |
| 3 | 205.8 | 10 | $7.06\times10^{-2}$ | | |
| 4 | 202.7 | 15 | $6.95\times10^{-2}$ | $5\times10^{-5}$ | 0.02 |
| 5 | 199.5 | 20 | $6.84\times10^{-2}$ | | |
| 6 | 196.3 | 25 | $6.73\times10^{-2}$ | | |
| 7 | 193.1 | 30 | $6.62\times10^{-2}$ | | |

In Eq. (78) $c^{(mn)} = H_{\mathrm{FMO}}^{(mn)}\ (m \neq n)$ and $V_0^{(n)} = H_{\mathrm{FMO}}^{(nn)}\ (n = 1, \cdots, 7)$ are the diagonal and off-diagonal elements of the Hamiltonian matrix $H_{\mathrm{FMO}}$ [taken from Ref. [66]]

$$H_{\mathrm{FMO}} = \begin{pmatrix} 0 & -62 & 17 & 8 & -1 & -9 & 28 \\ -62 & 175 & -57 & -5 & -70 & -19 & 6 \\ 17 & -57 & 260 & -4 & -2 & 32 & 1 \\ 8 & -5 & -4 & 280 & 6 & -8 & -106 \\ -1 & -70 & -2 & 6 & 320 & 40 & 2 \\ -9 & -19 & 32 & -8 & 40 & 360 & 13 \\ 28 & 6 & 1 & -106 & 2 & 13 & 420 \end{pmatrix} \left(\text{unit: cm}^{-1}\right) \ ; \quad (123)$$

the intersection $R^{(mn)}\ (m \neq n)$ is the element in the $m$-th row and $n$-th column of the matrix

$$R_{\mathrm{FMO}} = \begin{pmatrix} 0 & 224 & 157 & 112 & 99 & 93 & 94 \\ 224 & 0 & 102 & 67 & 65 & 66 & 72 \\ 157 & 102 & 0 & 34 & 48 & 54 & 65 \\ 112 & 67 & 34 & 0 & 62 & 65 & 76 \\ 99 & 65 & 48 & 62 & 0 & 69 & 84 \\ 93 & 66 & 54 & 65 & 69 & 0 & 100 \\ 94 & 72 & 65 & 76 & 84 & 100 & 0 \end{pmatrix} \left(\text{unit: au}\right) \ . \quad (124)$$

**Table 3. Parameters [in Eq. (79)] for the six spin-boson models in Ref.** [45]

| model | $\Delta$ / au | $\varepsilon$ / au | $\beta$ / au | $\omega_c$ / au | $\alpha$ / au |
|---|---|---|---|---|---|
| I | 0.1 | 0 | 1 | 0.25 | 0.09 |
| II | 0.1 | 0 | 50 | 0.25 | 0.09 |
| III | 0.1 | 1 | 50 | 0.25 | 0.1 |
| IV | 0.1 | 1 | 2.5 | 0.1 | 0.1 |
| V | 0.1 | 0 | 10 | 0.1 | 2 |
| VI | 0.1 | 5 | 1 | 0.2 | 4 |

In Eq. (79), $M_{(j)}=1$, $\omega_{(j,1)}=\omega_{(j,2)}=\omega_j$, $R_{eq,(j)}^{(1)}=-R_{eq,(j)}^{(2)}=\frac{c_j}{M_{(j)}\omega_j^2}$, $j=1,\cdots N_b$ where $N_b$ is the number of the degrees of freedom of the harmonic bath, $V_0^{(1)}=-\sum_{j=1}^{N_b}\frac{c_j^2}{2M_j\omega_j^2}+\varepsilon$, $V_0^{(2)}=-\sum_{j=1}^{N_b}\frac{c_j^2}{2M_j\omega_j^2}-\varepsilon$, $\Delta^{(12)}=\Delta$. Here $N_b=100$ frequencies $\{\omega_j\}$ and coupling constants $\{c_j\}$ are selected from the spectral density of the exponentially damped Ohmic form

$$J(\omega)=\frac{\pi}{2}\sum_{j=1}^{N_b}\frac{c_j^2}{\omega_j}\delta\left(\omega-\omega_j\right)=\frac{\pi}{2}\alpha\omega e^{-\omega/\omega_c} \quad . \tag{125}$$

We then employ the discretization strategy in Ref. [67] for Eq. (125). The frequencies $\{\omega_j\}$ are chosen as

$$\omega_j=-\omega_c\log\left(1-\frac{j}{N_b+1}\right),\quad j=1,\cdots,N_b \tag{126}$$

then the coupling constants $\{c_j\}$ are determined as

$$c_j=\sqrt{\frac{\alpha\omega_j^2\omega_c}{N_b+1}},\quad j=1,\cdots,N_b \quad . \tag{127}$$

**Table 4. Parameters [in Eq. (80)] for the seven-electronic-state coupled to multidimensional Morse oscillators**

In Eq. (80), $M_{(j)} = 1, \quad j = 1, \cdots N_b$ with $N_b = 50$;

$$\left.\begin{aligned} \omega_{(j,n)} &= -\omega_c \left[1 - 0.015(n-1)\right] \log\left(1 - \frac{j}{N_b + 1}\right) \\ R_{\mathrm{eq},(j)}^{(n)} &= \frac{1}{M_{(j)}\omega_{(j,1)}} \sqrt{\frac{2\eta}{\hbar(N_b+1)}} + 5(n-1) \\ D_{(j,n)} &= 0.0728\left[1 - 0.015(n-1)\right] \\ \alpha_{1,(j)}^{(mn)} &= 5\times10^{-5} \\ \alpha_{2,(j)}^{(mn)} &= 0.02 \end{aligned}\right\} \left(j = 1, \cdots, N_b, \quad n = 1, \cdots, 7\right) \quad (128)$$

with $\omega_c = 212.2\ \mathrm{cm}^{-1}$, $\eta = 35\ \mathrm{cm}^{-1}$; $V_0^{(n)} = H_{\mathrm{FMO}}^{(nn)}$ $(n = 1, \cdots, 7)$ and $c_{(j)}^{(mn)} = H_{\mathrm{FMO}}^{(mn)}$ $(m \neq n)$ are the diagonal and off-diagonal elements of the Hamiltonian matrix given in Eq. (123), respectively; $\left\{R_{(j)}^{(mn)} = R^{(mn)}; \ j = 1, \cdots, N_b, \quad m \neq n\right\}$ are elements of the matrix in Eq. (124).

**Fig. 1.** Diabatic potential matrix elements for the three two-state models in Table 1.

**Fig. 2.** Comparison between the "middle" and conventional thermostat schemes for first-order expansion method of diabatic MES-PIMD at the inverse temperature $\beta = 63000$ for Model b in Table 1. ($P = 64$ beads are used.) Results for the average potential and kinetic energy, heat capacity, and coherence length are plotted as functions of the time interval $\Delta t$. Atomic units (au) are used.

**Fig. 3.** Results produced by adiabatic MES-PIMD and three methods [diagonalization, hyperbolic function, and first-order expansion] of diabatic MES-PIMD on different physical properties (average potential and kinetic energy, heat capacity, and coherence length) as functions of the number of path integral beads $P$ (at the inverse temperature $\beta = 63000$ for Model b in Table 1). Atomic units (au) are used. Statistical error bars are included. Exact results obtained by DVR are plotted as the references.

**Fig. 4.** Comparison of converged results yielded by MES-PIMD to exact data in a wide range of the inverse temperature from $\beta = 17000$ to $\beta = 63000$ for three models listed in Table 1. Statistical error bars for MES-PIMD results are included. "(a)-MES-PIMD" represents the numerical results produced by MES-PIMD for Model a in Table 1; "(a)-exact" stands for the exact quantum results obtained by DVR for Model a in Table 1; *etc.*

**Fig. 5.** Comparison of converged results yielded by MES-PIMD to exact data in a wide range of $\Delta R = R_{eq}^{(2)} - R_{eq}^{(1)}$ in the suite of models listed in Table 1. (Note that the intersection of two potential energy surfaces $R^{(12)}$ changes as the minimum of that ground state and that of the excited state $\Delta R = R_{eq}^{(2)} - R_{eq}^{(1)}$ varies.) Three inverse temperatures are studied. Statistical error bars for MES-PIMD results are included. "beta=17000 MES-PIMD" represents the numerical results produced

by MES-PIMD at $\beta = 17000$; "beta=17000 exact" stands for the exact quantum results obtained by DVR at $\beta = 17000$; *etc.*

**Fig. 6.** Diabatic potential matrix elements for the seven-state model in Table 2. Panel (a): diagonal elements; (b): four typical off-diagonal elements (other off-diagonal elements not shown).

**Fig. 7.** Same as Fig. 2, but for the 1-D seven-state model in Table 2 at $T = 70\ \text{K}$. ($P = 64$ beads are used.)

**Fig. 8.** Same as Fig. 3, but for the 1-D seven-state model in Table 2 at $T = 70\ \text{K}$.

**Fig. 9.** Comparison of converged results yielded by MES-PIMD to exact data at the temperature in the range $T = 50 \sim 250\ \text{K}$ for the 1-D seven-state model in Table 2. Statistical error bars for MES-PIMD results are included.

**Fig. 10.** Results produced by adiabatic MES-PIMD and three methods [diagonalization, hyperbolic function, and first-order expansion] of diabatic MES-PIMD on different physical properties (average potential and kinetic energy, heat capacity, and coherence length) as functions of the number of path integral beads $P$ at the inverse temperature $\beta = 1$ for spin-boson model I in Table 3. Atomic units (au) are used.

**Fig. 11.** Same as Fig. 11, but for spin-boson model II in Table 3 at the inverse temperature $\beta = 50$.

**Fig. 12.** Same as Fig. 11, but for spin-boson model III in Table 3 at the inverse temperature $\beta = 50$.

**Fig. 13.** Same as Fig. 11, but for spin-boson model IV in Table 3 at the inverse temperature $\beta = 2.5$.

**Fig. 14.** Same as Fig. 11, but for spin-boson model V in Table 3 at the inverse temperature $\beta = 10$.

**Fig. 15.** Same as Fig. 11, but for spin-boson model VI in Table 3 at the inverse temperature $\beta = 1$.

**Fig. 16.** Results produced by adiabatic MES-PIMD and three methods [diagonalization, hyperbolic function, and first-order expansion] of diabatic MES-PIMD on different physical properties (average

potential and kinetic energy, heat capacity, and coherence length) as functions of the number of path integral beads $P$ for the seven-state model in Table 4 at $T = 100\ \mathrm{K}$.

**Fig. 17.** Same as Fig. 8, but for the 1-D seven-state model in Table 2 except that the couplings between different electronic states are increased by a factor of 10. [I.e., $\left\{c^{(mn)} = H_{\mathrm{FMO}}^{(mn)}\ (m \neq n)\right\}$ are ten times of their original values in Eq. (123).] The temperature is also $T = 70\ \mathrm{K}$.

**Fig. 18.** The model system and all parameters are the same as those of Fig. 3. We use the diagonalization method to sample configurations of the path integral beads. The estimators derived in all the three splitting methods are employed for evaluating the same physical property. Since the configurational distribution of the beads is the same, the comparison demonstrates the numerical behaviors of the different estimators.

**Fig. 19.** Same as Fig. 18. But the configurational distribution of the path integral beads is obtained by the first-order expansion method instead. Fig. 19 should be compared to Fig. 18 and Fig. 3.

**Fig. 20.** The coherence length as a function of the difference between the minimum of that ground state and that of the excited state $\Delta R = R_{eq}^{(2)} - R_{eq}^{(1)}$ with other parameters fixed in the suite of models listed in Table 1 at the inverse temperature $\beta = 27000$. Both MES-PIMD and exact (DVR) results are demonstrated.

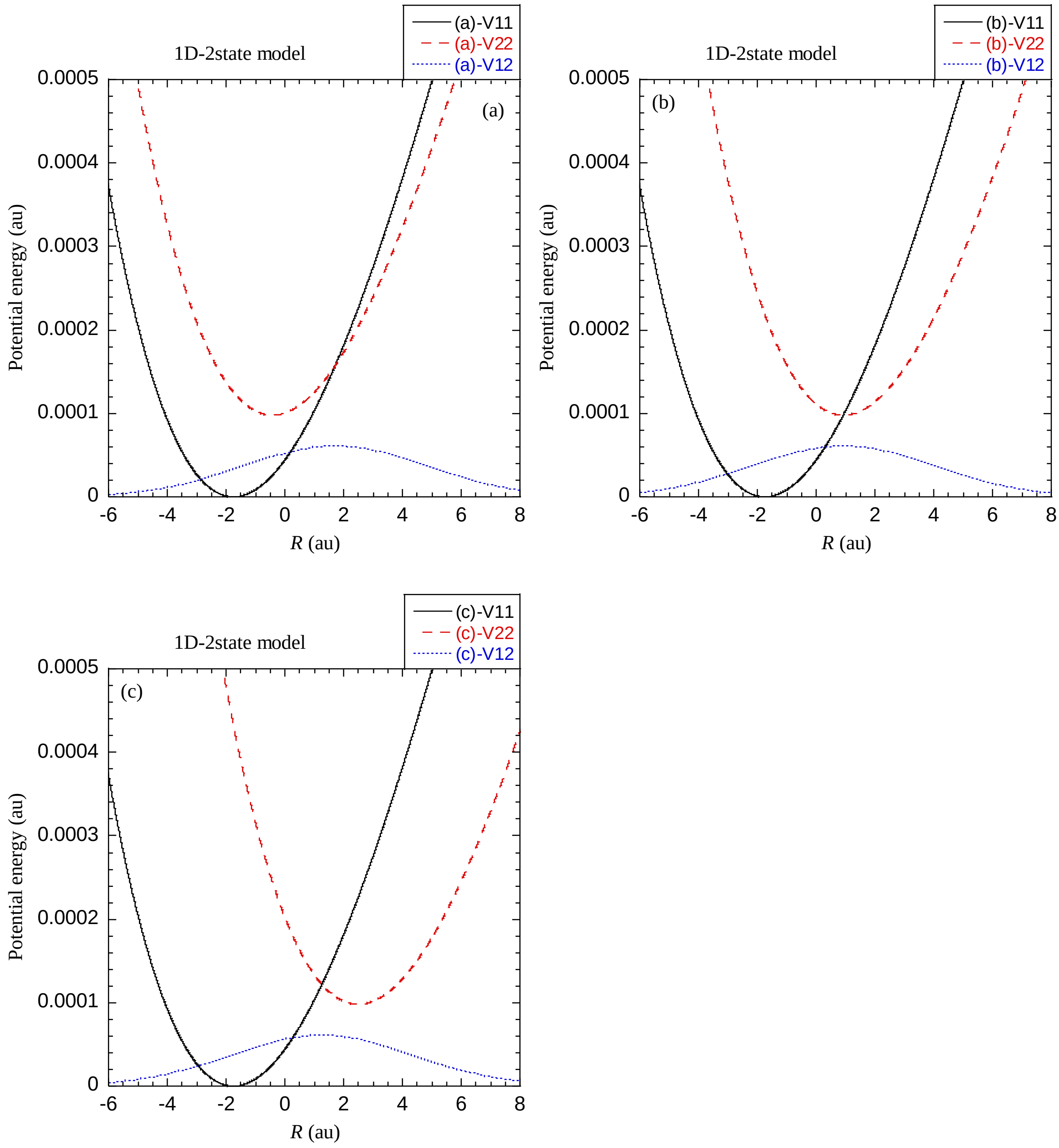

**Fig. 1**

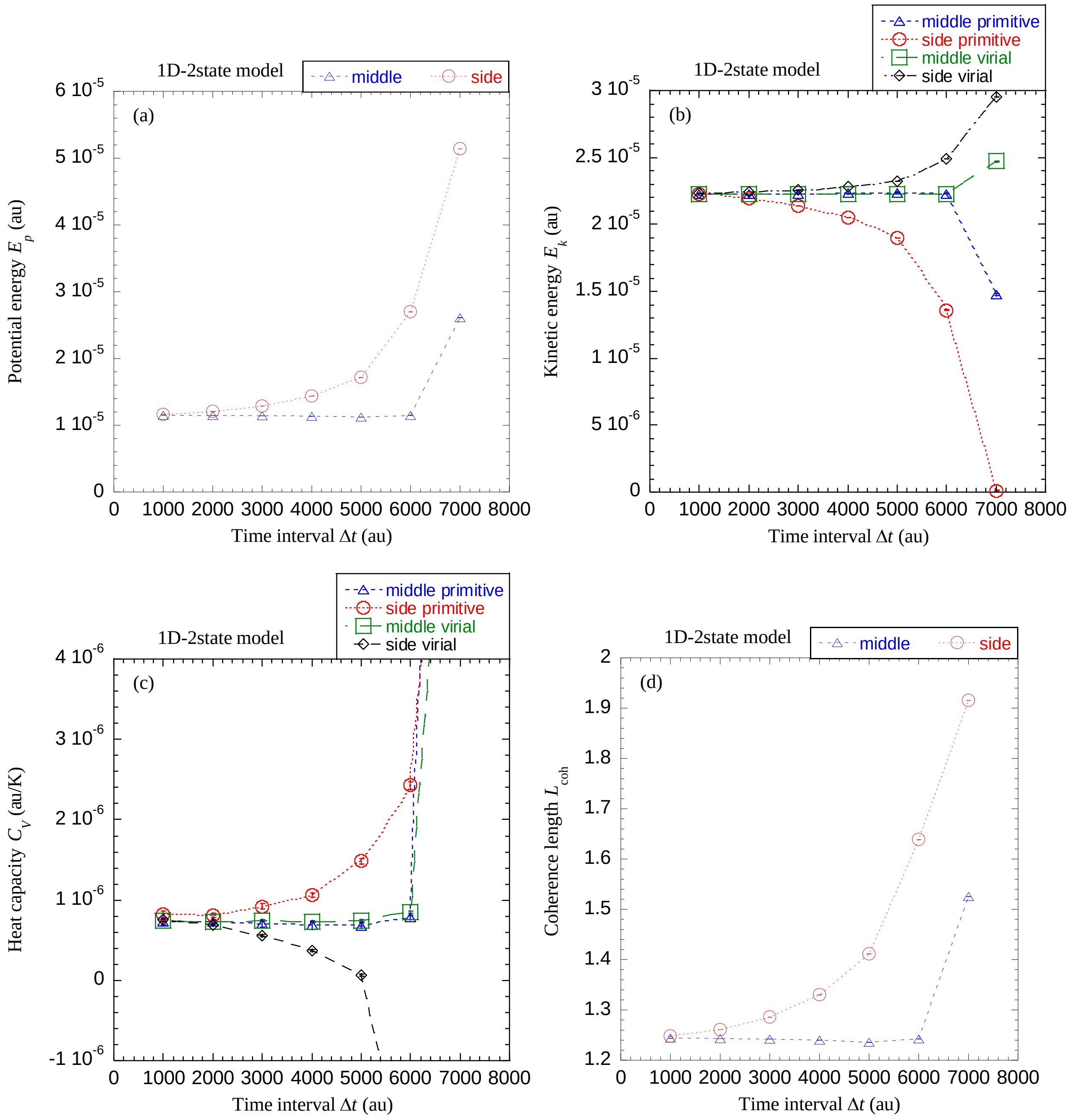

1D-2state model
middle
side
(a)
Potential energy $E_p$ (au)
Time interval Δt (au)
1D-2state model
middle primitive
side primitive
middle virial
side virial
(b)
Kinetic energy $E_k$ (au)
Time interval Δt (au)
1D-2state model
middle primitive
side primitive
middle virial
side virial
(c)
Heat capacity $C_V$ (au/K)
Time interval Δt (au)
1D-2state model
middle
side
(d)
Coherence length $L_{coh}$
Time interval Δt (au)

**Fig. 2**

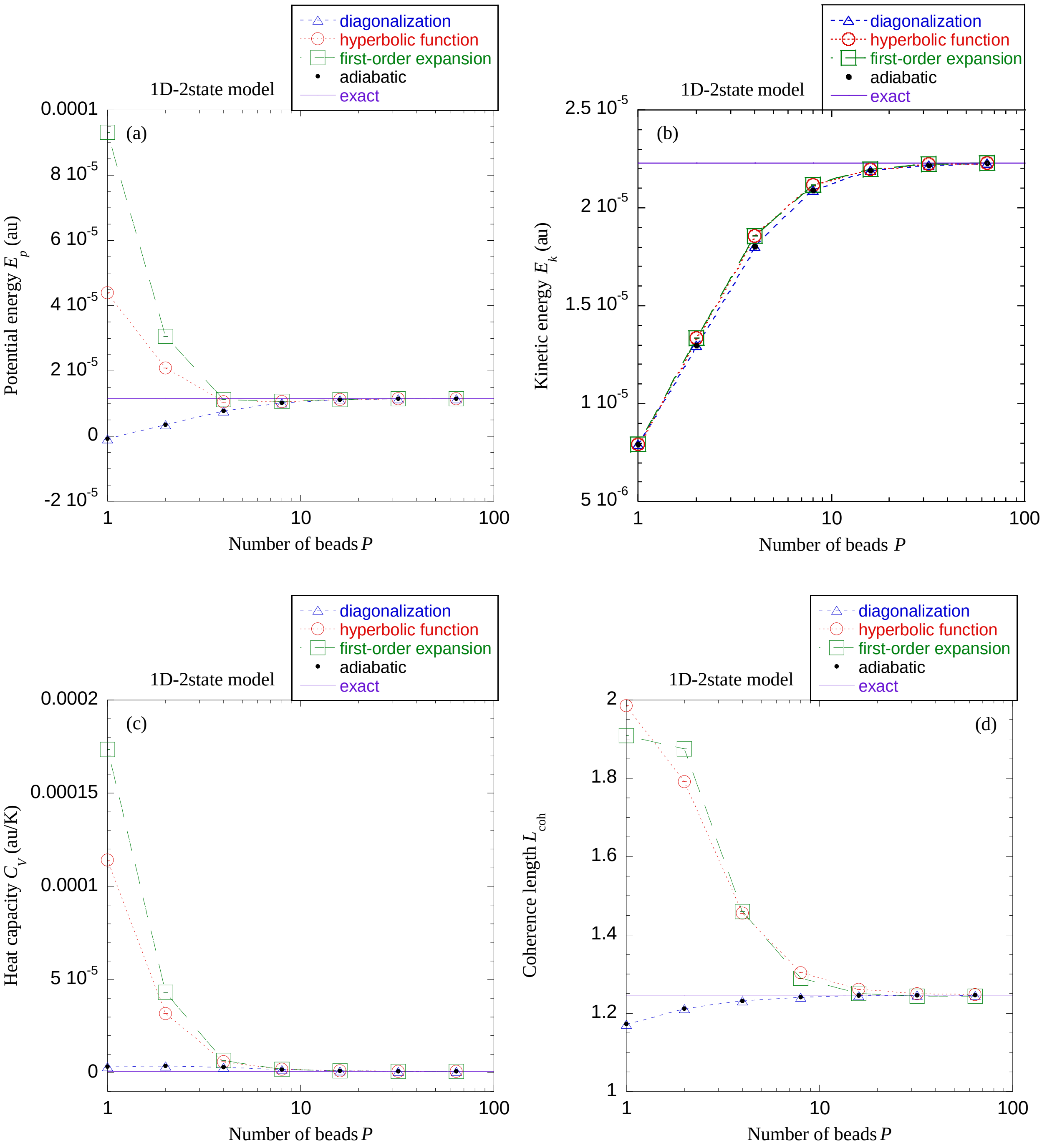

1D-2state model
diagonalization
hyperbolic function
first-order expansion
adiabatic
exact
(a)
(b)
(c)
(d)
Potential energy $E_p$ (au)
Kinetic energy $E_k$ (au)
Heat capacity $C_V$ (au/K)
Coherence length $L_{coh}$
Number of beads $P$

**Fig. 3**

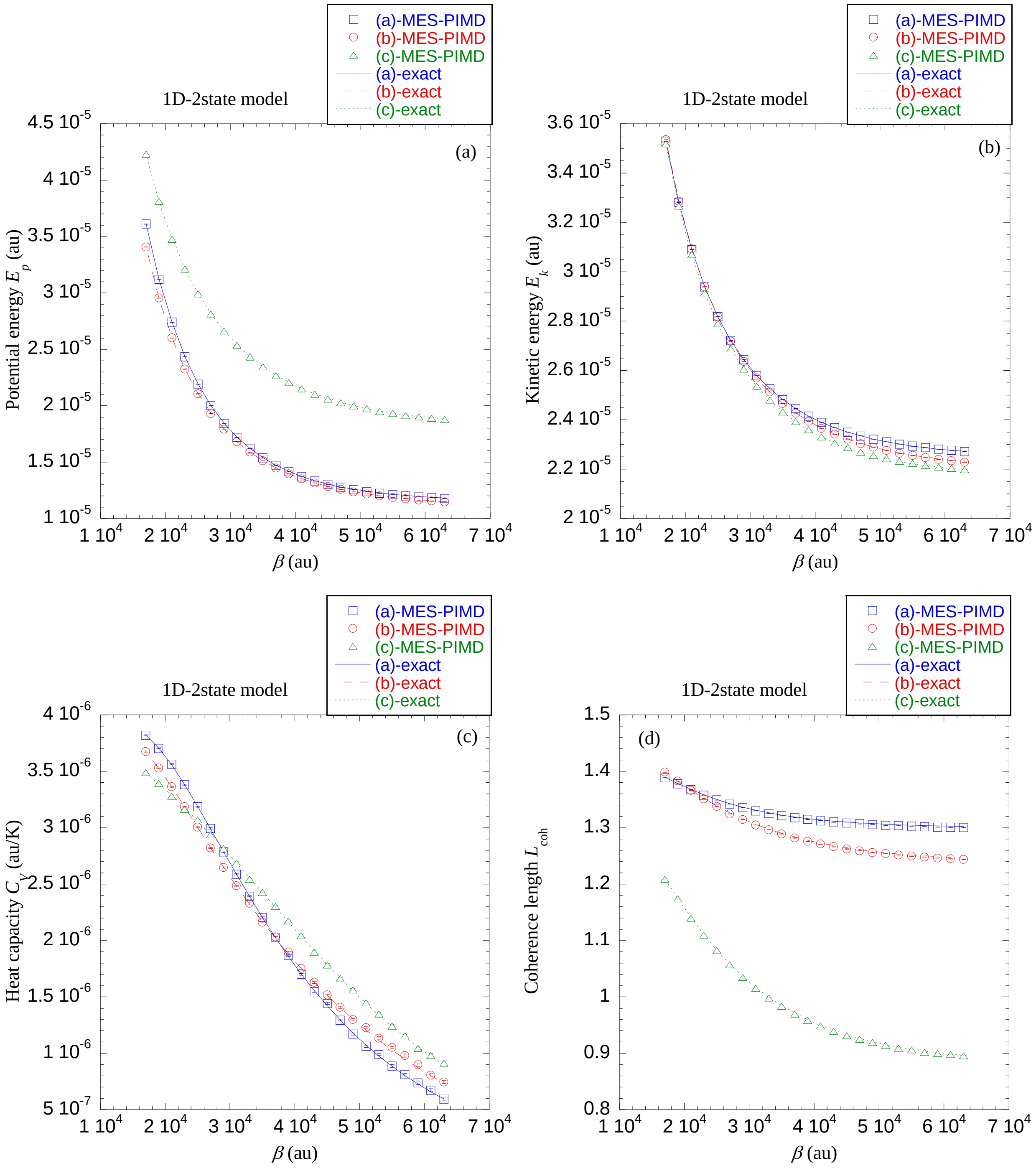

**Fig. 4**

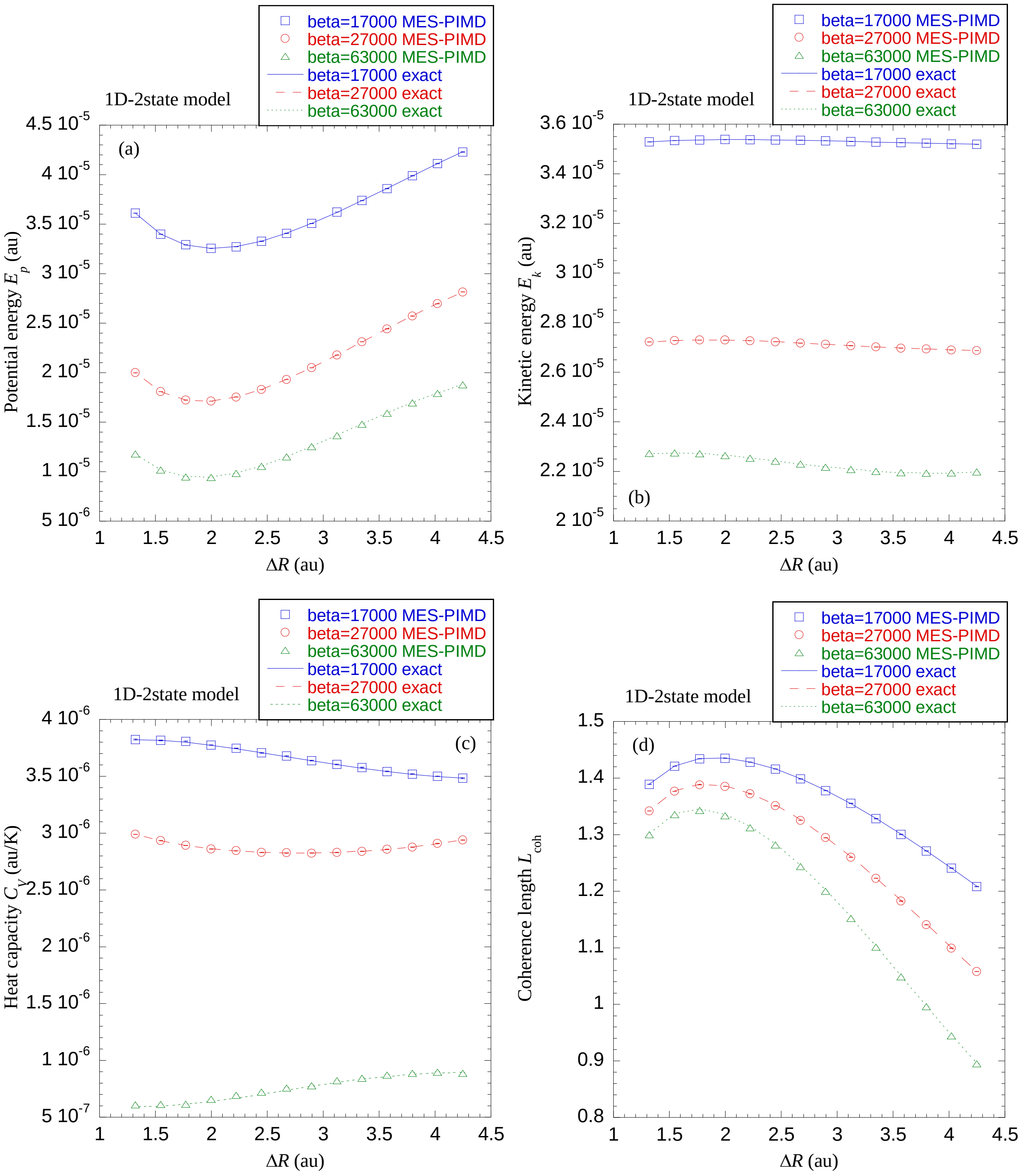

**Fig. 5**

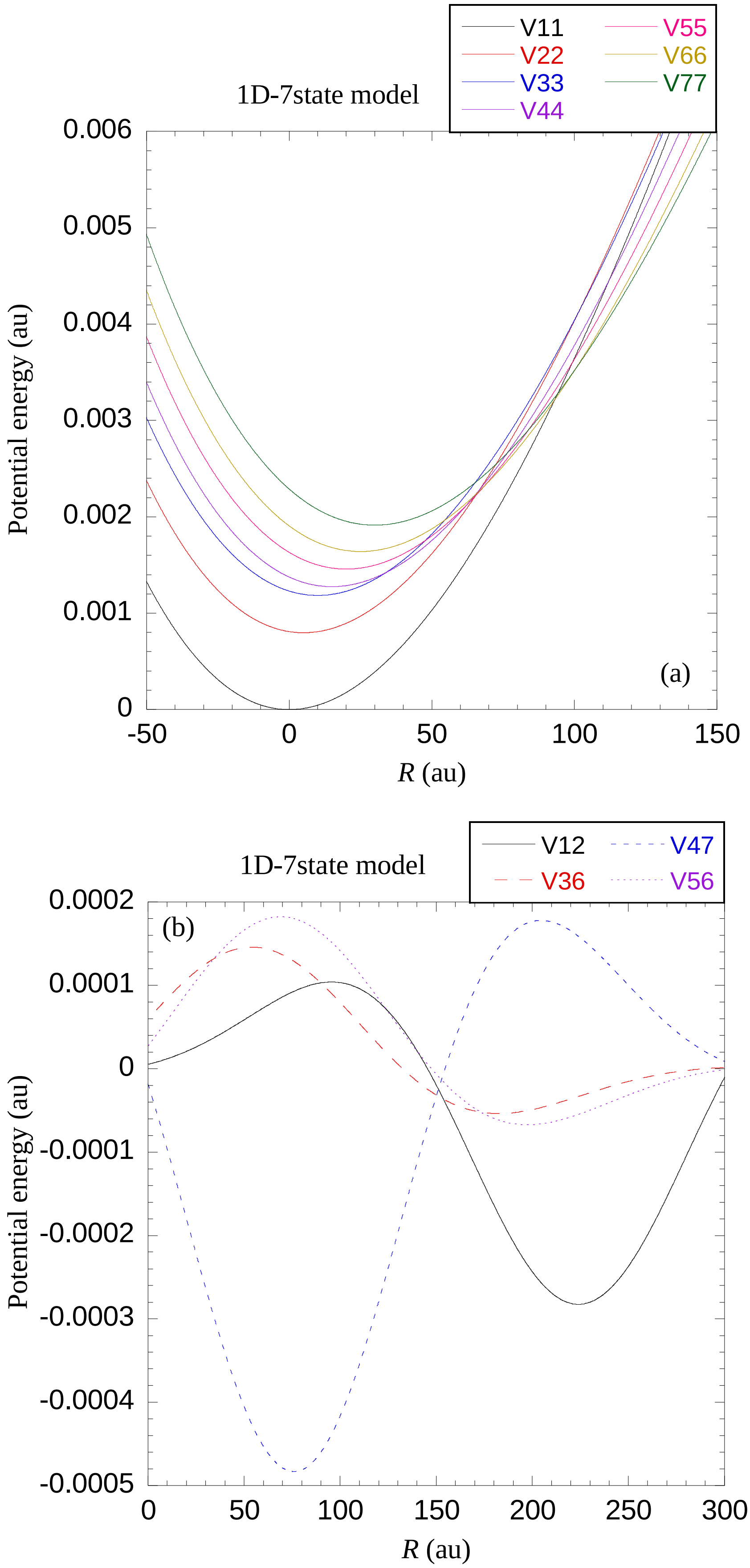

**Fig. 6**

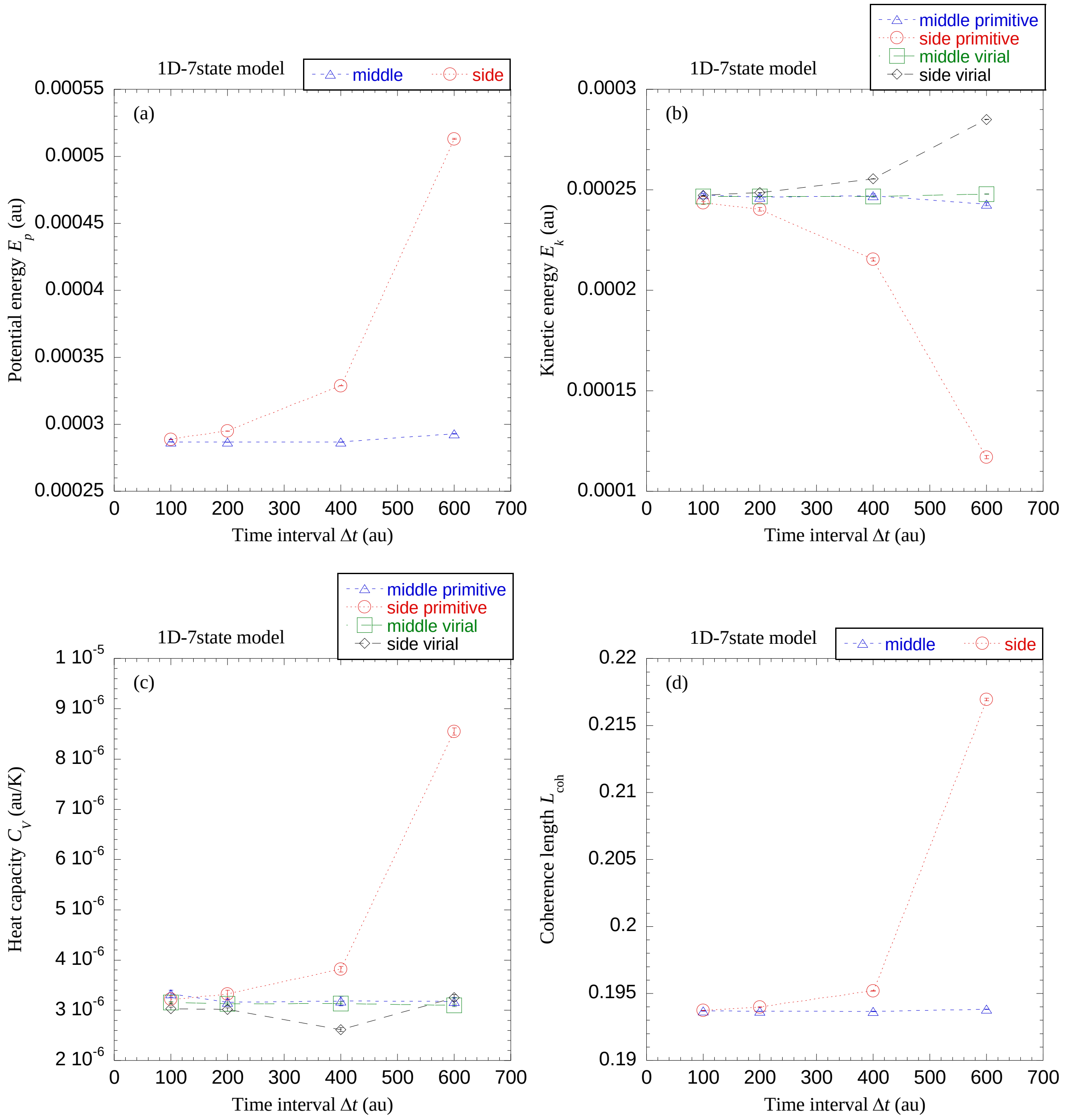

**Fig. 7**

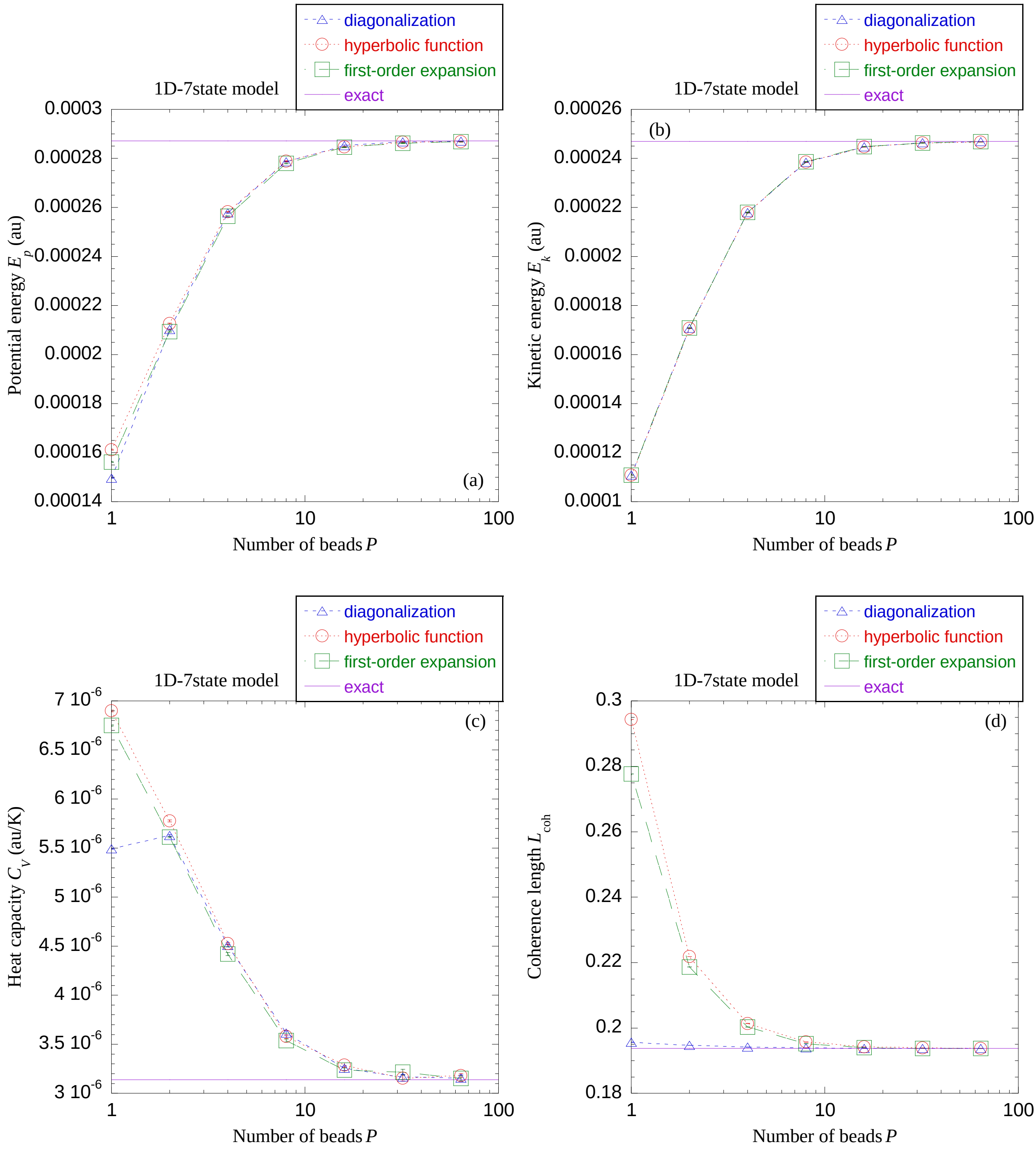

**Fig. 8**

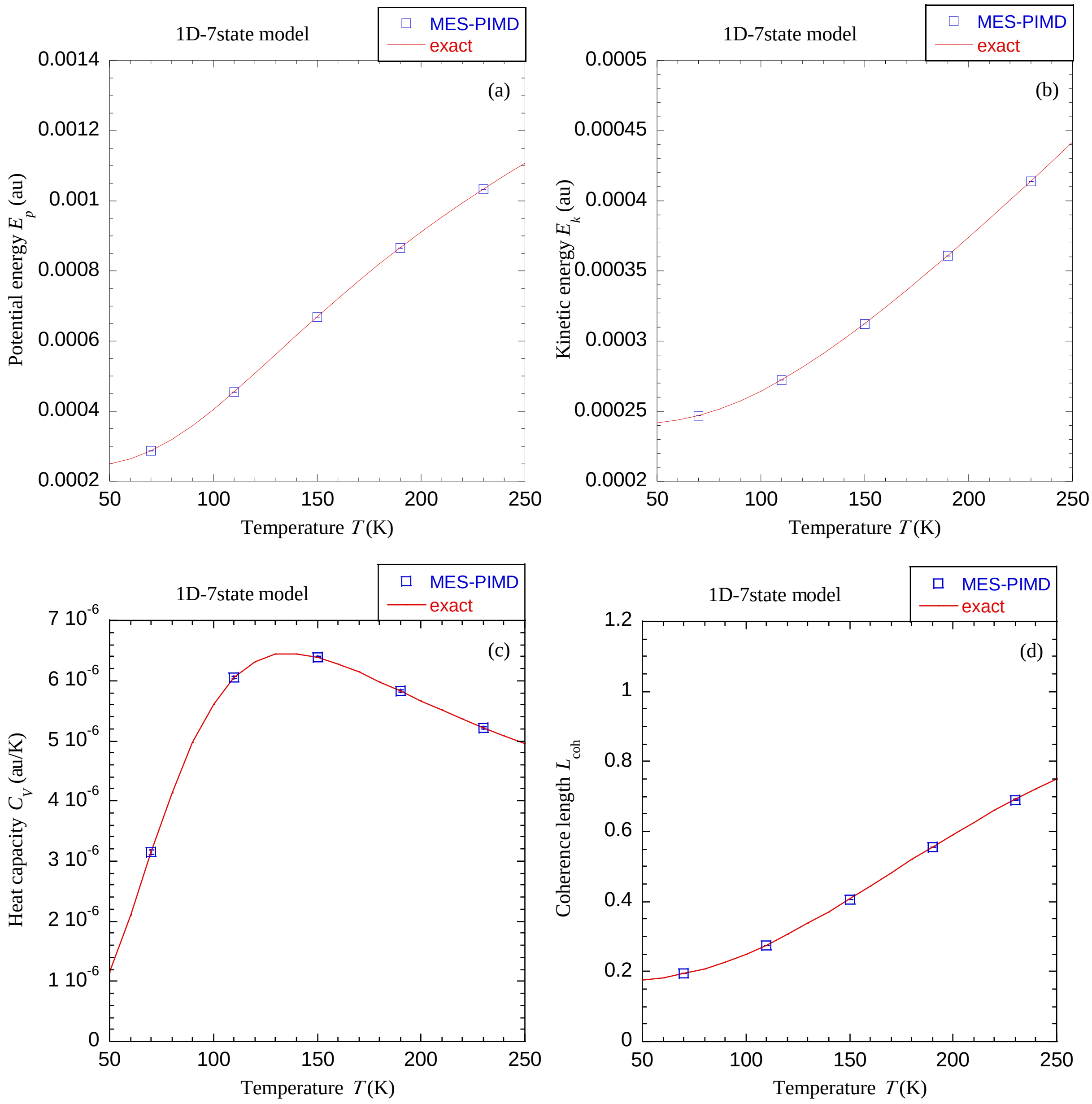

**Fig. 9**

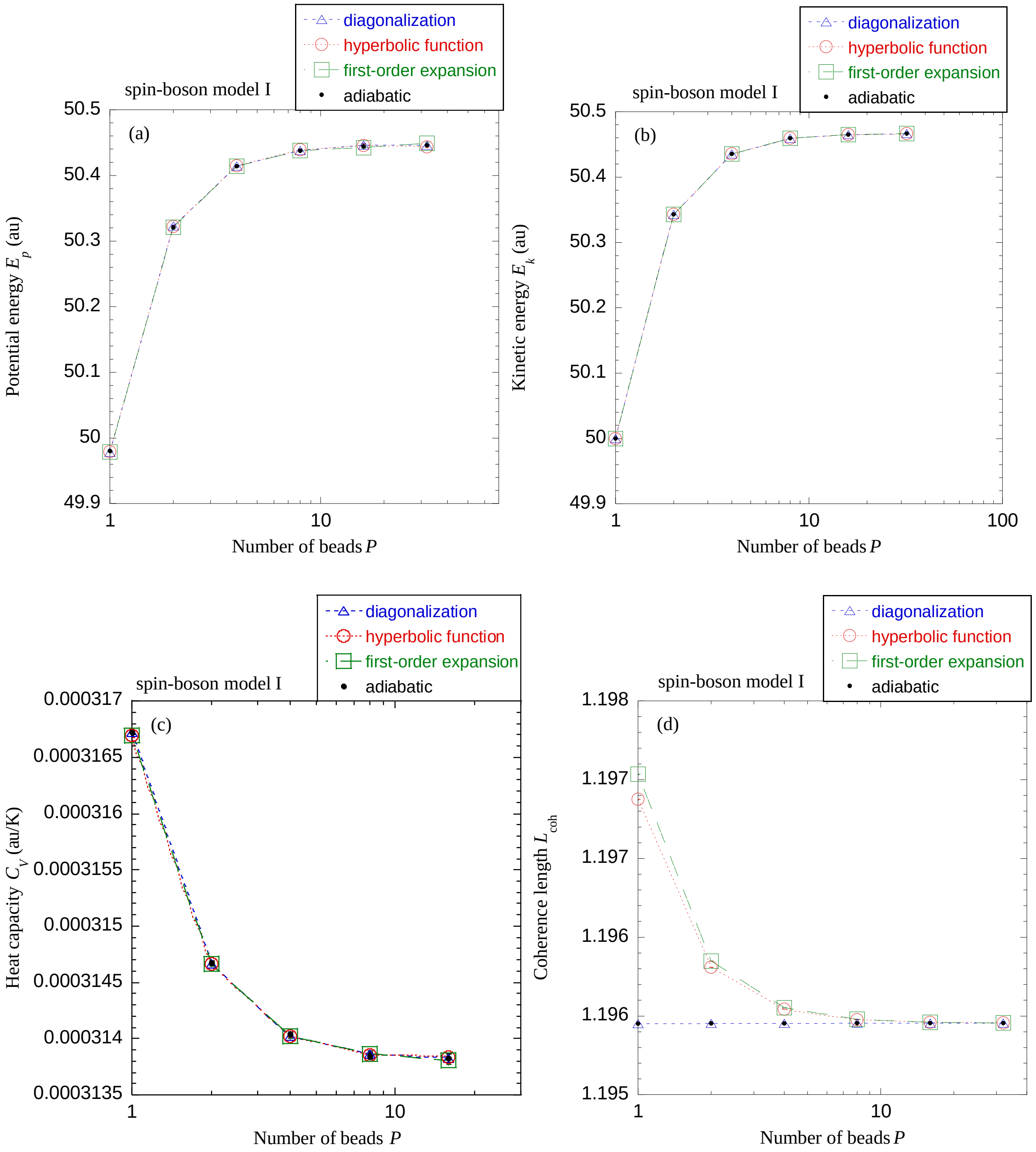
diagonalization
hyperbolic function
first-order expansion
adiabatic
spin-boson model I
(a)
Potential energy $E_p$ (au)
Number of beads $P$
(b)
Kinetic energy $E_k$ (au)
(c)
Heat capacity $C_V$ (au/K)
(d)
Coherence length $L_{coh}$

**Fig. 10**

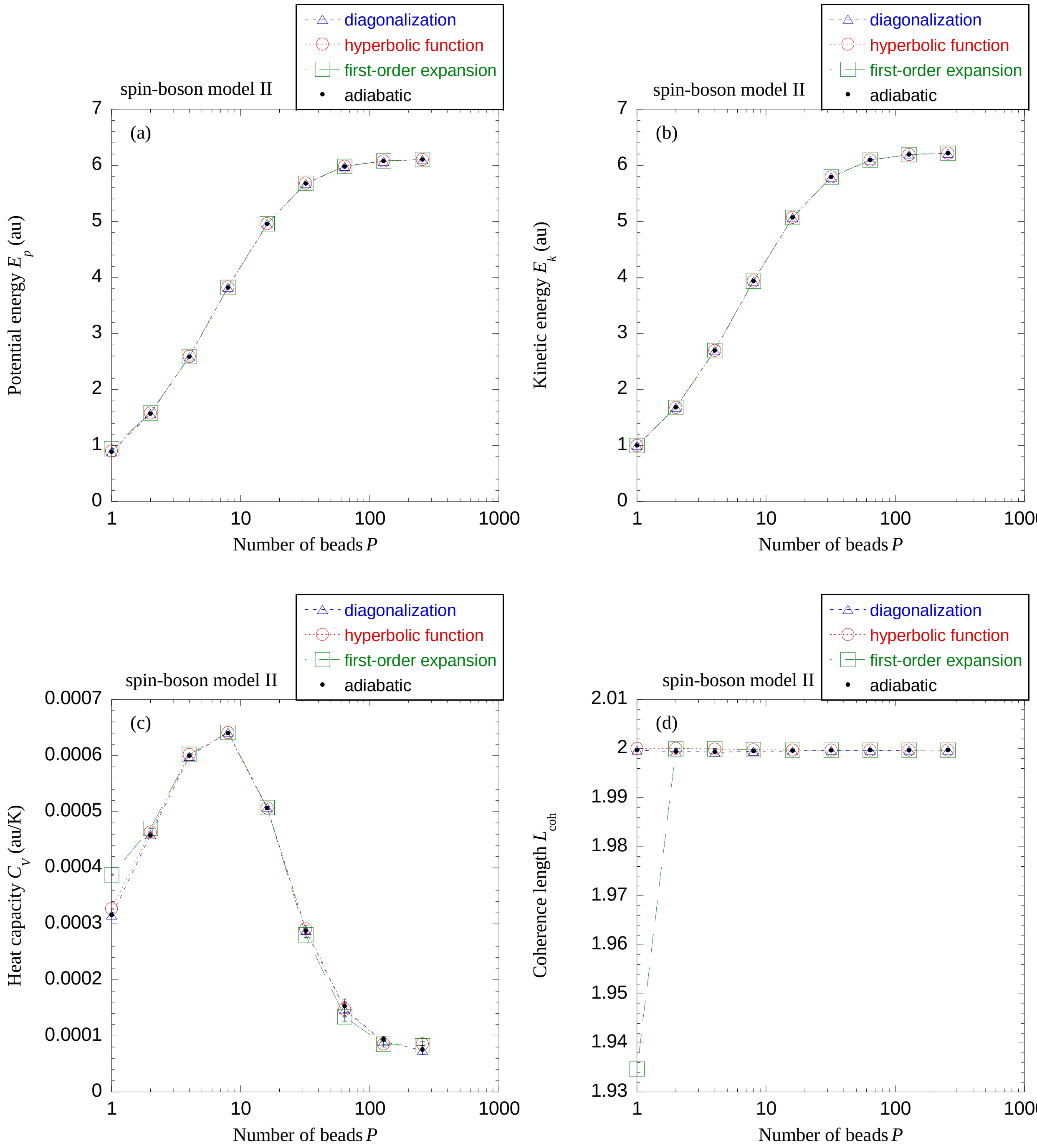
diagonalization
hyperbolic function
first-order expansion
adiabatic
spin-boson model II
(a)
Potential energy $E_p$ (au)
Number of beads $P$
(b)
Kinetic energy $E_k$ (au)
(c)
Heat capacity $C_V$ (au/K)
(d)
Coherence length $L_{coh}$

**Fig. 11**

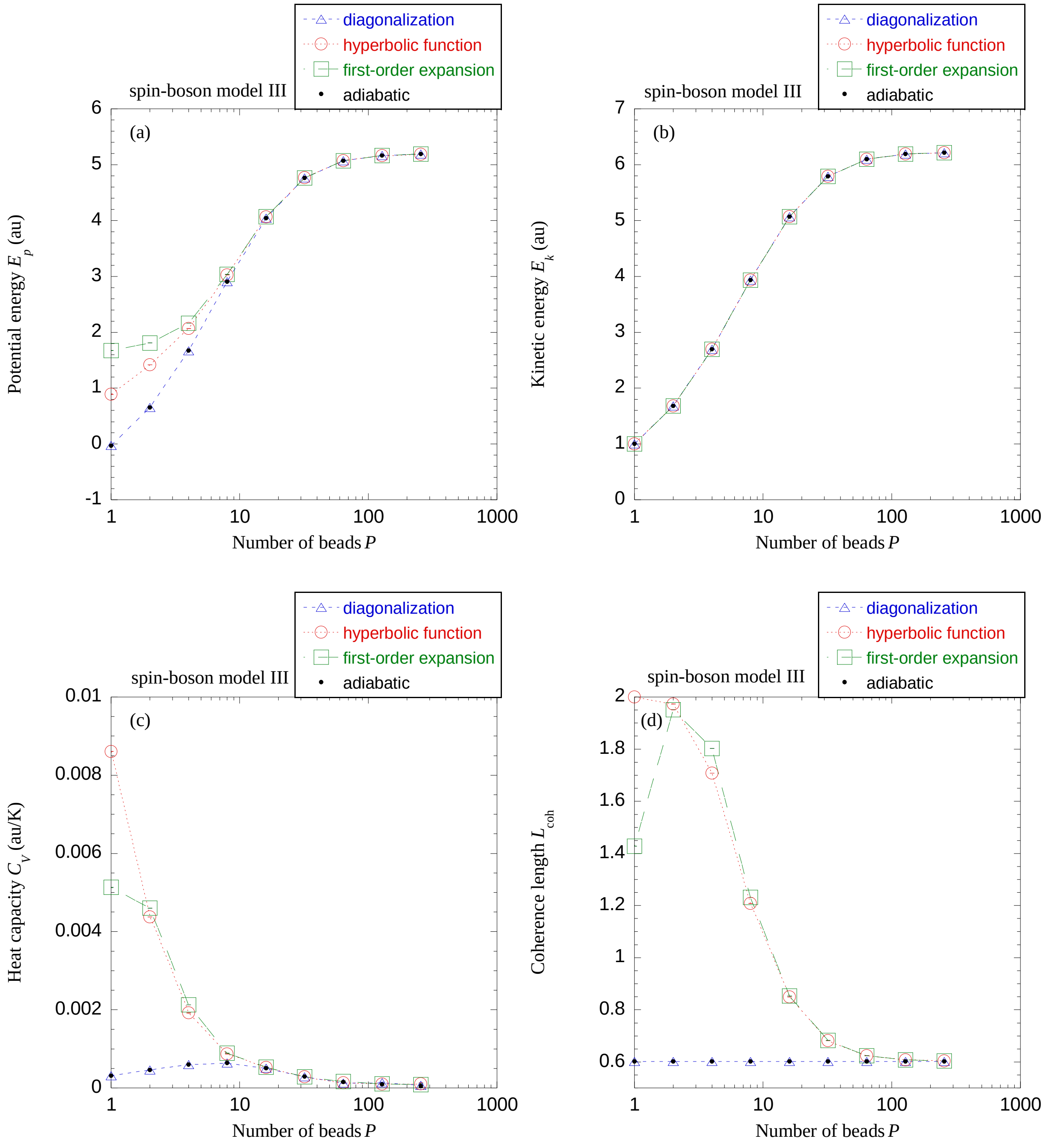
diagonalization
hyperbolic function
first-order expansion
adiabatic
spin-boson model III
(a)
Potential energy $E_p$ (au)
Number of beads $P$
(b)
Kinetic energy $E_k$ (au)
(c)
Heat capacity $C_V$ (au/K)
(d)
Coherence length $L_{coh}$

**Fig. 12**

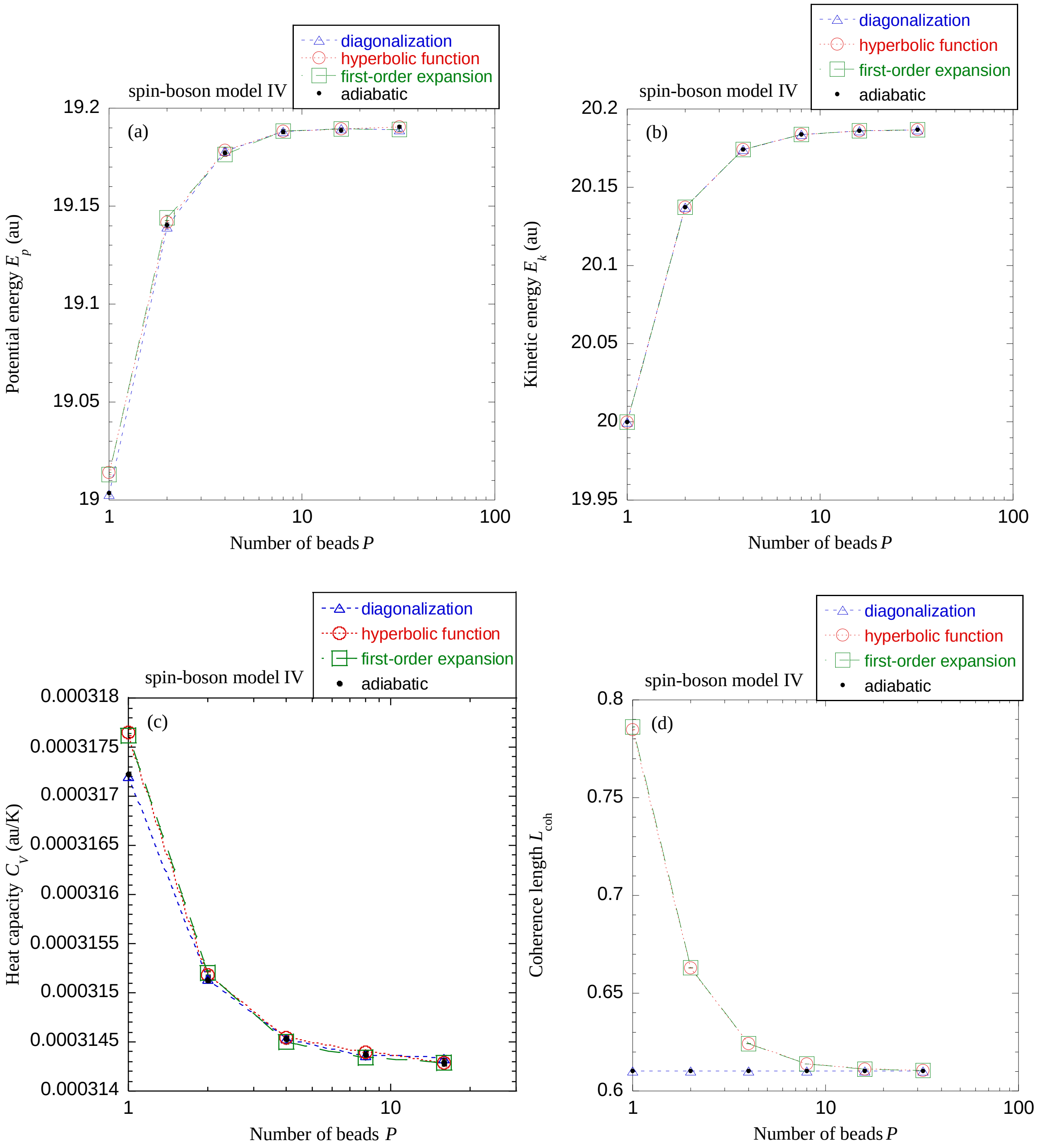

**Fig. 13**

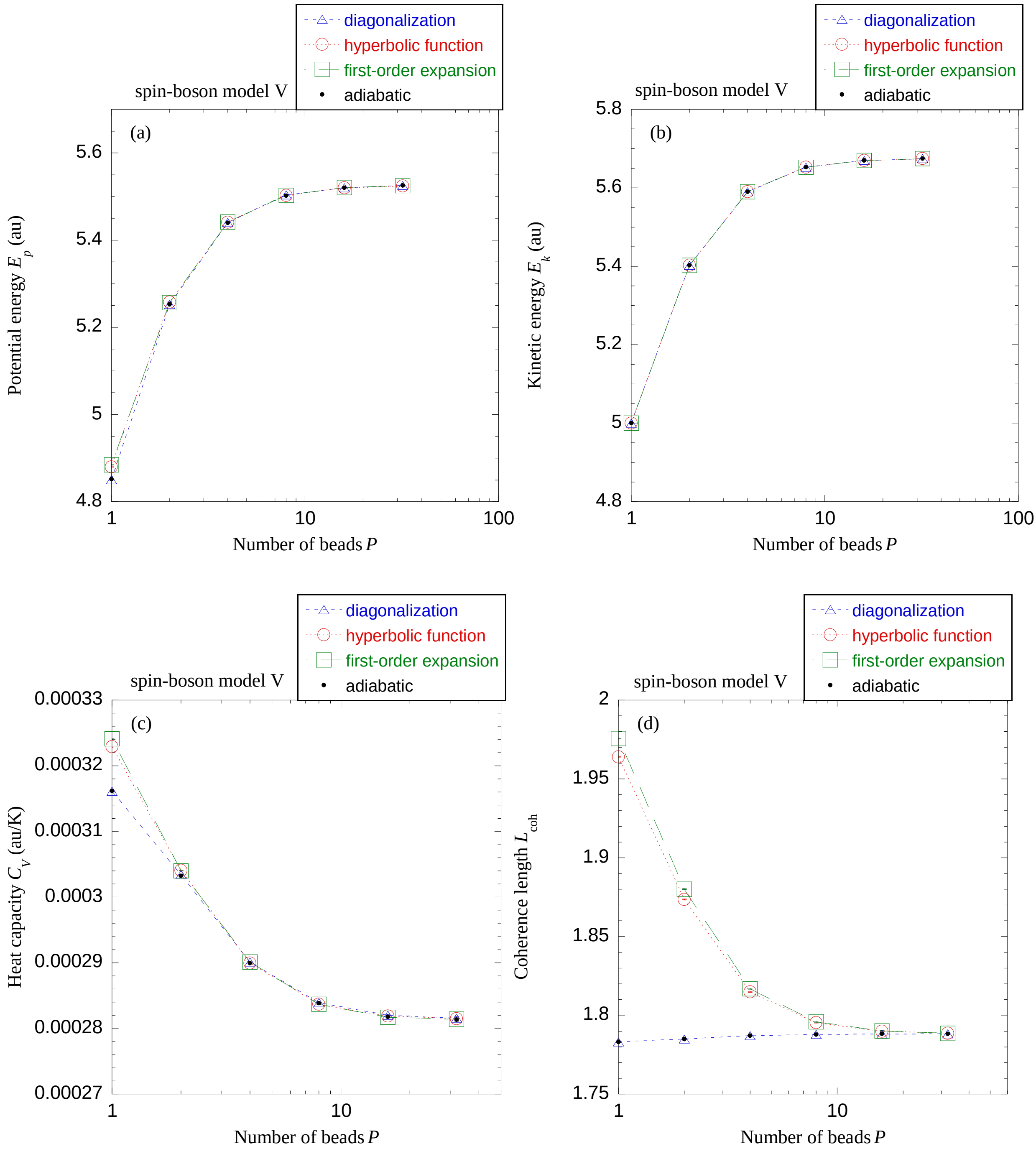

**Fig. 14**

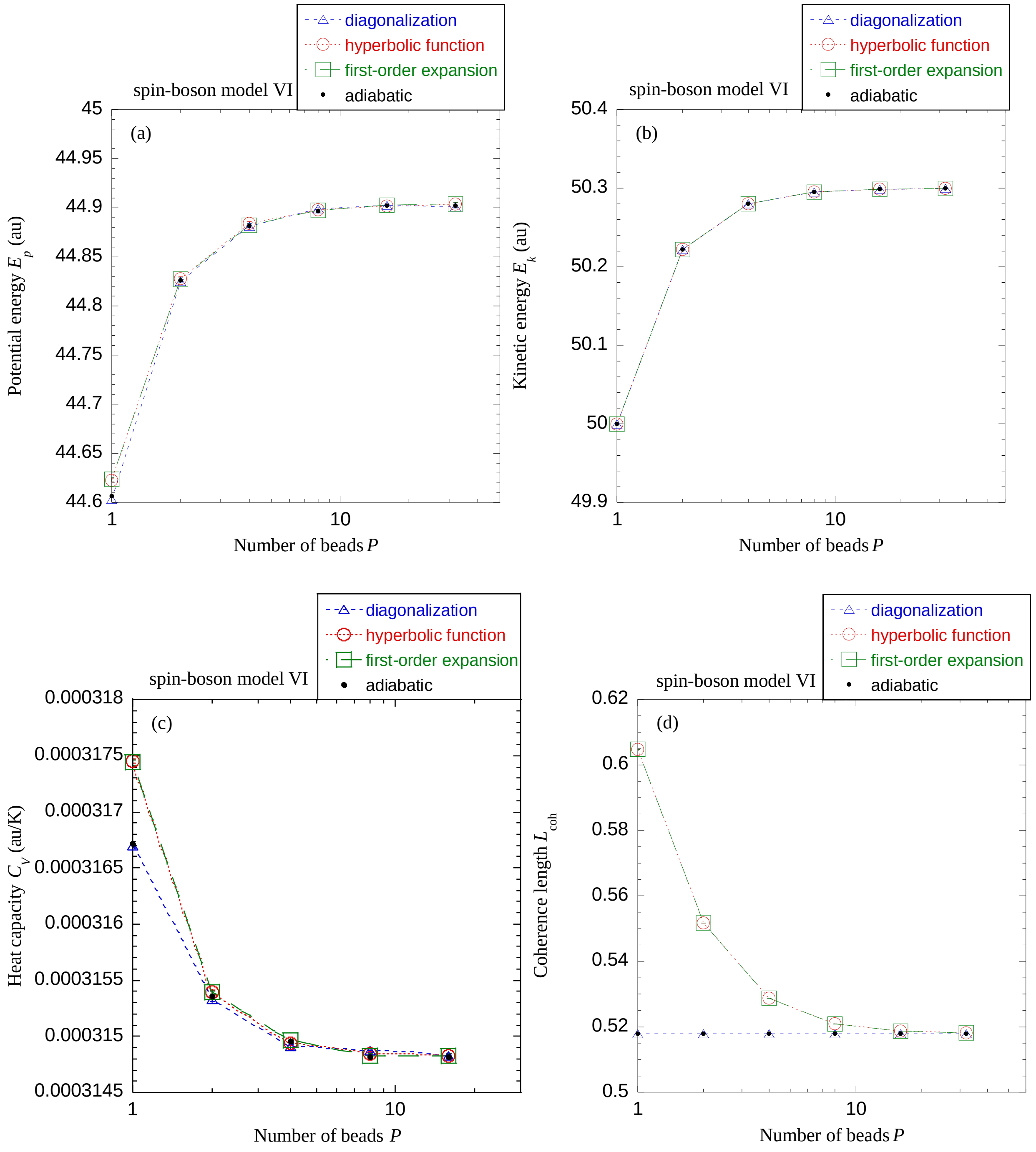

diagonalization
hyperbolic function
first-order expansion
adiabatic
spin-boson model VI
(a)
Potential energy $E_p$ (au)
Number of beads $P$
(b)
Kinetic energy $E_k$ (au)
(c)
Heat capacity $C_V$ (au/K)
(d)
Coherence length $L_{coh}$

**Fig. 15**

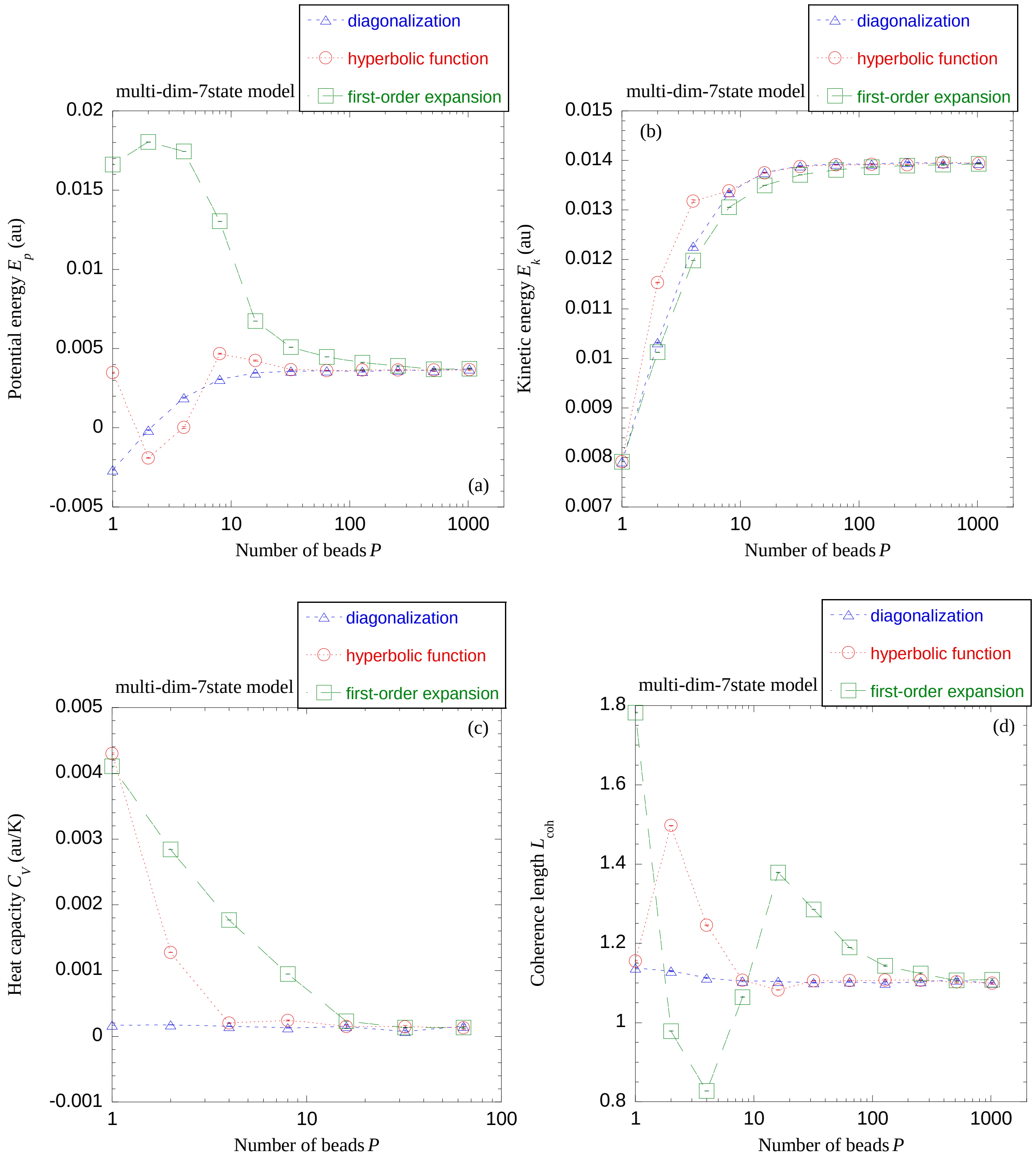

**Fig. 16**

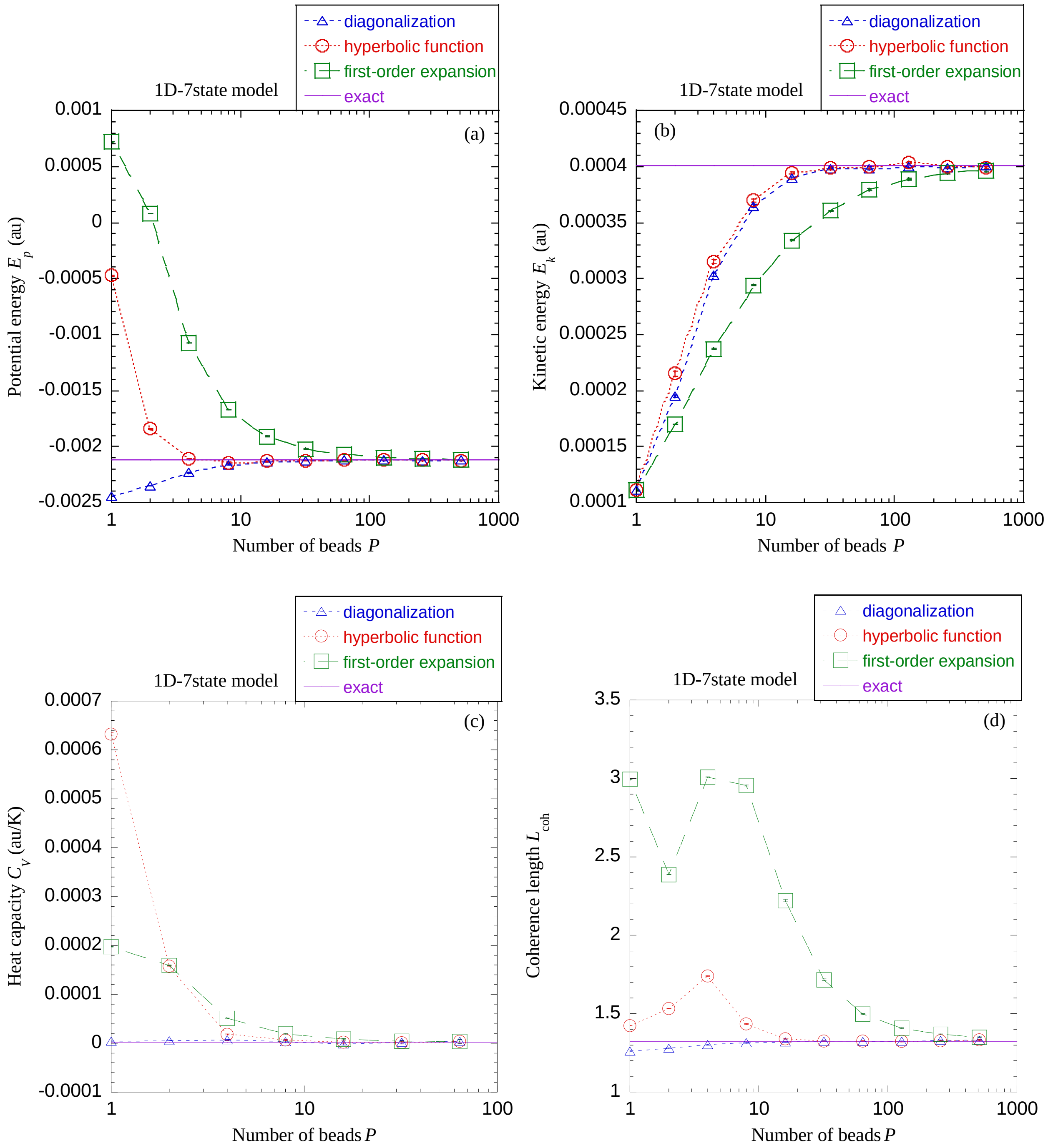

**Fig. 17**

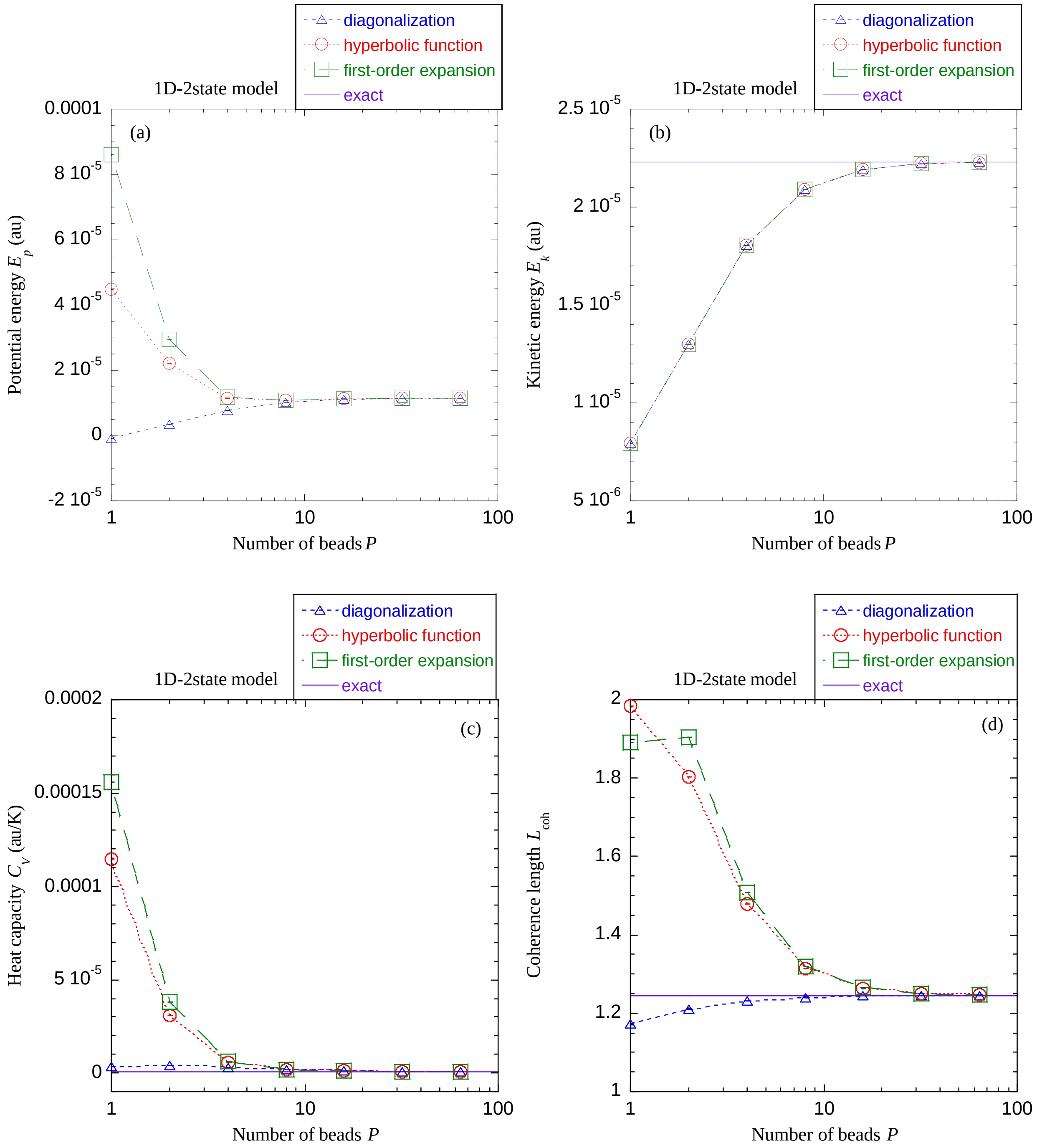

diagonalization
hyperbolic function
first-order expansion
exact
1D-2state model
(a)
Potential energy $E_p$ (au)
Number of beads $P$
(b)
Kinetic energy $E_k$ (au)
Number of beads $P$
(c)
Heat capacity $C_V$ (au/K)
Number of beads $P$
(d)
Coherence length $L_{coh}$
Number of beads $P$

**Fig. 18**

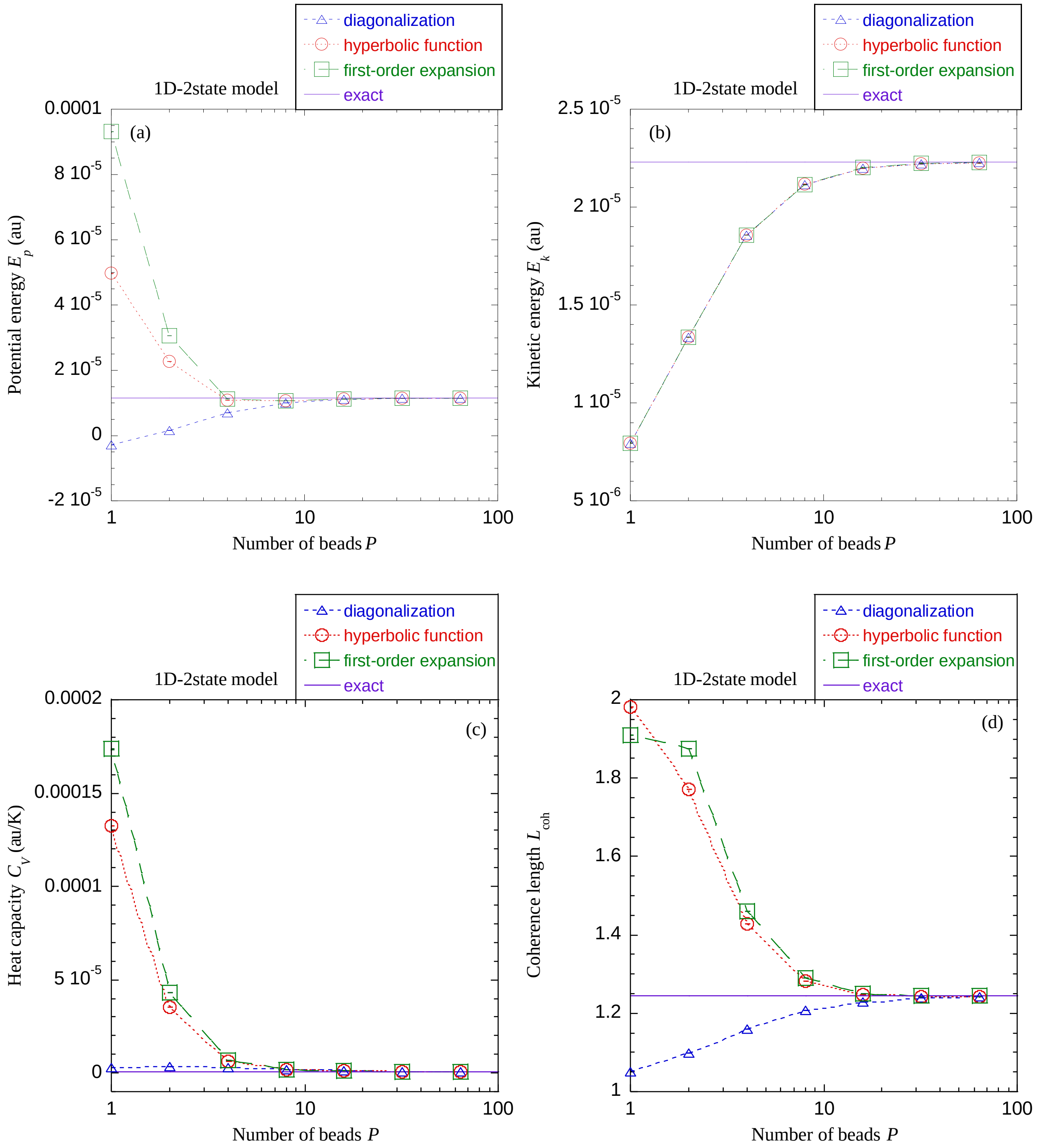
diagonalization
hyperbolic function
first-order expansion
exact
1D-2state model
(a)
Potential energy $E_p$ (au)
Number of beads $P$
(b)
Kinetic energy $E_k$ (au)
(c)
Heat capacity $C_V$ (au/K)
(d)
Coherence length $L_{coh}$

**Fig. 19**

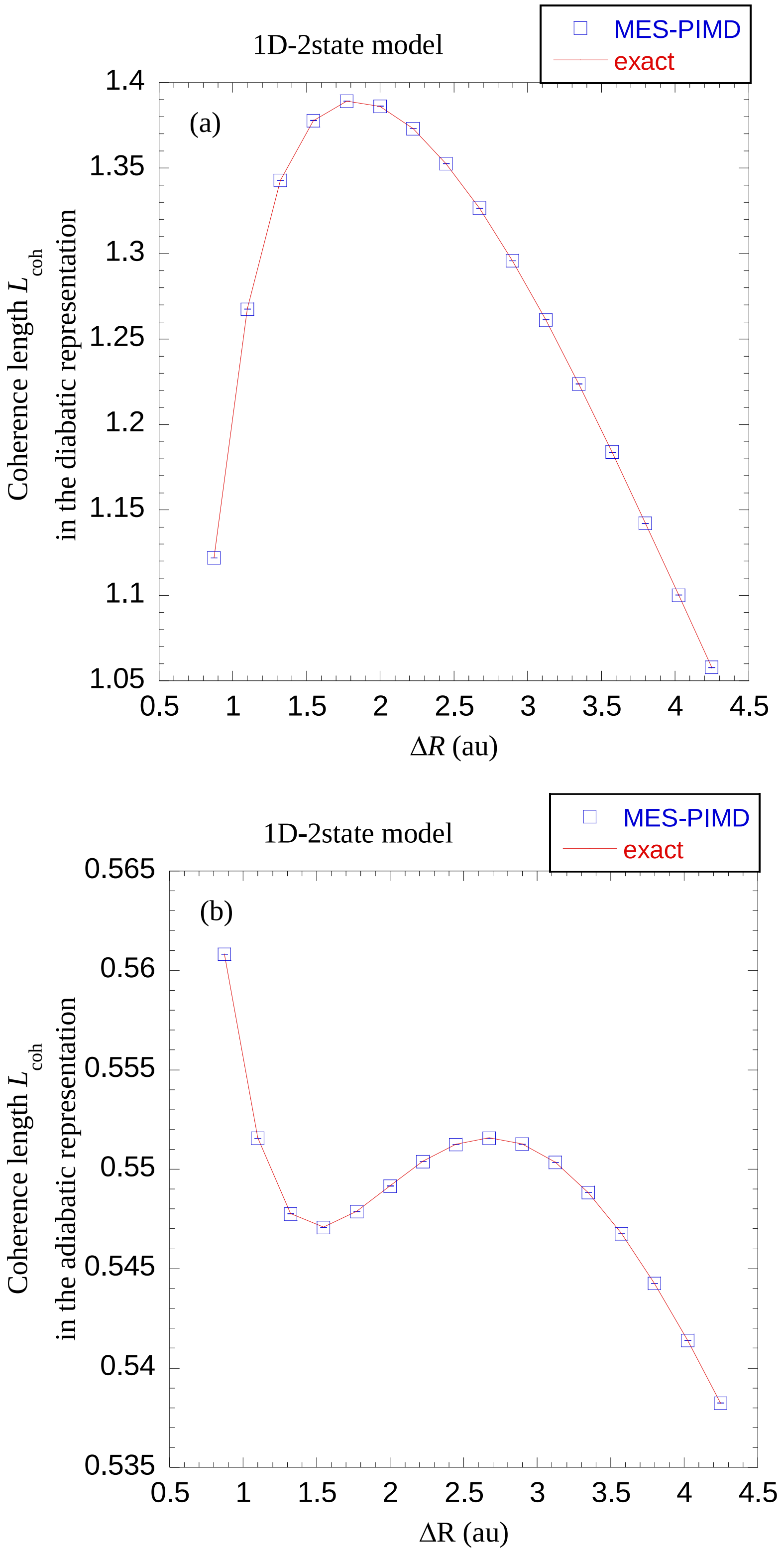

**Fig. 20**